\documentclass{notes}
\AddToHook{cmd/appendix/before}{%
  \crefalias{section}{appendix}%
  \crefalias{subsection}{subappendix}%
}
\title{Functional inequalities \\from the fundamental limits of Bayesian estimation}
\author{
Shuchen Li \\ Yale University \\ \texttt{shuchen.li@yale.edu}
\and
Ilias Zadik \\ Yale University \\ \texttt{ilias.zadik@yale.edu}
}
\date{}

\begin{document}

\maketitle
\begin{abstract}
In this work, we develop a Bayesian information-theoretic framework for proving
functional inequalities for log-concave measures. Motivated by the work of \citet{KlartagOrdentlich}, we establish a unified equivalence between $\Phi$-Sobolev inequalities and Bayesian
estimation lower bounds. We prove that for a large class of smooth convex functions $\Phi$, a log-concave measure $\mu$ satisfies a $\Phi$-Sobolev inequality with constant $C>0$, \emph{if and only if} for $X \sim \mu$, any $t>0,$ and any real-valued measurable $f$, the Bayes risk of estimating $f(X)$ with respect to the $\Phi$-Bregman loss from the Gaussian additive observation $Y_t=\sqrt tX+G, G \sim \cN(0,I_n)$ is at least a $1/(1+Ct)$-fraction of the performance of the trivial estimator. 

We then demonstrate the usefulness of this equivalence by combining
it with direct information-theoretic arguments, such as
testing-to-estimation reductions, to obtain old and new functional inequalities. Our applications include the following. \begin{itemize}
    \item 
For the Poincar\'e inequality, corresponding to
    $\Phi(x)=(x-1)^2$, we give a simple static
    information-theoretic proof that the Poincar\'e constant of an isotropic log-concave measure on $\mathbb{R}^n$ is $O(n^{1/2}).$ This bound was first proven via stochastic localization in the seminal work of \citet{LeeVempala}.

\item For the log-Sobolev inequality, corresponding to
    $\Phi(x)=x\log x$, we prove that all smooth $K$-subgaussian
    log-concave measures $\mu$ satisfy $\LS(\mu) \lesssim K^2n^{1/4}$. This surpasses the recent state-of-the-art bound of $O(K^2n^{1/2})$ by \citet{BizeulLSI}. We remark that \citet{OpenAI-LSI}, in a parallel and independent work, claimed a proof of the optimal bound $\LS(\mu) \lesssim K^2$, as part of their October 6, 2026 release of 722 papers.

\item Finally, we give a Bayesian reformulation of the common structure
underlying the three very recent proofs of the Kannan--Lov\'asz--Simonovits (KLS) conjecture, which
establish a dimension-free Poincar\'e bound for isotropic
log-concave measures~\cite{SongZhang2026,BizeulKlartagLehec2026,
BalasubramanianKasiviswanathan2026}. In particular, tools from the study of the ``low-degree MMSE'' in computational complexity of  statistical inference allow us to motivate and recast some of the key steps and quantities used in these proofs. 
\end{itemize}This work contributes to a broader research program that uses
ideas from Bayesian statistics and information theory as \emph{proof
tools} in probability theory and related areas.

\end{abstract}

\tableofcontents

\section{Introduction}
\label{sec:introduction}

Functional inequalities have been very impactful in high-dimensional probability, theoretical computer science and convex geometry. For instance, they have been instrumental in our understanding of mixing times of Markov chains and diffusions \cite{BakryEmery,DiaconisSaloffCoste1996,BakryGentilLedoux}, the concentration of high-dimensional distributions \cite{GromovMilman1983,BobkovLedoux1997,Ledoux2001}, and the complexity of sampling from log-concave measures~\cite{FriezeKannan1999,LovaszVempala2007,VempalaWibisono2019}.
Two central examples are the family of Poincar\'e inequalities, which compared the variance of a function by its gradient norm, and of log-Sobolev inequalities, which controls instead the entropy of a function \cite{Gross1975}. 

In this work, we will focus on a more general class of functional inequalities, called $\Phi$-Sobolev inequalities, introduced in \cite{Chafai2004}. To define them, let $J\subseteq\R$ be an open interval and let $\Phi:J\to\R$ be smooth with $\Phi''>0$ and $1/\Phi''$ concave. Their so-called \emph{$\Phi$-entropy functional} is given by
\[
 \Ent_\mu^\Phi(f)=\E_\mu\Phi(f)-\Phi(\E_\mu f).
\]This allows us to define the following general class of Sobolev inequalities.\begin{restatable}[$\Phi$-Sobolev inequality]{definition}{phisobolevdef}
We say that a measure $\mu$ satisfies the \emph{$\Phi$-Sobolev inequality} with constant $C$ if it holds
\begin{equation}\label{eq:phi-functional}
 \Ent_\mu^\Phi(f)\leq\frac C2\E_\mu\bigl[\Phi''(f)\|\nabla f\|_2^2\bigr]
\end{equation}
for every $f\in\cA_0^J$\restatablefootnote{$\cA_0^J$ is the set of all shifted compactly supported smooth functions with values in $J$, see \cref{sec:background_functional_ineqs} for the exact definition.}.
\end{restatable}Special important cases of these inequalities are the Poincar\'e and log-Sobolev inequalities mentioned above.

\paragraph{The Poincar\'e inequality and the KLS conjecture} Setting $\Phi(z)=(z-1)^2$ on $J=\R$, the $\Phi$-entropy is $\var_\mu(f)$ and $\Phi''=2$, so \eqref{eq:phi-functional} becomes the celebrated Poincar\'e inequality formally defined as follows.
\begin{restatable}[Poincar\'e inequality]{definition}{poincaredef}
We say that a probability measure $\mu$ over \(\R^n\) satisfies the \emph{Poincar\'e inequality} with constant $C$ if it holds
\begin{equation}\label{eq:Poincare_ineq}
 \var_\mu(f)\leq C\E_\mu\|\nabla f\|_2^2
\end{equation}
for every $f\in\cA_0^\R$.
The \emph{Poincar\'e constant} $\CP(\mu)$ is the least $C\geq0$ such that $\mu$ satisfies the Poincar\'e inequality with constant $C$.
\end{restatable}
Poincar\'e inequalities have been studied extensively for decades
in analysis \cite{Poincare1890,PayneWeinberger1960,Cheeger1970,BrascampLieb1976}, probability \cite{GromovMilman1983,BobkovLedoux1997,Bobkov1999,Ledoux2001,BakryGentilLedoux} and theory of computer science \cite{SinclairJerrum1989,DiaconisStroock1991,Sinclair92,JSTV2004,LovaszSimonovits1993,LovaszVempala2007,AnariLiuOveisGharan2020}. A central application is to sampling: for instance, the Poincar\'e constant controls the convergence rate of the corresponding Langevin diffusion to its invariant measure $\mu$. Under suitable smoothness and oracle-access assumptions, this connection yields Markov-chain sampling algorithms with complexity polynomial in the dimension $n$, the Poincar\'e constant $\CP(\mu)$, and the relevant smoothness, initialization, and accuracy parameters
\cite[Theorem~4 and Corollary~7]{ChenChewiSalimWibisono}.

A particularly important setting is that of log-concave measures.
The celebrated Kannan--Lov\'asz--Simonovits (KLS) conjecture
\cite{KLS1995} asserts that there
exists a universal constant $C>0$ such that every isotropic
log-concave probability measure $\mu$ on $\mathbb R^n$ satisfies
$\CP(\mu)\leq C$, independently of the dimension.

Progress toward this conjecture has driven substantial developments
in high-dimensional geometry and sampling.
Using a localization lemma (called ``needle decomposition'') that reduces suitable high-dimensional
integral inequalities to one-dimensional inequalities,
\citet{KLS1995} established $\CP(\mu)\lesssim n$.
The bound was later improved by Bobkov~\cite{Bobkov2007} combined with
the ``thin-shell'' bound of Gu\'edon and Milman~\cite{GuedonMilman2011} to get
$\CP(\mu)\lesssim n^{5/6}$
\cite{Bobkov2007,GuedonMilman2011}.
In a seminal work, \citet{Eldan2013} introduced the method of stochastic localization, obtaining
$\CP(\mu)\lesssim n^{2/3}\log(2+n)$ and providing a central tool
for subsequent advances.
Developing this approach even further, \citet{LeeVempala} first proved
$\CP(\mu)\lesssim n^{1/2}$, which remained the state of the art until the landmark result of \citet{Chen2021}, who obtained
the first subpolynomial bound $\CP(\mu)\leq n^{o(1)}$.

Subsequently, \citet{KlartagLehec2022} established the first polylogarithmic bound \(\CP(\mu) \lesssim (\log n)^{10}\), which was sharpened by \citet{JambulapatiLeeVempala2022} (\(\CP(\mu)\lesssim(\log n)^{6.4452}\)),
\citet{Klartag2023} (\(\CP(\mu)\lesssim\log n\)) and \citet{Letwin2026} (\(\CP(\mu)\lesssim\sqrt{\log n}\)). Finally, in October 2026 the dimension-free bound conjectured by Kannan, Lov\'asz and Simonovits was established in three very recent works
\cite{SongZhang2026,BizeulKlartagLehec2026,
BalasubramanianKasiviswanathan2026}.

\paragraph{Log-Sobolev inequality and conjecture by Bizeul}

Now, if one uses instead $\Phi(z)=z\log z$ on $J=\R_{>0}$, then $\Phi''(z)=1/z$ and the $\Phi$-entropy is the \emph{entropy}
\[
 \Ent_\mu(f):=\E_\mu[f\log f]-\E_\mu f\log\E_\mu f,
\]
with the convention $0\log0=0$.
Then \eqref{eq:phi-functional} becomes the mentioned log-Sobolev inequality after the substitution $f=g^2$.

\begin{restatable}[Log-Sobolev inequality]{definition}{lsidef}
We say that a probability measure $\mu$ over \(\R^n\) satisfies the \emph{log-Sobolev inequality} (LSI) with constant $C$ if it holds
\begin{equation}\label{eq:lsi-convention}
 \Ent_\mu(f^2)\leq 2C\E_\mu\|\nabla f\|_2^2
\end{equation}
for every $f\in\cA_0^{\R_{>0}}$.
The \emph{log-Sobolev constant} $\LS(\mu)$ is the least $C\geq0$ such that $\mu$ satisfies the log-Sobolev inequality with constant $C$.
\end{restatable}

Since the foundational work of \citet{Gross1975}, logarithmic Sobolev
inequalities have played a central role in analysis
\cite{BakryEmery,Carlen1991,OttoVillani}, probability
\cite{HolleyStroock1987,BobkovGotze1999,BobkovLedoux2000,BakryGentilLedoux},
and theoretical computer science
\cite{DiaconisSaloffCoste1996,FriezeKannan1999,JSTV2004,VempalaWibisono2019}.
For instance, they strengthen Poincar\'e inequalities and imply exponential
decay of relative entropy along the Langevin diffusion
\cite[Chapter~5]{BakryGentilLedoux}.

For general log-concave measures, a natural relatively recent analogue of the KLS
conjecture, posed by \citet{BizeulLSI}, asks whether an additional necessary subgaussian behavior of all
one-dimensional marginals suffices for a dimension-free
log-Sobolev inequality for all isotropic log-concave measures.
More precisely, suppose that $X\sim\mu$ satisfies
\begin{align}\label{eq:subG}\mathbb E e^{\langle u,X-\mathbb EX\rangle}
\leq e^{K^2\|u\|_2^2/2}\end{align} for every $u\in\mathbb R^n$.
The conjecture by
\citet[Conjecture~2]{BizeulLSI} predicts
$\LS(\mu)\lesssim K^2$. A classical result of \citet{Bobkov1999} gives
$\LS(\mu)\lesssim K^2n$.
Building on concentration estimates obtained through stochastic
localization, \citet[Theorem~4]{BizeulLSI} improved this to
$\LS(\mu)\lesssim K^2n^{1/2}$.
In the present work, we propose a general proof framework, that establishes the bound
$\LS(\mu)\lesssim K^2n^{1/4}$.

\subsection{Main contribution}

In this work, motivated by the important prior work \cite{KlartagOrdentlich} we develop a Bayesian estimation framework to prove functional inequalities. Our first main result is a general equivalence that translates any $\Phi$-Sobolev inequality to a Bayesian estimation task. Specifically, we prove that for any log-concave measure and any convex function $\Phi$ as above, a measure $\mu$ satisfies a $\Phi$-Sobolev
inequality if and only if a family of Bayesian estimation tasks under Gaussian noise is impossible to be solved. 

We then turn this equivalence into a proof method: we use direct information-theoretic arguments
to prove these Bayesian tasks are impossible and hence yield functional inequalities for
the underlying measure.

\paragraph{The Bayesian estimation task} 
In this framework, we fix a function $\Phi$ as above. The measure of interest $\mu$ serves as the ``prior''
distribution of an unknown signal $X$. Specifically, assume a statistician observes the
signal through a Gaussian additive model,
\begin{align}\label{eq:GAM}
 X\sim\mu,\qquad Y_t=\sqrt t\,X+G,\qquad
 G\sim\gamma_n=\cN(0,I_n)\ \text{independent of }X,
\end{align}
where $t\geq0$ is the signal-to-noise ratio controlling how strong the signal $X$ is in the observation $Y_t$.
Now for each scalar observable $f: \mathbb{R}^n \rightarrow \mathbb{R}$, the task of the statistician is to estimate the one-dimensional quantity $f(X)$ from $Y_t$. 

The exact loss we measure success will change depending on the convex function $\Phi$ and it will be the Bregman loss associated with $\Phi$ (see \cref{dfn:bregman} for exact definition). We remark that for $\Phi(z)=(z-1)^2$ this is the usual quadratic loss, while for $\Phi(z)=z\log z$ it becomes the entropy loss. Of special importance to us will be the optimal Bregman loss achieved by any estimator given any SNR $t$ and any scalar function $f$, which we denote by $\mathcal R_\Phi(t,f)$ (see \cref{dfn:bregman-optimal} for exact definition).

\subsubsection{The main equivalence}
As mentioned above, our first main result is a general equivalence for $\Phi$-Sobolev inequalities and the above Bayesian estimation tasks. Specifically we prove the following.
\begin{theorem}[Informal version of \cref{thm:general-equivalence}]\label{thm:informal-equivalence}
    For any smooth $\Phi$ with $\Phi''>0$ and $1/\Phi''$ concave, and any log-concave measure $\mu$ the following are equivalent:
\begin{itemize}
\item[(a)] The measure $\mu$ satisfies a $\Phi$-Sobolev inequality \eqref{eq:phi-functional} with constant $C.$
\item[(b)] Consider the GAM \eqref{eq:GAM} for prior $\mu$. Then for every $t>0$ and every $f\in\cA_0^J$, the optimal Bregman loss w.r.t.\ $\Phi$ satisfies
\[
 \mathcal R_\Phi(t,f)\ge
 \frac{\operatorname{Ent}_\mu^\Phi(f)}{1+Ct}.
\]
\end{itemize}
\end{theorem} 

We remark here that one should interpret $\operatorname{Ent}_\mu^\Phi(f)$ as \emph{the statistically trivial Bregman loss} associated with using as an estimator the average of $f$ under the prior $\mu$; that is the loss of an estimator ignoring the observations $Y_t$ (see \cref{rem:trivial_loss} for a more detailed explanation). Hence, the second part of the equivalence should be understood as saying that the optimal loss is at least, roughly, a $1/Ct$ fraction of the trivial Bregman loss. 

Moreover, it is worth noticing a useful corollary of our equivalence, which helps operationalize this framework. If one proves that for some $b>0,$
$\mathcal R_\Phi(T,f)\ge b\,\operatorname{Ent}_\mu^\Phi(f)$
for every $f$ at a fixed SNR $T$, then $\mu$ automatically satisfies the $\Phi$-Sobolev inequality with constant
at most $(bT)^{-1}$.

As mentioned above the equivalence is motivated and based on the works of \citet{KlartagOrdentlich}, and multiple versions of it are already known or can be relatively easily extracted from established results in the literature. Therefore, our contribution should be more understood as a unified treatment of the general case. 

In more detail, first, under the stated assumptions on $\Phi$, the implication from the $\Phi$-Sobolev inequality to the risk bound follows from established contraction estimates for $\Phi$-divergences under the Gaussian channel \cite{ChenChewiSalimWibisono,MitraWibisono,KlartagOrdentlich}, together with the conditional $\Phi$-entropy decomposition; see also the posterior formulation in \cite{ShiTianZhang}. The reverse direction appears not to be known in that generality. The Poincar\'e case follows explicitly from the bounds on \cite[Theorem 1.1]{KlartagOrdentlich}, but also the variance-conservation argument in \cite[Section 3.1.1]{ChenEldan} using the Bayesian interpretation of stochastic localization by \citet{ElAlaouiMontanari}. The log-Sobolev case follows by combining a lower bound on the expected posterior entropy with the Bakry--\'Emery inequality for the strongly log-concave Gaussian posteriors; this combination is an easy exercise for example if one uses the entropy decomposition and posterior estimate in \cite[Section 4, Lemma 33]{BizeulLSI}. In both cases, a uniform risk lower bound over the relevant class of labels at a single positive signal-to-noise ratio suffices to recover the corresponding functional inequality. Moreover, assuming a technical conjecture, \citet{KlartagOrdentlich} establish the reverse direction for $\Phi(z)=z^\lambda-1$ with $1<\lambda<2$. We prove the reverse implication with a direct argument based on the posterior of the GAM, which avoids proving their technical conjecture and covers every convex $\Phi$ under the stated assumptions.

Stepping away from Gaussian channels, in a very influential work \citet{Raginsky2016} establishes very related implications from Bayesian estimation tasks to (channel-associated variants of) $\Phi$-Sobolev inequalities for discrete channels, under suitable assumptions on $\Phi$. In some sense, one can interpret our result to aim to make a similar case for the case of log-concave measures and Gaussian Bayesian estimation tasks.

\subsubsection{Functional inequalities from estimation lower bounds}

In the remainder of this work, we use the equivalence as a tool for proving functional inequalities. Although several components of the equivalence are established in the literature, to the best of our knowledge, the reverse implication has seen only relatively limited operational use to prove such inequalities via information theory. Our focus is on obtaining functional inequalities from direct information-theoretic lower bounds on Bayesian estimation risk.

That being said, an important precedent is the aforementioned stochastic localization method \cite{Eldan2013}, a central tool for proving functional inequalities that admits a Bayesian interpretation according to the important observation of \citet{ElAlaouiMontanari}, and therefore can be interpreted as an information-theory tool to prove functional inequalities. We explore how other, perhaps more classical (e.g., the I-MMSE formula \cite{GuoShamaiVerdu} or testing lower bounds), methods from information theory and statistical estimation can be used for the same purpose. In particular, we show how direct estimates at a single signal-to-noise ratio (also called ``static'' methods, as opposed to the dynamic nature of stochastic localization) can yield functional inequalities through the reverse implication. Our aim is to demonstrate the usefulness of this approach and hopefully reveal even more the deep connection between techniques from convex geometry, sampling, and information theory.

The proposed high-level proof strategy consists of two steps:
\begin{quote}
\begin{itemize}
    \item[(i)] \emph{Use direct information-theoretic arguments to establish a uniform lower bound on the Bayes risk for the estimation tasks in part~(b) of \cref{thm:informal-equivalence}, with prior $\mu$, at a single positive signal-to-noise ratio.}
    \item[(ii)] \emph{Apply the reverse implication of the equivalence to deduce the corresponding $\Phi$-Sobolev inequality for $\mu$.}
\end{itemize}
\end{quote}

\paragraph{The detection-to-estimation reduction}
A classical tool in information theory and statistics for establishing estimation lower bounds, such as those in part~(i), is what we refer to as a \emph{detection-to-estimation reduction}: one can obtain estimation lower bounds by controlling the statistical distinguishability of a planted model from a suitable null model, where the estimation lower bounds are trivial (or, at least, much easier) to obtain. This approach follows the classical statistical principle of deriving estimation lower bounds from testing lower bounds, exemplified by the classical Le Cam's method and Fano's inequality \cite{Yu1997}. This approach is rather useful, as many more tools are available in the information-theory community to prove detection lower bounds rather than estimation lower bounds, such as (truncated) second-moment estimates for the squared norm of the likelihood ratio $\mathrm{d} P_t/\mathrm{d} Q$, which we later use. 

More specifically, our goal is to show that the observation $Y_t=\sqrt{t}X+G$, with $X\sim\mu$ and $G\sim\gamma_n$ independent, provides limited information about $f(X)$, as measured by the Bayes risk $\mathcal R_\Phi(t,f)$, uniformly over the relevant class of labels $f$. Write $P_t=\Law(X,Y_t)$ for the planted prior-observation law. One can then aim to define a null prior-observation law $Q_t$ which is close to $P_t$ but also under which the estimation lower bound is easier to prove. 

For instance, a useful and classical null model, employed throughout our applications but also in the literature (see e.g., \cite[Section 3.6]{Zadik2026}), retains the prior $X\sim\mu$ but replaces the observation by an independent standard Gaussian. Its observation law is therefore $\gamma_n$, independent of $t$; in this case we will simply refer to this null law by $Q$. \emph{Under this null model}, observing $Y_t$ provides no information about $X$, so the optimal risk is exactly the trivial risk (see \cref{rem:trivial_loss}):
\[
\inf_h \mathbb E_{\mu\otimes \gamma_n}
\bigl[B_\Phi(f(X),h(Y))\bigr]
=
\operatorname{Ent}_\mu^\Phi(f).
\]So if we have also established, for example, that $P_t$ and $Q$ are sufficiently close to each other, then the property would be also true for the planted law as well, completing the desired proof. 

The general strategy is then to do the following steps.
\begin{enumerate}[label=(Step~\arabic*), ref=Step~\arabic*, leftmargin=*]
  \item\label{step:1} Define a null model $Q_t$, under which the optimal risk of estimating $f(X)$ is large enough.
  \item\label{step:2} Use a quantitative detection-estimation reduction to reduce the desired lower bound on $\mathcal R_\Phi(t,f)$ under \(P_t\) to an appropriate closeness between $P_t$ and $Q_t$.
  \item\label{step:3} Bound a suitable divergence between the planted law $P_t$ and an appropriately defined null law $Q_t$ to establish the desired closeness for \ref{step:2}.
\end{enumerate}

\paragraph{Poincar\'e inequalities}
Our first application of this proof framework is to prove Poincar\'e inequalities for log-concave measures. 
In that case, our first result (\cref{cor:poincare-mmse}) combines our equivalence for $\Phi(z)=(z-1)^2$ with Cheeger's inequality \cite{Cheeger1970} to prove that if for a single SNR $T>0$ and a constant $b>0$ it holds \[\mmse_T(\mathbf1_A)\geq b\] for every set $A$ with $\mu(A)=1/2$, then $\CP(\mu)\lesssim1/(b^2T)$.

We next use this sufficient condition to show how one can obtain short information-theoretic proofs of two classical bounds toward the KLS conjecture (\cref{thm:intro-kls-lv}). Specifically, let us fix a log-concave probability measure $\mu$ and write $\Sigma=\operatorname{Cov}(\mu)$. First, we use a direct approach; a single application of Pinsker's inequality (\cref{lem:intro-mmse-comparison}) combined with Jensen's inequality (\cref{lem:mi-null}), yields an MMSE lower bound which recovers the classical estimate of Kannan--Lov\'asz--Simonovits \cite{KLS1995} (see part (i) of \cref{thm:intro-kls-lv}):
\[
 C_{\mathrm P}(\mu)\lesssim\operatorname{tr}\Sigma.
\] We remark that their original argument to establish this bound uses a geometric ``needle decomposition'' localization approach, reducing the isoperimetric problem to one-dimensional inequalities. An alternative proof, based on the Pr\'ekopa--Leindler inequality, was subsequently given by \citet[Theorem~1.2]{Bobkov1999}. To the best of our understanding, both these approaches use substantially different ingredients from our argument.

Next, we sharpen the MMSE lower bound using the detection-to-estimation reduction described above. In that case, we consider the simple null model mentioned above with joint law $\mu\otimes\gamma_n$, under which estimating any balanced indicator $\mathbf1_A$ has error exactly $1/4$ (establishing \ref{step:1} of the detection-estimation reduction). Combining Pinsker's inequality (\cref{lem:pinsker-mmse}) with the entropy contraction implied by the Gaussian log-Sobolev inequality yields, for every $q\in(0,1]$ and every $A$ with $\mu(A)=1/2$,
\[
 \mmse_{qt}(\mathbf1_A)
 \geq \frac14-\frac q2
 \left[\log 2+\log\bigl(1+\chi^2(Y_t\Vert \gamma_n)\bigr)\right],
\]
see \cref{lem:intro-mmse-comparison}.\footnote{%
Here and below, we use a random vector to denote its law. For instance, $\chi^2(Y_t\Vert\gamma_n)$ is the $\chi^2$-divergence between the law of $Y_t$ and $\gamma_n$, see \cref{sec:bayesian-estimation}.} This inequality establishes \ref{step:2} of our reduction as now a universal bound on $\chi^2(Y_t\Vert \gamma_n)$ gives, upon choosing a sufficiently small constant $q$, a uniform positive MMSE lower bound at SNR $qt$. Our criterion then yields $C_{\mathrm P}(\mu)\lesssim 1/t$.

Now, in terms of bounding the $\chi^2$-divergence, likelihood-ratio second moments for Gaussian additive models admit known explicit simple formulas in terms of the overlap between independent samples from the prior; see \cite[Proposition~2.5]{KuniskyWeinBandeira} and \cite[Lemma~6.3]{Zadik2026}. Therefore, we then apply these formulas\footnote{Specifically, a conditional version of this second-moment calculation, to exclude a ``bad'' rare event.} to obtain that $\chi^2(Y_t\Vert \gamma_n)=O(1)$ for $t \lesssim 1/\sqrt{\operatorname{tr}(\Sigma^2)}$, which improves the preceding bound to (see part (ii) of \cref{thm:intro-kls-lv}):
\[
 C_{\mathrm P}(\mu)\lesssim
 \sqrt{\operatorname{tr}(\Sigma^2)},
\]
recovering the bound of \citet{LeeVempala}. We remark that historically this bound remained the best known general ``KLS'' bound for almost four years until Chen's breakthrough \cite{Chen2021}. Their proof uses substantially different technical ingredients than ours: Lee and Vempala control the covariance process along stochastic localization, whereas our argument uses direct ``static'' information inequalities and conditional second-moment estimates at a fixed signal-to-noise ratio.

\paragraph{Log-Sobolev inequalities: part I}

Our second and main application concerns obtaining log-Sobolev inequalities via our framework.
In particular, we assume that $\mu$ is a log-concave measure that has a smooth convex potential and satisfies subgaussianity \eqref{eq:subG}.
In that case, our first result (\cref{cor:lsi-information}) combines our main equivalence for $\Phi(z)=z\log z$ with a key characterization relating the contraction of the relative entropy\footnote{%
For probability measures $\zeta\ll\eta$, the relative entropy, or Kullback--Leibler divergence, is $\KL(\zeta\Vert\eta)=\int\log\frac{\mathrm{d}\zeta}{\mathrm{d}\eta}\,\mathrm{d}\zeta$, and $\KL(\zeta\Vert\eta)=\infty$ otherwise; see \cref{def:kl}.}
and mutual information by \citet{PolyanskiyWu}, to prove that if for a single SNR $T>0$ and a constant $b>0$ it holds\footnote{%
We recall that for a randomized binary label $U$, the Markov chain $U-X-Y_t$ gives $I(U;X\mid Y_t)=H(U\mid Y_t)-H(U\mid X)$, where $H$ is the Shannon entropy of the binary random variable $U$.
We direct the reader to \cref{def:entropy,def:mutual-information} for exact definitions.}
\[I(U;X|Y_T)\geq b I(U;X)\]
for every randomized binary label $U$ of $X$ (see \cref{dfn:random_label} for the exact definition), then $C_{\mathrm{LS}}(\mu)\lesssim1/(bT)$.
Hence, a log-Sobolev constant bound is equivalent with impossibility of estimation of randomized binary labels of the signal $X$ in the mutual information error $I(U;X|Y_T)$.
We remark that while the mutual information estimation perspective helps with the intuition, from a technical standpoint an equivalent \cite[Theorem~4]{PolyanskiyWu} and useful way of checking the condition is that for every other prior $\zeta\ll \mu$ with bounded density $\mathrm{d}\zeta/\mathrm{d}\mu$, if $Y_{t,\zeta}$ is the GAM observation for prior $\zeta,$ then it holds $\KL(Y_{t,\zeta}\Vert Y_t)\leq (1-b)\KL(\zeta\Vert\mu)$ (see also \cref{lem:label-contraction}).

First, to apply the sufficient condition, we again use a detection-to-estimation reduction with the independent null model $\mu\otimes\gamma_n$, as above. Under this null model, the observation reveals no information about any randomized label, so the mutual-information error equals the trivial $I(U;X)$. This completes \ref{step:1} of the proposed reduction.

For \ref{step:2}, we set
\[
\Delta_t=\log\bigl(1+\chi^2(Y_t\Vert\gamma_n)\bigr),
\]
and prove in \cref{thm:output-information} that, for a universal constant $c>0$, every randomized binary label $U$ satisfies for $s =c\frac{t}{1+\Delta_t}$
\[
I(U;X\mid Y_s)\ge 10^{-3}I(U;X).
\] Our sufficient condition therefore gives
\[
C_{\mathrm{LS}}(\mu)\lesssim\frac{1+\Delta_t}{t}.
\]

Finally, for \ref{step:3}, after centering $\mu$, a direct second-moment calculation using directional subgaussianity (\cref{lem:output-chi2}) yields $\Delta_t\lesssim nK^4t^2$ whenever $tK^2$ is sufficiently small. Taking $t\asymp K^{-2}n^{-1/2}$ gives
\[
C_{\mathrm{LS}}(\mu)\lesssim K^2\sqrt{n},
\]which is the bound due to \citet{BizeulLSI}. Similar to above, Bizeul's proof is based on different ingredients than our proof, such as a two-level concentration estimate obtained through stochastic
localization and Milman's concentration-to-log-Sobolev principle
\cite[Section~2.4]{BizeulLSI}.

We highlight that, as opposed to the case of Poincar\'e inequalities above, the proof of \cref{thm:output-information} for \ref{step:2} of the reduction is not elementary and requires a substantial contraction argument, stated in \cref{lem:restricted-contraction}. While versions of this lemma have appeared already in the literature \cite[Theorem~1.1]{BartheMilman}, we are not aware of such a statement in that generality, and we think it could be of independent interest. 

The proof of the contraction lemma proceeds in two steps, first establishing it under a technical support condition on the two priors, and then extending it to all priors. The first part is a direct computation using entropy duality and the transportation inequality implied by a log-Sobolev inequality \cite[Theorem~1]{OttoVillani}. The extension is longer and more technical: it splits the likelihood ratio into moderate and large values, and uses, among other tools, the bounded-likelihood-ratio comparison of \cite[Theorem~9]{SasonVerdu2016} and Cheeger's inequality for the reversible posterior-resampling kernel \cite[Theorem~2.1]{LawlerSokal1988}.

\paragraph{Log-Sobolev inequalities: part II}

Next, we sharpen the mutual-information error lower bound by introducing
a new null model for the detection-to-estimation reduction.
For the GAM with prior $\mu$ and SNR $t>0$, denote the posterior law by
\[
\mu_{t,y}=\Law(X\mid Y_t=y).
\]

The construction begins with a posterior-averaging identity
(see \eqref{eq:posterior-average} below), a direct consequence of Bayes' rule
often referred to as a Nishimori identity: for a sample from the observation law $Y'\sim \Law(Y_t)$,
and a sample from the posterior $X'\sim\mu_{t,Y'}$ conditionally on $Y'$, the marginal
law of $X'$ equals the prior law $\mu$. This leads to the interesting consequence that the planted law can be generated
as the $(X,Y_t)$ marginal of the following experiment:
\begin{itemize}
    \item[(i)] Draw an observation $Y'\sim \Law(Y_t)$.
    \item[(ii)] Draw a posterior sample $X\sim\mu_{t,Y'}$.
    \item[(iii)] Draw $G\sim\gamma_n$ independently of $(Y',X)$
    and form the fresh GAM observation $Y_t=\sqrt{t}X+G$.
\end{itemize}

This representation suggests a family of candidate null models that could potentially be close to the planted law:
fix any probability measure
$\xi$ on $\mathbb{R}^n$ that could intuitively be close to the planted observation law $\Law(Y_t)$, and any $s\ge0$. Then, define the candidate null law as the $(X,Y_s)$ marginal of the following experiment:
\begin{itemize}
    \item[(i)] Draw $Y'\sim\xi$.
    \item[(ii)] Draw a posterior sample $X\sim\mu_{t,Y'}$.
    \item[(iii)] Draw $G\sim\gamma_n$ independently of $(Y',X)$
    and form the fresh GAM observation $Y_s=\sqrt{s}X+G$.
\end{itemize}
Equivalently, this is a GAM at SNR $s$ with prior
\[
\nu_{t,\xi}:=\int_{\mathbb{R}^n}\mu_{t,y}\,\xi(dy).
\]
In particular, choosing $\xi=\Law(Y_t)$ gives $\nu_{t,\xi}=\mu$ by
posterior averaging, and taking $s=0$ recovers the simple null law
$\mu\otimes\gamma_n$. Yet, other choices of $\xi$ and $s$ can in principle provide
other useful null models.

For such a null model to be useful, it must satisfy the estimation lower
bound (\ref{step:1}) required by the reduction. 
Since this null model is a GAM, by our
main equivalence, $\nu_{t,\xi}$ must satisfy a log-Sobolev inequality
with $sC_{\mathrm{LS}}(\nu_{t,\xi})\lesssim1.$

Since $\mu$ is log-concave, each posterior $\mu_{t,y}$ is
$t$-strongly log-concave, and the Bakry--\'Emery criterion gives
$C_{\mathrm{LS}}(\mu_{t,y})\le\frac1t.$ This suggests taking $\xi$ to be a Dirac mass, which immediately
provides the desired estimation lower bound whenever $s\lesssim t$. However, a single posterior appears not to provide the closeness to the
planted law required by \ref{step:2} of our reduction. We therefore consider another
natural choice, the standard Gaussian law $\xi=\gamma_n$, which
produces the key mixture for our results,
\begin{equation}\label{eq:reference-law}
    \nu_t
    :=\int_{\mathbb{R}^n}
    \mu_{t,y}\,\gamma_n(\mathrm{d}y).
\end{equation} Thus, for this choice of $\xi$, the null model is the joint law of $(X,Y_s)$, where
$X\sim\nu_t$ and $Y_s=\sqrt{s}X+G$, with $G\sim\gamma_n$ independent
of $X$. With $t$ fixed, we call this the \emph{null-posterior model}
at SNR $s$, on which we focus from now on.

Our first result verifies \ref{step:1} of the reduction: under the
null-posterior model, every randomized binary label $U$ of $X$ satisfies
\[
I(U;X\mid Y_s)
\ge \frac{t}{2(t+s)}\,I(U;X),
\qquad s>0,
\]
where both mutual informations are computed under the null law.
In particular, we also obtain the log-Sobolev bound $
C_{\mathrm{LS}}(\nu_t)\le\frac{2}{t}.$ See \cref{prop:reference-lsi} for the exact statement. The proof is based on a direct posterior analysis, combining Tweedie-type identities with the Gaussian
log-Sobolev inequality and the Bakry--\'Emery criterion.

For \ref{step:2}, we define
\[
\Delta'_t:=\log\bigl(1+\chi^2(\mu\Vert\nu_t)\bigr).
\]Notice that at the same SNR $s$ the planted and null-posterior models are now GAMs and therefore have
the same conditional observation law given $X$ differing only
in their priors. Consequently, their joint likelihood ratio is
$d\mu/d\nu_t(X)$, and the $\chi^2$-divergence between their joint
laws is exactly $\chi^2(\mu\Vert\nu_t)$. 

Applying our relative-entropy contraction estimate with reference
law $\nu_t$, together with $C_{\mathrm{LS}}(\nu_t)\le 2/t$,
we show in \cref{thm:reference-information}, for a universal constant $c>0$ and
\[
s=\frac{ct}{1+\Delta'_t},
\]
every randomized binary label $U$ of $X\sim\mu$ satisfies
\[
I(U;X\mid Y_s)\ge 10^{-3}I(U;X).
\]
Our sufficient condition therefore yields
\[
C_{\mathrm{LS}}(\mu)\lesssim\frac{1+\Delta'_t}{t}.
\] We remark that there is a non-trivial advantage of this comparison compared to the previous null model; as we discussed, applying the same
posterior-sampling kernel $y\mapsto\mu_{t,y}$ to $\Law(Y_t)$
and $\gamma_n$ gives $\mu$ and $\nu_t$, respectively:
\[
\E_{y\sim Y_t}\mu_{t,y}=\mu,
\qquad
\E_{y\sim\gamma_n}\mu_{t,y}=\nu_t.
\]
Thus, the data-processing inequality gives
\[
\chi^2(\mu\Vert\nu_t)
\le \chi^2(Y_t\Vert\gamma_n).
\]
The null-posterior comparison therefore never increases the
divergence cost, and the resulting log-Sobolev bound is at least
as strong, up to universal constants, as the bound obtained
using the simple null law.

Finally, for \ref{step:3}, in \cref{prop:likelihood-L2} via a Herbst argument using the log-Sobolev
inequality for $\nu_t$ we show
\[
\Delta'_t\lesssim nK^6t^3,
\qquad 0<t\le \frac{1}{2K^2}.
\]
Consequently,
\[
C_{\mathrm{LS}}(\mu)
\lesssim \frac{1}{t}+nK^6t^2.
\]
Choosing $t\asymp K^{-2}n^{-1/3}$, with a sufficiently small
universal proportionality constant, yields the improved log-Sobolev bound,
\[
C_{\mathrm{LS}}(\mu)\lesssim K^2n^{1/3}.
\]

To the best of our knowledge, the measure $\nu_t$ has received
relatively little attention in the literature on log-Sobolev
inequalities. Yet our estimates reveal two useful properties:
\[
C_{\mathrm{LS}}(\nu_t)\le \frac{2}{t},
\qquad
\log\bigl(1+\chi^2(\mu\Vert\nu_t)\bigr)
\lesssim nK^6t^3,
\]
where the second bound holds for $0<t\le 1/(2K^2)$.
These properties suggest that combining our information-theoretic
approach with further state-of-the-art tools from convex geometry could yield
stronger log-Sobolev bounds.

We indeed establish such an improvement by combining these properties
of $\nu_t$ with Milman's connection between Gaussian concentration
and log-Sobolev inequalities for log-concave measures
\cite{Milman2010}, and Bizeul's two-level concentration estimate
\cite[Corollary~1.2]{BizeulSubspace} for small radii, together with a subpolynomial bound on the KLS constant \cite{Chen2021}. Together, we prove in \cref{thm:subgaussian-lsi} these ingredients yield the further improved bound
\[
C_{\mathrm{LS}}(\mu)\lesssim K^2n^{1/4}.
\]In particular, we obtain a sharper large-radius concentration
bound for $\mu$ by transferring the Gaussian concentration of
$\nu_t$ through a change-of-measure argument controlled by
$\chi^2(\mu\Vert\nu_t)$, and optimizing $t$ as a function of
the radius.

\subsubsection{A Bayesian interpretation of the recent proofs of the KLS conjecture}

Over the last month, independently of the present work, there have been three proofs of the KLS conjecture, proving that for some universal constant $C>0,$ for all isotropic log-concave $\mu$, $\CP(\mu)=O(1)$ (see \cite{SongZhang2026,BizeulKlartagLehec2026,BalasubramanianKasiviswanathan2026} for the general case; see also \cite{MikulincerZadik2026} for the specific case of unconditional measures). A critical feature of all three proofs in the general case is that, at a high level, they bound and leverage the so-called mixed cumulants of $\mu$ of all orders, and then bound the Poincar\'e constant of $\mu$ in terms of these cumulants. We emphasize that this is the high-level strategy, and some proofs such as \cite{SongZhang2026} apply this approach in a recursive way, while some of the rest such as  \cite{BalasubramanianKasiviswanathan2026} apply this approach in a more direct manner. 

It is natural to ask how our Bayesian estimation framework relates
to these recent proofs of the KLS conjecture. Our final contribution
provides a natural statistical reformulation of a slight variant
of the proof of \citet{BalasubramanianKasiviswanathan2026},
giving an estimation-theoretic interpretation of several of its
key quantities and steps.

The starting point is our main equivalence: to establish the KLS
conjecture, it suffices to show that, for every isotropic
log-concave prior $\mu$, observing the GAM at a fixed universal
SNR leaves a fixed fraction of the variance of every binary label
unexplained. More precisely, we seek universal constants $t,c>0$
such that
\[
\mmse_t(\mathbf1_A)
\ge c\,\var_{\mu}(\mathbf1_A)
\]
for every measurable set $A$ and every such $\mu$.

While researchers in statistics and information theory have studied the MMSE for decades, to understand the connection with the current KLS proofs, it is beneficial to study a more recently introduced and potentially larger quantity called the \emph{degree-$D$ MMSE}: the minimum mean squared error
among polynomial estimators of total degree at most $D$ in the
observation. Of course for slightly regular measures, the case $D=\infty$ corresponds to the original MMSE. Now, over the last decade, an intense amount of work has been spent on lower bounding the $D$-degree MMSE in the field of computational--statistical
trade-offs in high-dimensional statistics (see \cite{SchrammWein2022}, \cite[Chapter 14]{Zadik2026}, \cite{WeinSurvey25}). What makes it relevant here is that an influential approach to lower-bounding this quantity,
developed in the seminal work of \citet{SchrammWein2022}, uses mixed cumulants involving
one copy of the target label and at most $D$ coordinates of the
signal. Through Jensen's inequality and a Hermite expansion, their
argument bounds the improvement over the trivial estimator by a
weighted sum of squared mixed cumulants.

This connection motivates cumulant bounds as a route to KLS and a natural hope would be to combine the cumulant bound obtained in the recent KLS proofs and use the Schramm and Wein bound \cite{SchrammWein2022} to obtain a $D$-degree MMSE lower bound for all $D>0.$
However, as we explain in Section~\ref{sec:SW_cumulants}, a direct application of the
Schramm--Wein bound does not yield the required uniform-in-$D$ MMSE lower bound. 
Yet, in the literature of $D$-degree MMSE it is known that the Schramm--Wein bound can be suboptimal at the fine grained detail required by such a result and tighter methods to control the $D$-degree MMSE have been established.

In Sections~\ref{sec:sohn1}, \ref{sec:sohn2}, we turn to a flexible ``certificate''
framework for degree-$D$ MMSE by \citet{SohnWein2025} which is known to obtain tighter $D$-degree MMSE lower bounds in certain cases. We show that a slight modification of the cumulant-to-Poincar\'e
step in the proof of \citet{BalasubramanianKasiviswanathan2026}
can naturally construct such certificates, yielding the required MMSE lower
bound uniformly in $D$. Combined with the cumulant estimates,
this recovers the KLS conjecture through our main equivalence.
\subsubsection{A broader ``Bayesian'' proof method}

This work contributes to a broader research program that uses
ideas from Bayesian statistics and information theory as \emph{proof
tools} in probability theory and related areas. Some such successful applications include:
\begin{itemize}
    \item
    In probabilistic combinatorics, a Bayesian proof of the
    spread lemma \cite{MosselNilesWeedSunZadik2025}, which
    recovers the fractional Kahn--Kalai theorem
    \cite{FrankstonKahnNarayananPark2021}. Here the underlying channel is a Bernoulli union channel rather than a GAM, but the spread lemma is directly recovered by an appropriate MMSE lower bound.
    Using an extension of this Bayesian technique, alongside a nice decomposition idea of \citet{Tran2026}, OpenAI also recently announced a proof of the ``second'' Kahn--Kalai conjecture \cite{OpenAISecondKahnKalai2026}. This proof was also part of their October 6, 2026 release.

    \item
    In the theory of  generic chaining,
    new Bayesian proofs of Talagrand's majorizing-measure
    theorem \cite{Zadik2026MMT} and the Bernoulli theorem
    of Bednorz and Lata{\l}a \cite{LiuZadik2026Bernoulli}. In the former, the key steps are obtained by an analysis of the MMSE curve of a GAM, while in the latter the focus switches to a capped MMSE curve of a Cauchy additive channel.
\end{itemize}
In the present paper, we extend this Bayesian program to proving functional
inequalities for log-concave measures, building on the already established such connections developed in prior work such as in
\cite{Raginsky2016,ElAlaouiMontanari,KlartagOrdentlich}.

\subsection{Concurrent and independent work by OpenAI}
In the last weeks before the completion of the present work, \citet{OpenAI-LSI} announced a proof that every centered log-concave, $K$-subgaussian probability measure $\mu$ on $\R^n$ satisfies $C_{\mathrm{LS}}(\mu)\lesssim K^2$, with a universal constant independent of the dimension.
This claims to resolve the conjecture of \citet{BizeulLSI}, and in particular to improve our bound $C_{\mathrm{LS}}(\mu)\lesssim K^2n^{1/4}$ in \cref{thm:subgaussian-lsi} to a dimension-free one.

Interestingly, while the authors have not verified OpenAI's proof, it appears that among other tools, it also leverages Bayesian estimation identities, such as Gaussian posterior
prediction and mutual-information estimates, e.g., an entropy-loss comparison \cite[Lemma 4.2]{OpenAI-LSI} and its application of LSI
to posterior densities \cite[Proposition~7.2]{OpenAI-LSI}.

\subsection{Statement on the use of generative AI}

The authors began developing this Bayesian approach to prove functional inequalities in July 2026.
Without AI assistance, they obtained the new information-theoretic
proof of the original KLS bound (part~(i) of
\cref{thm:intro-kls-lv}). Subsequent interactions with AI models
led to an extension of the method to the Lee--Vempala bound
(part~(ii) of \cref{thm:intro-kls-lv}).

The authors then explored together with the help of AI-generative models applications to other functional
inequalities, particularly log-Sobolev inequalities. These discussions led to the general equivalence
(\cref{thm:general-equivalence}) and a new information-theoretic
proof of Bizeul's bound (\cref{cor:lsi-quarter}), using a
detection-to-estimation reduction with the simple Gaussian null law.
Through this stage, the core proof ideas were primarily
developed by the authors, with the exception of the ideas underlying the proof of a key contraction lemma (\cref{lem:restricted-contraction}). Yet, a subsequent key suggestion generated by
AI was to replace the Gaussian null with the null-posterior
law. Combined with the detection-to-estimation framework,
this led to the improved log-Sobolev bounds
(\cref{cor:lsi-sixth,thm:subgaussian-lsi}).

Finally, the Bayesian reformulation of the recent proofs of the KLS
conjecture (\cref{sec:cumulant-perspective}) also involved both
author and AI core ideas. AI models helped explain the
existing proofs to the authors and generated the proof ideas
for \cref{prop:cumulant-dual-norm}. The main ideas connecting
these arguments to low-degree MMSE were developed primarily
by the authors. The AI models used are GPT-5.6 Sol, GPT-6 Sol, GPT-6.1 Sol, GPT-6 Astra, and Claude Opus 5.5 and Fable 5.1.

The authors devoted substantial time to revising and clarifying
the exposition of all arguments throughout the paper. The authors take full responsibility for the content and accuracy of this work.

\subsection*{Acknowledgments}

The second author thanks Sinho Chewi and Dan Mikulincer for
helpful discussions. He is especially grateful to Sinho Chewi
for suggesting to him to apply the Bayesian approach to Bizeul's
conjecture~\cite{BizeulLSI}.

\subsection{Notation}We call a probability measure $\mu$ on $\R^n$ log-concave if $\mu(\mathrm{d}x)=e^{-V(x)}\mathrm{d}x$ with \(\nabla^2 V \succeq 0\).
Standard notions from analysis and information theory are recalled in \cref{app:background}.
We write $\R_{>0}=(0,+\infty)$.
For nonnegative quantities $A$ and $B$, we write $A\lesssim B$ or $B\gtrsim A$ if $A\leq cB$ for a universal constant $c>0$, and $A\asymp B$ if $A\lesssim B$ and $B\lesssim A$.
All logarithms are natural.

\section{Getting started}

In this section, we offer the required formal background to state our main results.
\subsection{Functional inequalities}\label{sec:background_functional_ineqs}
Let $C_c^\infty(\R^n)$ denote the set of smooth functions $g:\R^n\to\R$ with compact support (\cref{def:smooth}).
For a (possibly unbounded) open interval $J\subseteq\R$, we test functional inequalities on the class
\[
 \cA_0^J=\bigl\{f=c+g:\ c\in\R,\ g\in C_c^\infty(\R^n),\ f(\R^n)\subseteq J\bigr\}
\]
of smooth functions with values in $J$ that are constant outside a compact set.
For $f=c+g\in\cA_0^J$, the range $f(\R^n)=f(\supp\,g)\cup\{c\}$ is compact, since $\supp\,g$ is compact and $f$ is continuous.
Hence $f$ is bounded, and so is $\nabla f=\nabla g$, since it is continuous and vanishes outside the compact set $\supp\,g$.
Therefore, all the expectations in the $\Phi$-Sobolev, Poincar\'e, and log-Sobolev inequalities \eqref{eq:phi-functional}, \eqref{eq:Poincare_ineq}, and \eqref{eq:lsi-convention} are finite for every probability measure.
Testing the log-Sobolev inequality on positive functions is convenient for the proofs below, and loses nothing, see \cref{rem:test-functions}.

\begin{remark}[Broader classes of functions]\label{rem:test-functions}
We test the Poincar\'e, log-Sobolev, and $\Phi$-Sobolev inequalities on $\cA_0^\R$, $\cA_0^{\R_{>0}}$, and $\cA_0^J$, respectively, so that both sides are always finite.
As usual, these inequalities extend by standard approximation arguments to every function, with values in $J$ for the $\Phi$-Sobolev inequality, for which both sides are well defined and finite, and we use these extensions without further comment, see e.g., \cite[Chapters~3 and~7]{BakryGentilLedoux}.
For log-concave \(\mu\), the Poincar\'e and log-Sobolev inequalities extend to the weighted Sobolev space $H^1(\mu)$, see e.g., \cite[Proposition~27]{BartheKlartag2020}.
\end{remark}
\subsection{Bayesian estimation}\label{sec:bayesian-estimation}

Now fix a measure $\mu$. Here we interpret the measure as the prior on the following Bayesian recovery task, called Gaussian channel or Gaussian additive model (GAM). Even if we defined the GAM in the introduction, we redefine it here, alongside its Markov kernel.

\begin{definition}[Gaussian additive model (GAM)]\label{dfn:GAM}
    For a ``signal'' $X\sim\mu$ a statistician observes the noisy Gaussian observation $Y_t=\sqrt tX+G$, where the ``noise'' $G\sim\cN(0,I_n)$ is independent of $X$ and $t>0$ is called the signal-to-noise ratio (SNR). Equivalently, given $X=x$, the observation $Y_t$ has law $K_{t,x}=K_t(x,\cdot):=\cN(\sqrt t\,x,I_n)$, and we call the Markov kernel $K_t$ the Gaussian channel at SNR $t$. The goal of the statistician is to estimate (or ``recover'') $X$ from $Y_t.$
\end{definition}A key aspect of a Bayesian task is (a) which features of $X$ does one aim to estimate, (b) which error metric should one adopt to measure the success of estimation. For part (a), in this work, we focus on one-dimensional functions of $X$, i.e., the goal is to estimate $f(X) $ for a given function $f: \R^n\to J.$ In terms of part (b), i.e., the error metric, we parameterize the risk based on an arbitrary function $\Phi:J\to\R$. Given $\Phi,$ we use as error metric the \emph{Bregman divergence generated by $\Phi$} defined by
\begin{align}\label{dfn:bregman}
 B_\Phi(z,a)=\Phi(z)-\Phi(a)-\Phi'(a)(z-a).
\end{align}This allows us to define the following crucial object.

\begin{definition}\label{dfn:bregman-optimal}
    For any real-valued function $f: \R^n\to J$, we call the $\Phi$-optimal risk (or the $\Phi$-Bayes risk) of estimating $f(X)$ from $Y_t$ to be
\[
 \cR_\Phi(t,f):=\inf_h\E B_\Phi(f(X),h(Y_t)),
\]
where the infimum is over all measurable estimators $h:\R^n\to J$ based on $Y_t$.
\end{definition}In words, $\cR_\Phi(t,f)$ measures how small can the error $B_{\Phi}$ be if the statistician uses the optimal estimator to estimate $f(X)$ from $Y_t$.

\begin{remark}\label{rem:trivial_loss} It is easy to observe that for the ``trivial'' constant estimator $h(Y_t)=\E_{\mu} f(X),$ the error $\E B_\Phi(f(X),h(Y_t))=\Ent_\mu^\Phi(f)$, and for this reason we refer to $\Ent_\mu^\Phi(f)$ as the $\Phi$-trivial risk. In particular, the $\Phi$-optimal risk is upper bounded by the $\Phi$-trivial risk, i.e., it always holds \[ \cR_\Phi(t,f) \leq \Ent_\mu^\Phi(f).\]
\end{remark}

\paragraph{Observation laws and the planted model} For a probability measure $\zeta$ on $\R^n$, we denote by
\[
 \zeta K_t:=\int K_{t,x}\,\zeta(\mathrm{d}x)
\]
the law of the observation $Y_t$ when $X\sim\zeta$.
We call the joint law $P_t:=\Law(X,Y_t)$ of the signal $X\sim\mu$ and the observation the \emph{planted model}, so that $Y_t\sim\mu K_t$, while the pure noise $Y_0=G$ has the standard Gaussian law $\gamma_n=\cN(0,I_n)$.
When no confusion can arise, we also use a random vector to denote its law, for instance $\chi^2(Y_t\Vert\gamma_n)=\chi^2(\mu K_t\Vert\gamma_n)$, and we write $\E_{y\sim Y_t}h(y)=\E h(Y_t)$.

\section{Main result: a general equivalence}

Our main equivalence is as follows.
\begin{theorem}[Functional inequalities as Bayes-risk lower bounds]\label{thm:general-equivalence}
Fix any log-concave $\mu$ and $C>0$. Fix also any open interval $J$ and $\Phi:J\to\R$ smooth with $\Phi''>0$ and $1/\Phi''$ concave. Then, the following are equivalent.
\begin{enumerate}[label=(\roman*)]
\item The measure $\mu$ satisfies the $\Phi$-Sobolev inequality \eqref{eq:phi-functional} with constant $C$.
\item For every $f\in\cA_0^J$ and every $t>0$, for the Gaussian additive model (\cref{dfn:GAM}) it holds
\begin{equation}\label{eq:phi-risk}
 \cR_\Phi(t,f)\geq\frac{\Ent_\mu^\Phi(f)}{1+Ct}.
\end{equation}In words, the $\Phi$-optimal risk is at least a $(1+Ct)^{-1}$-fraction of the $\Phi$-trivial risk.\end{enumerate}
More generally, if there are some $b>0$ and $T>0$, such that for every $f\in\cA_0^J$,
\[
 \cR_\Phi(T,f)\geq b\Ent_\mu^\Phi(f),\]
then $\mu$ satisfies the $\Phi$-Sobolev inequality \eqref{eq:phi-functional} with $C=\frac{1}{bT}$.
\end{theorem}
The proof is given in \cref{sec:equivalence}.

\section{Applications}
\subsection{Poincar\'e inequalities}
To obtain Poincar\'e inequalities from our main equivalence we have to set $\Phi(z)=(z-1)^2$. In that case, it is easy to see that the $\Phi$-optimal risk becomes the usual (Euclidean) minimum mean-square error, 
\begin{equation}\label{eq:dfn_mmse_PI}
\mmse_t(f)=\inf_h\E(f(X)-h(Y_t))^2.\end{equation} \cref{thm:general-equivalence} then gives the following.

\begin{corollary}[Poincar\'e--MMSE equivalence]\label{cor:poincare-mmse}
For any log-concave $\mu$ and $C>0$, the following are equivalent.
\begin{enumerate}[label=(\roman*)]
\item The measure $\mu$ satisfies the Poincar\'e inequality \eqref{eq:Poincare_ineq} with constant $C$, that is, $\CP(\mu)\leq C$.
\item Fix any $t>0$ and any $f\in\cA_0^\R$. Then for the GAM (\cref{dfn:GAM}) with SNR $t$ it holds,
\begin{equation}\label{eq:mmse-retention}
 \mmse_t(f)\geq\frac{\var_\mu(f)}{1+Ct}.
\end{equation}
\end{enumerate}
Moreover, if for some $T,b>0$,
$\mmse_T(f)\geq b\var_\mu(f)$ for every $f\in\cA_0^\R$, then $\CP(\mu)\lesssim1/(bT)$.

Finally, it suffices to check ``balanced indicators'' $f$, in the following sense. If for some $T,b>0$, it holds that $\mmse_T(\mathbf1_A)\geq b$ for every Borel set $A$ with $\mu(A)=1/2$, then $\CP(\mu)\lesssim1/(b^2T)$. 
\end{corollary}

\begin{remark}
We remark that the last sufficient statement is in fact implicit in some stochastic localization proofs of isoperimetric bounds \cite{Eldan2013,LeeVempala,Chen2021}.
Indeed, as mentioned in the introduction, the localized measure at time $t$ of say \citet{LeeVempala} is exactly the posterior $\mu_{t,Y_t}$ \cite{ElAlaouiMontanari}.
Now, for a set $A$ with $\mu(A)=1/2$, the stochastic localization arguments give that $\mu_{T,Y_T}(A)$ stays bounded away from $0$ and $1$ with probability bounded below, up to a time $T$ at which the covariance of the localized measures is still controlled (see e.g., \cite[Lemma~29]{LeeVempala}, \cite[Lemma~4]{Chen2021}, or \cite[Lemma~4.1]{Eldan2013}). 
They then use it to lower bound by a constant the $\E[\mu_{T,Y_T}(A)(1-\mu_{T,Y_T}(A))]$ (recall that this is exactly $\mmse_T(\mathbf1_A)=\E[\mu_{T,Y_T}(A)(1-\mu_{T,Y_T}(A))]$), and the isoperimetry of the $T$-strongly log-concave posteriors then allows one to lower bound the boundary measure of $A$ \cite[Lemma~31]{LeeVempala}, as already noted in \cite[Remark~4.3]{Eldan2013}, yielding the Poincar\'e constant upper bounds. The above strategy can therefore be clearly rearranged to yield the above sufficient condition.
\end{remark}
We now use \cref{cor:poincare-mmse} to obtain Poincar\'e inequalities for log-concave measures, via lower bounding the MMSE using information-theory tools.
Specifically, a simple application of Pinsker's inequality (\cref{fact:pinsker}) gives the following lemma, which we prove in \cref{sec:pinsker-mmse}. 
\begin{lemma}[MMSE and mutual information]\label{lem:pinsker-mmse}
Let $\mu$ be a probability measure on $\R^n$, let $t>0$, and let $A$ be a Borel set.
Then
\[
 \mmse_t(\mathbf1_A)\geq\var_\mu(\mathbf1_A)-\frac12I(\mathbf1_A(X);Y_t),
\]
where $I(\mathbf1_A(X);Y_t)$ is the mutual information (\cref{def:mutual-information}) between the indicator $\mathbf1_A(X)$ and the Gaussian observation $Y_t$.
\end{lemma}
Hence, one can derive a Poincar\'e inequality by proving that $Y_t$ carries a sufficiently small amount of information about the label $\mathbf1_A(X)$.

We now bound the mutual information by comparing the Gaussian observation $Y_t$ with the ``pure noise'' observation $Y_0\sim\gamma_n$. Specifically, we follow the detection-estimation reduction mentioned in the introduction, where the planted model $P_t$ is compared with the ``null model'' $\mu\otimes\gamma_n$, the joint law of $(X,Y_0)$. We achieve such a reduction by bounding the mutual information first via a trivial Kullback--Leibler bound between the distributions of $Y_t$ and $Y_0$, and then via a slightly more sophisticated bound using the log-Sobolev inequality of the Gaussian measure alongside Jensen's inequality. The above yield the following two MMSE lower bounds.

\begin{lemma}[MMSE lower bounds from the null model]\label{lem:intro-mmse-comparison}
Let $\mu$ be a probability measure on $\R^n$ with $\Sigma=\cov(X)$, and let $A$ be a Borel set.
Then for every $t>0$,
\[
 \mmse_t(\mathbf1_A)\geq\var_\mu(\mathbf1_A)-\frac t4\tr(\Sigma),
\]
and for every $t>0$ and $0<q\leq1$,
\[
 \mmse_{qt}(\mathbf1_A)\geq\var_\mu(\mathbf1_A)-\frac q2\Bigl(\log2+\log\bigl(1+\chi^2(Y_t\Vert\gamma_n)\bigr)\Bigr).
\]
\end{lemma}
We obtain the first bound of \cref{lem:intro-mmse-comparison} in the proof of \cref{prop:kls}, and the second bound in \cref{rem:label-information} together with \cref{lem:kl-chi2}.
It turns out that the above inequalities together with the equivalence \cref{cor:poincare-mmse} yield two key results in the literature of bounding the Poincar\'e constant of log-concave measures: the original bound obtained by Kannan, Lov\'asz and Simonovits when they stated the so-called KLS conjecture, and the bound obtained by \citet{LeeVempala}. We remark that, given the above lemma, the proof of the bound from \citet{KLS1995} is immediate and the bound from Lee--Vempala follows from a relatively routine conditional $\chi^2$ (or second moment) computation. As mentioned in the introduction no needle decomposition, localization argument or stochastic calculus is used. 

\begin{theorem}\label{thm:intro-kls-lv}
Every log-concave $\mu$ with $\Sigma=\cov(X)$ satisfies the following bounds.
\begin{enumerate}[label=(\roman*)]
\item $\CP(\mu)\lesssim\tr(\Sigma)$. In particular, for isotropic measures $\mu,$ $\CP(\mu)\lesssim n.$ This result is due to \citet{KLS1995}.
\item $\CP(\mu)\lesssim\sqrt{\tr(\Sigma^2)}$. In particular, for isotropic measures $\mu,$ $\CP(\mu)\lesssim n^{1/2}.$ This result is due to \citet{LeeVempala}.
\end{enumerate}
\end{theorem}

\subsection{Log-Sobolev inequalities}\label{sec:lsi-applications}

We now turn to the implications of our main equivalence for log-Sobolev inequalities. To present it we first need the definition of a randomized binary label.
\begin{definition}\label{dfn:random_label}
    For a GAM (\cref{dfn:GAM}), a randomized binary label of $X$ is a random variable $U\in\{0,1\}$ such that $\P(U=1\mid X)$ is a measurable function of $X$ and the sequence $U-X-Y_t$ is a Markov chain.
\end{definition}
Now for $\Phi(z)=z\log z$, \cref{thm:general-equivalence} gives the following equivalence between the log-Sobolev inequality and the impossibility of estimation of randomized binary labels of the input, measured in mutual information.

\begin{corollary}[Log-Sobolev--mutual information equivalence]\label{cor:lsi-information}
For any log-concave $\mu$ and $S>0$, the following are equivalent.
\begin{enumerate}[label=(\roman*)]
\item The measure $\mu$ satisfies the log-Sobolev inequality \eqref{eq:lsi-convention} with constant $S$, that is, $\LS(\mu)\leq S$.

\item Fix any $t>0$. For the GAM (\cref{dfn:GAM}) with SNR $t$ and every randomized binary label \(U\in \{0, 1\}\) of $X$, it holds
\begin{equation}\label{eq:info-retention}
 I(U;X\mid Y_t)\geq\frac{I(U;X)}{1+St}.
\end{equation}
\end{enumerate}
Moreover, if for some $T,b>0$, every randomized binary label \(U\) of $X$ satisfies $I(U;X\mid Y_T)\geq bI(U;X)$, then $\LS(\mu)\leq\frac1{bT}$.
\end{corollary}
The proof combines \cref{thm:general-equivalence} with the characterization of the contraction coefficient of relative entropy through the mutual information of labels \cite[Theorem~4]{PolyanskiyWu}.

We now use \cref{cor:lsi-information} to obtain log-Sobolev inequalities for log-concave measures.
As in the introduction, we run the detection-to-estimation reduction with two different null models, which give the bounds $\LS(\mu)\lesssim K^2n^{1/2}$ and $\LS(\mu)\lesssim K^2n^{1/3}$.
For these two bounds, we assume that $\mu$ is log-concave and subgaussian \eqref{eq:subG}.
Translation does not change $\LS(\mu)$, and dilation by $1/K$ divides it by $K^2$, so we assume without loss of generality that $\E X=0$ and, for every $u\in\R^n$,
\begin{equation}\label{eq:normalized}
 \E e^{\langle u,X\rangle}\leq e^{\|u\|_2^2/2}.
\end{equation}

\paragraph{The Gaussian null model and the $n^{1/2}$ bound}
As we use the null model $\mu\otimes\gamma_n$, under which the observation $Y_0\sim\gamma_n$ is independent of $X$, \ref{step:1} is immediate as every randomized binary label $U$ of $X$ satisfies $I(U;X\mid Y_0)=I(U;X)$.

For \ref{step:2}, the following result allows us to show that the result follows if we can bound the $\chi^2$ divergence between the observation $Y_t$ and the pure noise $Y_0 \sim \gamma_n$.
\begin{theorem}\label{thm:output-information}
Let $t>0$, suppose that $\Delta_t=\log(1+\chi^2(Y_t\Vert\gamma_n))<\infty$, and let $s=t/(1000(1+\Delta_t))$.
Let $X\sim\mu$, and let $Y_s=\sqrt sX+G$ be the GAM observation of $X$ at SNR $s$.
For every randomized binary label $U$ of $X$,
\begin{equation}\label{eq:output-entropy}
 I(U;X\mid Y_s)\geq\frac{I(U;X)}{10^3}.
\end{equation}
In particular,
\[
 \LS(\mu)\lesssim\frac1t\bigl(1+\log(1+\chi^2(Y_t\Vert\gamma_n))\bigr).
\]
\end{theorem}
We prove \cref{thm:output-information} in \cref{sec:proof-output-information}.
The proof of this step is rather elaborate and follows via a key contraction lemma (\cref{lem:restricted-contraction}) of independent interest.

Finally, for \ref{step:3}, we have via a direct bound that $\log(1+\chi^2(Y_t\Vert\gamma_n))\leq nt^2$ for $0<t\leq1/\sqrt2$ (see \cref{lem:output-chi2}), and choosing $t=n^{-1/2}$ in \cref{thm:output-information} gives the following bound.
\begin{corollary}\label{cor:lsi-quarter}
The log-Sobolev constant satisfies $\LS(\mu)\lesssim n^{1/2}$.
\end{corollary}
We prove \cref{cor:lsi-quarter} in \cref{sec:proof-lsi-quarter}.
We repeat here that this is the bound originally obtained by \citet{BizeulLSI}; we remark that he states it in his work for the unsquared constant $\rho_{\mathrm{LS}}(\mu)=\sqrt{\LS(\mu)}$, as $\rho_{\mathrm{LS}}(\mu)\lesssim n^{1/4}$.

\paragraph{The null-posterior model and the $n^{1/3}$ bound}
Recall from the introduction that the null is a GAM with prior $\nu_t=\E_{y\sim\gamma_n}\mu_{t,y}$; specifically the null-posterior model $Q_s$, which is the law of $(X,Y_s)$ with $X\sim\nu_t$ and $Y_s=\sqrt sX+G$.

To prove \ref{step:1}, which is a non-trivial step under this null measure, we show that the prior of the null-posterior model $\nu_t$ has a log-Sobolev constant at most $2/t.$
\begin{proposition}\label{prop:reference-lsi}
Let $t>0$, let $X\sim\nu_t$, and for $s>0$ let $Y_s=\sqrt sX+G$ be the GAM observation of $X$ at SNR $s$. 
For every $s>0$ and every randomized binary label $U$ of $X$,
\begin{equation}\label{eq:null-information}
 I(U;X\mid Y_s)\geq\frac t{2(t+s)}\,I(U;X).
\end{equation}
Moreover, $\LS(\nu_t)\leq2/t$.
\end{proposition}
We prove \cref{prop:reference-lsi} in \cref{sec:reference-lsi}.
For \ref{step:2}, the planted and null-posterior models at SNR $s$ share the channel $K_s$, so their likelihood ratio simplifies to
\[
 \frac{\mathrm{d}P_s}{\mathrm{d}Q_s}(x,y)=\frac{\mathrm{d}\mu}{\mathrm{d}\nu_t}(x),
\]
and hence, for every $s\geq0$,
\[
 \chi^2(P_s\Vert Q_s)=\chi^2(\mu\Vert\nu_t).
\]
Comparing the two models then gives the following theorem, allowing us to prove that impossibility of detection implies impossibility of estimation.
\begin{theorem}\label{thm:reference-information}
Let $t>0$, suppose that $\Delta'_t=\log(1+\chi^2(\mu\Vert\nu_t))<\infty$, and fix \[s=t/(1000(1+\Delta'_t)).\]
Let $X\sim\mu$, and let $Y_s=\sqrt sX+G$ be the GAM observation of $X$ at SNR $s$.
For every randomized binary label $U$ of $X$, 
\begin{equation}\label{eq:reference-entropy}
 I(U;X\mid Y_s)\geq\frac{I(U;X)}{10^3}.
\end{equation}
In particular,
\[
 \LS(\mu)\lesssim\frac1t\bigl(1+\log(1+\chi^2(\mu\Vert\nu_t))\bigr).
\]
\end{theorem}
We prove \cref{thm:reference-information} in \cref{sec:reference-information}.
The posterior kernel $y\mapsto\mu_{t,y}$ maps $\mu K_t$ to $\mu$ and $\gamma_n$ to $\nu_t$, so the data processing inequality (\cref{lem:data-processing}) gives $\chi^2(\mu\Vert\nu_t)\leq\chi^2(Y_t\Vert\gamma_n)$, and this reduction result obtained via the null-posterior model is at least as strong as the result obtained by the Gaussian null model (\cref{thm:output-information}).

In fact, establishing also \ref{step:3}, the planted and null-posterior models are much closer: we can prove $\log(1+\chi^2(\mu\Vert\nu_t))\lesssim nt^3$ for $0<t\leq1/2$ (see \cref{prop:likelihood-L2}).
Choosing $t=\tfrac12n^{-1/3}$ in \cref{thm:reference-information} then gives the following bound.
\begin{corollary}\label{cor:lsi-sixth}
The log-Sobolev constant satisfies $\LS(\mu)\lesssim n^{1/3}$.
\end{corollary}
We prove \cref{cor:lsi-sixth} in \cref{sec:proof-lsi-sixth}.

\paragraph{Improving the bound}Here, we aim to combine the intriguing properties of $\nu_t$ with tools beyond information-theory tools to strengthen the log-Sobolev constant bound even further.
Specifically, we use a result by \citet{Milman2010} showing that Gaussian concentration implies a log-Sobolev inequality for log-concave measures, which reduces the task to proving Gaussian concentration for $\mu$.
Transferring concentration from $\nu_t$ to $\mu$, via a simple change of measure argument where $t$ is chosen separately at each radius, and combining it with a two-level concentration bound of \citet[Corollary~1.2]{BizeulSubspace} and any subpolynomial bound on the KLS constant gives the following bound, proved in \cref{sec:improvement}.
In terms of the unsquared constant, this bound is $\rho_{\mathrm{LS}}(\mu)\lesssim n^{1/8}$.

\begin{theorem}[Log-Sobolev constant of subgaussian log-concave laws]\label{thm:subgaussian-lsi}
Let $n>1$ and $X\sim\mu$, where $\mu(\mathrm{d}x)=e^{-V(x)}\mathrm{d}x$ with $V\in C^2(\R^n)$ convex, and suppose that $X$ satisfies \eqref{eq:normalized}.
Then $\LS(\mu)\lesssim n^{1/4}$.
\end{theorem}

\section{A recasting of the proof of the KLS conjecture from the Bayesian viewpoint}
\label{sec:cumulant-perspective}

During October 2026, there have been three proofs of the so-called KLS conjecture, proving that for some universal constant $C>0,$ for all isotropic log-concave $\mu$, $\CP(\mu)=O(1)$ (see \cite{SongZhang2026,BizeulKlartagLehec2026,BalasubramanianKasiviswanathan2026} for the general case, and \cite{MikulincerZadik2026} for the specific case of unconditional measures). A critical feature of all three proofs in the general case is that, at a high level, they bound and leverage the so-called mixed cumulants of $\mu$ of all orders, and then bound the Poincar\'e constant of $\mu$ in terms of these cumulants. We emphasize that this is the high-level strategy, and some proofs such as \cite{SongZhang2026} apply this approach in a recursive way, while some of the rest such as  \cite{BalasubramanianKasiviswanathan2026} apply this approach in a more direct manner.

In this section, we first present a slight variation of the third proof of the KLS conjecture by \citet{BalasubramanianKasiviswanathan2026}. Then we show how one can canonically recast this successful approach from the Bayesian estimation viewpoint presented in this paper, and statistically interpret and motivate the key quantities and steps.

Throughout this section, for tensors $S,T$ of order $d$ over $\R^n$, let
\[
 \langle S,T\rangle=\sum_{i_1,\ldots,i_d=1}^nS_{i_1\ldots i_d}T_{i_1\ldots i_d}
\]
be the Hilbert--Schmidt inner product, and let $\|T\|_{\mathrm{HS}}^2=\langle T,T\rangle$ be the Hilbert--Schmidt norm.

\subsection{The structure of (a slight variant of) the proof by \texorpdfstring{\citet{BalasubramanianKasiviswanathan2026}}{Balasubramanian and Kasiviswanathan}}

\paragraph{Cumulant bounds}
The first key step in the proof is to obtain tight cumulant bounds.
Fix any centered $f\in L^2(\mu)$ with $\E f(X)^2=1$.
For $d\geq1$, let $\kappa_d^\mu(f)$ be the symmetric cumulant tensor
\begin{equation}\label{eq:cumulant-tilt}
 \kappa_d^\mu(f)
 :=\left.\nabla_z^d
 \frac{\E[f(X)e^{\langle z,X\rangle}]}
      {\E e^{\langle z,X\rangle}}\right|_{z=0}.
\end{equation}
Its $(i_1,\ldots,i_d)$ entry is the mixed cumulant $\operatorname{Cum}(f(X),X_{i_1},\ldots,X_{i_d})$ of $\mu$ involving the function $f$.
An equivalent definition, used in the three proofs, is through the Appell tensors $A_d^\mu$ \cite{Anshelevich-AppellPolynomials}, which are defined by the expansion
\begin{equation}\label{eq:appell-expansion}
 \frac{e^{\langle z,x\rangle}}{\E e^{\langle z,X\rangle}}
 =\sum_{d\geq0}\frac{\langle A_d^\mu(x),z^{\otimes d}\rangle}{d!}.
\end{equation}
In terms of the Appell tensors, the cumulant tensor is
\[
 \kappa_d^\mu(f)=\E[f(X)A_d^\mu(X)].
\]

The key estimate of the three works is that the cumulants grow at most exponentially, in the sense that
\begin{equation}\label{eq:cumulant-geometric}
 \frac{\|\kappa_d^\mu(f)\|_{\mathrm{HS}}}{d!}
 \leq R^d\qquad d\geq1,
\end{equation}
with a universal constant $R$, uniformly over all isotropic log-concave priors $\mu$ and all such functions $f$.
For the exact statements, see \cite[Lemma~8.18, equation~(229), and Proposition~9.26]{SongZhang2026}, \cite[Theorem~4.1, Proposition~7.1, and the proof of Theorem~1.1 in Section~7]{BizeulKlartagLehec2026}, and \cite[Theorem~4.1, with the coefficient-norm definitions following equation~(3.1)]{BalasubramanianKasiviswanathan2026}.

\paragraph{Dual Sobolev norm bound}
The three proofs then deduce the KLS conjecture from \eqref{eq:cumulant-geometric}.
We present a small variant of this implication, from the dual form of the Poincar\'e inequality.

First, we may assume without loss of generality that the isotropic log-concave measure $\mu(\mathrm{d}x)=e^{-V(x)}\mathrm{d}x$ is regular (see e.g., \cite[Section~2]{BizeulKlartagLehec2026}), in the sense that $V\in C^\infty(\R^n)$, all derivatives of $V$ grow at most polynomially, and there exists \(0<a\leq M<\infty\), such that for all \(x\in \R^n\), 
\begin{equation}\label{eq:appD-regularity}
 aI_n\preceq\nabla^2V(x)\preceq MI_n. 
\end{equation}
To introduce the dual form of the Poincar\'e inequality, we first need to define the following notion. 
Let $H^1(\mu)$ be the Sobolev space w.r.t.\ the measure $\mu$ (see e.g., \cite[Chapter~4]{EvansGariepy}), i.e.,
\[
 H^1(\mu) = \{h\in L^2(\mu):\ \E_\mu\|\nabla h\|_2^2<\infty\},
\]
where $\nabla h$ is the weak gradient of $h$,
which is the ordinary gradient for continuously differentiable $h$. 
The space $H^1(\mu)$ is equipped with the Sobolev norm $\|h\|_{H^1(\mu)}^2=\E_\mu h^2+\E_\mu\|\nabla h\|_2^2$.
For $f\in L^2(\mu)$ with $\E_\mu f=0$, we define the dual Sobolev norm
\begin{equation}\label{eq:dual-sobolev}
 \|f\|_{H^{-1}(\mu)}
 =\sup\bigl\{\E_\mu[fh]:\ h\in H^1(\mu),\ \E_\mu\|\nabla h\|_2^2\leq1\bigr\}\in[0,\infty].
\end{equation}
The reason for \(\E_\mu f=0\) is that otherwise taking \(h\) to be an arbitrarily large constant function would make the supremum infinite.
Note that this is a slight abuse of notation, since $\|\cdot\|_{H^{-1}(\mu)}$ is the dual norm of the seminorm $h\mapsto(\E_\mu\|\nabla h\|_2^2)^{1/2}$ rather than of the Sobolev norm $\|\cdot\|_{H^1(\mu)}$.

Now we introduce the dual form of the Poincar\'e inequality. 
\begin{definition}[Dual Poincar\'e inequality]\label{def:dual-poincare}
Let $\mu$ be a probability measure over $\R^n$ and let $C\geq0$.
The measure $\mu$ satisfies the dual Poincar\'e inequality with constant $C$ if every $f\in L^2(\mu)$ with $\E_\mu f=0$ satisfies
\begin{equation}\label{eq:dual-poincare}
 \|f\|_{H^{-1}(\mu)}^2\leq C\E_\mu f^2.
\end{equation}
\end{definition}
The Poincar\'e inequality and its dual are equivalent, with the same constant (see e.g., \cite{OttoVillani,BartheKlartag2020}).
This follows directly from norm duality. 
We include an elementary proof of \cref{lem:dual-poincare} in \cref{app:dual-poincare}.

\begin{lemma}[Equivalence of the Poincar\'e inequality and its dual]\label{lem:dual-poincare}
Let $\mu(\mathrm{d}x)=e^{-V(x)}\mathrm{d}x$ with $V\in C^\infty(\R^n)$.
Then for every $C\geq0$, we have $\CP(\mu)\leq C$ if and only if $\mu$ satisfies the dual Poincar\'e inequality \eqref{eq:dual-poincare} with constant $C$.
In particular,
\begin{equation}\label{eq:dual-poincare-sup}
 \CP(\mu)=\sup\bigl\{\|f\|_{H^{-1}(\mu)}^2:\ f\in L^2(\mu),\ \E_\mu f=0,\ \E_\mu f^2=1\bigr\}.
\end{equation}
\end{lemma}

Hence, to bound the Poincar\'e constant of $\mu$, it suffices to upper bound $\|f\|_{H^{-1}(\mu)}$ for every centered $f\in L^2(\mu)$ with $\E_\mu f^2=1$.
The dual Sobolev norm is computed by solving a Poisson equation.
Let
\[
 \cL=\Delta-\langle\nabla V,\nabla\rangle
\]
be the Laplace operator w.r.t.\ $\mu$. 
The next lemma shows that $\|f\|_{H^{-1}(\mu)}^2=\E_\mu\|\nabla u\|_2^2$, where $u$ is the solution of the Poisson equation $-\cL u=f$.
Since $\cL u=e^V\nabla\cdot(e^{-V}\nabla u)$, by integration by parts, for smooth $u$ and compactly supported smooth $h$,
\[
 \begin{aligned}
 \E_\mu[(-\cL u)h]
 &=-\int h\,\nabla\cdot(e^{-V}\nabla u)\,\mathrm{d}x\\
 &=\int e^{-V}\langle\nabla u,\nabla h\rangle\,\mathrm{d}x
 =\E_\mu\langle\nabla u,\nabla h\rangle,
 \end{aligned}
\]
which gives the weak formulation \eqref{eq:poisson-weak}.
This Poisson characterization of the dual Sobolev norm is standard (see e.g., \cite[Section~7]{OttoVillani}).
For completeness, we include a proof in \cref{app:dual-poincare}.
Our proof constructs $u$ by the Riesz representation theorem on the centered Sobolev space with inner product $(g,h)\mapsto\E_\mu\langle\nabla g,\nabla h\rangle$, and then identifies the dual norm by the Cauchy--Schwarz inequality.

\begin{lemma}[Dual Sobolev norm via the Poisson equation]\label{lem:poisson-dual-norm}
Let $\mu(\mathrm{d}x)=e^{-V(x)}\mathrm{d}x$ with $V\in C^{\infty}(\R^n)$ and $\CP(\mu)<\infty$.
Then for every $f\in L^2(\mu)$ with $\E_\mu f=0$, there is a unique $u\in H^1(\mu)$ with $\E_\mu u=0$ such that for all $h\in H^1(\mu)$,
\begin{equation}\label{eq:poisson-weak}
 \E_\mu\langle\nabla u,\nabla h\rangle=\E_\mu[fh],
\end{equation}
that is, a unique centered weak solution of the Poisson equation $-\cL u=f$, which we denote by $(-\cL)^{-1}f$.
Moreover,
\begin{equation}\label{eq:poisson-dual-norm}
 \|f\|_{H^{-1}(\mu)}^2=\E_\mu[fu]=\E_\mu\|\nabla u\|_2^2.
\end{equation}
\end{lemma}

For a centered $f\in L^2(\mu)$, we write
\begin{equation}\label{eq:poisson}
 U_f=\nabla(-\cL)^{-1}f
\end{equation}
for the gradient of the solution of the Poisson equation, so that $\|f\|_{H^{-1}(\mu)}^2=\E_\mu\|U_f\|_2^2$ by \cref{lem:poisson-dual-norm}.

The crucial step is that the cumulant bound \eqref{eq:cumulant-geometric} for every $d$ and $f$ implies $\|f\|_{H^{-1}(\mu)}^2\lesssim R^2$, which completes the proof of the KLS conjecture by \cref{lem:dual-poincare}.
To the best of our knowledge, this implication does not appear explicitly in any of the three proofs, but it follows from the analytic steps of the third proof \cite[Sections~5--7]{BalasubramanianKasiviswanathan2026}.

\begin{proposition}[Dual Sobolev norm from mixed cumulants]\label{prop:cumulant-dual-norm}
Let $\mu$ be regular, with constants $a,M$ as in \eqref{eq:appD-regularity}.
Then for every centered $f\in L^2(\mu)$ with \(\E_\mu f^2 =1\) and every $s>0$,
\begin{equation}\label{eq:cumulant-energy-series}
 \|f\|_{H^{-1}(\mu)}^2\leq
 s+
 \sum_{d\geq1}
 \frac{\|\kappa_d^\mu(f)\|_{\mathrm{HS}}^2}
      {(d!)^2s^{d-1}},
\end{equation}
where the right-hand side may be infinite.
In particular, if \eqref{eq:cumulant-geometric} holds, then taking $s=10R^2$ gives $\CP(\mu)\lesssim R^2$.
\end{proposition}
Note that the bound \eqref{eq:cumulant-energy-series} does not depend on the constants $a$ and $M$ in \eqref{eq:appD-regularity}. The regularity of $\mu$ is used only in the proof.

\begin{proof}
By \cref{lem:poisson-dual-norm}, $\|f\|_{H^{-1}(\mu)}^2=\E_\mu\|U_f\|_2^2$, and we split $U_f$ into its mean, which is a cumulant of $\mu$, and its centered part, which we control by repeatedly solving Poisson equations for tensor fields.

We first introduce the notation for tensor fields.
A tensor field $H$ of order $m$ is a map from $\R^n$ to the tensors of order $m$ over $\R^n$, with components $H_{i_1\ldots i_m}:\R^n\to\R$ for $i_1,\ldots,i_m\in[n]$.
A tensor $T$ of order $m$ is symmetric if $T_{i_{\sigma(1)}\ldots i_{\sigma(m)}}=T_{i_1\ldots i_m}$ for all indices $i_1,\ldots,i_m$ and every permutation $\sigma$ of $[m]$, and a tensor field is symmetric if all its values are symmetric tensors.
A tensor field $H$ of order $m$ belongs to $H^1(\mu)$ if each of its components does, and then $\nabla H$ is the tensor field of order $m+1$ with components $(\nabla H)_{ki_1\ldots i_m}=\partial_kH_{i_1\ldots i_m}$, where $\partial_k$ is the weak partial derivative.
Following \cite[Section~5]{BalasubramanianKasiviswanathan2026}, a symmetric tensor field $H\in H^1(\mu)$ is \emph{compatible} if $\nabla H$ is symmetric in all indices.
Moreover, if $\psi\in H^1(\mu)$ is compatible and $\nabla\psi\in H^1(\mu)$, then $\nabla\psi$ is compatible, since $\nabla^2\psi$ is symmetric in its first two indices because weak derivatives commute, and in its last $m+1$ indices because $\nabla\psi$ is symmetric, where $m$ is the order of $\psi$.

With this notation, we define tensor fields $U_m$ by iterating the construction of $U_f$ from $f$: each $U_{m+1}$ is the gradient of the solution of a Poisson equation whose right-hand side is the centered part of $U_m$.
Starting with $U_0=f$, for $m\geq0$ let $\psi_m\in H^1(\mu)$ be a compatible tensor field of order $m$ with $\E_\mu\psi_m=0$ such that, for every compatible $H\in H^1(\mu)$ of order $m$,
\begin{equation}\label{eq:bk-tensor-poisson}
 \E_\mu\langle\nabla\psi_m,\nabla H\rangle
 =\E_\mu\langle U_m-\E_\mu U_m,H\rangle,
\end{equation}
and set $U_{m+1}=\nabla\psi_m$.
The equation \eqref{eq:bk-tensor-poisson} is a weak Poisson equation for the centered part of $U_m$.
We require $\psi_m$ to be compatible so that $U_{m+1}=\nabla\psi_m$ is again a compatible tensor field.
We require \eqref{eq:bk-tensor-poisson} only for compatible $H$, since all test fields to which we apply it below are compatible.
The existence and uniqueness of such a $\psi_m$ follow from the Riesz representation theorem on the space of centered compatible tensor fields of order $m$ in $H^1(\mu)$, equipped with the inner product $(H,K)\mapsto\E_\mu\langle\nabla H,\nabla K\rangle$, similarly as in \cref{lem:poisson-dual-norm}.

We now relate $U_1$ to the dual Sobolev norm.
For $m=0$, every $H\in H^1(\mu)$ is compatible, since $H$ and $\nabla H$ have order at most $1$ and are thus trivially symmetric, and $U_0-\E_\mu U_0=f$, so \eqref{eq:bk-tensor-poisson} is the weak Poisson equation \eqref{eq:poisson-weak}.
Hence $\psi_0=(-\cL)^{-1}f$ by the uniqueness in \cref{lem:poisson-dual-norm}, so $U_1=U_f$, and \cref{lem:poisson-dual-norm} gives
\begin{equation}\label{eq:bk-series-energy}
 \|f\|_{H^{-1}(\mu)}^2=\E_\mu\|U_1\|_2^2.
\end{equation}

It turns out that the means of the fields $U_d$ are related to the cumulants of $\mu$, that is, we will show that for every $d\geq1$,
\begin{equation}\label{eq:bk-series-coefficients}
 \E_\mu U_d=\frac{\kappa_d^\mu(f)}{d!}.
\end{equation}
Fix a symmetric tensor $T$ of order $d$ and let $p(x)=\langle T,A_d^\mu(x)\rangle/d!$.
We first show that $\nabla^dp=T$ and that $\E_\mu\nabla^jp=0$ for $0\leq j<d$.
Both claims are linear in $T$, and every symmetric tensor of order $d$ is a linear combination of tensors $z^{\otimes d}$ with $z\in\R^n$, so we may assume $T=z^{\otimes d}$.
For $k\geq0$, let $q_k(x)=\langle A_k^\mu(x),z^{\otimes k}\rangle/k!$, so that $p=q_d$.
Replacing $z$ by $\lambda z$ with $\lambda\in\R$ in the definition \eqref{eq:appell-expansion} of the Appell tensors gives
\begin{equation}\label{eq:appell-lambda}
 \frac{e^{\lambda\langle z,x\rangle}}{\E_\mu e^{\lambda\langle z,X\rangle}}
 =\sum_{k\geq0}\lambda^kq_k(x).
\end{equation}
Taking the gradient in $x$ multiplies the left-hand side of \eqref{eq:appell-lambda} by $\lambda z$, so comparing the coefficients of $\lambda^k$ on both sides gives $\nabla q_k=z\,q_{k-1}$ for $k\geq1$.
Applying this $j$ times gives, for $0\leq j\leq d$,
\[
 \nabla^jp=\nabla^jq_d=z^{\otimes j}q_{d-j}.
\]
Setting $\lambda=0$ in \eqref{eq:appell-lambda} gives $q_0=1$, so $\nabla^dp=z^{\otimes d}=T$.
Integrating \eqref{eq:appell-lambda} against $\mu$ in $x$ turns the left-hand side into $1$ for every $\lambda$, so $\E_\mu q_k=0$ for $k\geq1$, and hence $\E_\mu\nabla^jp=z^{\otimes j}\E_\mu q_{d-j}=0$ for $j<d$.
The fields $\nabla^jp$ are compatible since the derivatives of the polynomial $p$ commute, and they belong to $H^1(\mu)$ since the log-concave measure $\mu$ has finite moments of all orders.
For $0\leq j<d$, the weak Poisson equation \eqref{eq:bk-tensor-poisson} with $m=j$ and $H=\nabla^jp$ gives
\[
 \E_\mu\langle U_{j+1},\nabla^{j+1}p\rangle
 =\E_\mu\langle U_j-\E_\mu U_j,\nabla^jp\rangle
 =\E_\mu\langle U_j,\nabla^jp\rangle,
\]
where the last equality holds since $\E_\mu\nabla^jp=0$.
Starting from $\kappa_d^\mu(f)=\E_\mu[fA_d^\mu]$ and $U_0=f$, and applying this identity for $j=0,\ldots,d-1$, we obtain
\[
 \frac{\langle\kappa_d^\mu(f),T\rangle}{d!}
 =\E_\mu[fp]
 =\E_\mu\langle U_1,\nabla p\rangle
 =\cdots
 =\E_\mu\langle U_d,\nabla^dp\rangle
 =\langle\E_\mu U_d,T\rangle.
\]
Both $\kappa_d^\mu(f)$ and $\E_\mu U_d$ are symmetric, because $U_d=\nabla\psi_{d-1}$ with $\psi_{d-1}$ compatible, so \eqref{eq:bk-series-coefficients} follows since $T$ is an arbitrary symmetric tensor.

With the means identified, we reduce \eqref{eq:cumulant-energy-series} to a bound on the centered part of $U_1$.
For $m\geq0$, let $v_m=\E_\mu\|U_m-\E_\mu U_m\|_{\mathrm{HS}}^2$ be the variance of $U_m$, and for $m\geq1$, let $c_m=\|\E_\mu U_m\|_{\mathrm{HS}}^2$, which equals $\|\kappa_m^\mu(f)\|_{\mathrm{HS}}^2/(m!)^2$ by \eqref{eq:bk-series-coefficients}.
Then $v_0=\E_\mu f^2=1$, and for $m\geq1$,
\begin{equation}\label{eq:bk-series-split}
 \E_\mu\|U_m\|_{\mathrm{HS}}^2
 =\E_\mu\|U_m-\E_\mu U_m\|_{\mathrm{HS}}^2+\|\E_\mu U_m\|_{\mathrm{HS}}^2
 =v_m+c_m.
\end{equation}
Hence, by \eqref{eq:bk-series-energy} and \eqref{eq:bk-series-split} with $m=1$, it suffices to show
\begin{equation}\label{eq:bk-series-scalar}
 v_1\leq s+\sum_{m\geq2}\frac{c_m}{s^{m-1}},
\end{equation}
since then
\[
 \|f\|_{H^{-1}(\mu)}^2
 =v_1+c_1
 \leq s+\sum_{m\geq1}\frac{c_m}{s^{m-1}}
 =s+\sum_{d\geq1}\frac{\|\kappa_d^\mu(f)\|_{\mathrm{HS}}^2}{(d!)^2s^{d-1}}.
\]

To bound $v_1$, we control the gradient of each field by the centered part of the previous field.
The following curvature estimate from \cite[Lemmas~5.1 and~6.1]{BalasubramanianKasiviswanathan2026} is the main analytic input of this proof. For every $m\geq0$,
\begin{equation}\label{eq:bk-tensor-curvature}
 \E_\mu\|\nabla U_{m+1}\|_{\mathrm{HS}}^2+a\E_\mu\|U_{m+1}\|_{\mathrm{HS}}^2
 \leq\E_\mu\|U_m-\E_\mu U_m\|_{\mathrm{HS}}^2,
\end{equation}
which also shows that $U_{m+1}\in H^1(\mu)$ (for $m=0$, it follows from Bochner's formula).
Since $\psi_m$ is compatible and $U_{m+1}=\nabla\psi_m\in H^1(\mu)$, the field $U_{m+1}$ is also compatible.

We then derive a recursion for the variances $v_m$, which forces them to grow quickly when $v_1$ is large.
Let $m\geq1$.
The field $U_m$ is compatible and belongs to $H^1(\mu)$, so we may take $H=U_m$ in the weak Poisson equation \eqref{eq:bk-tensor-poisson}, whose right-hand side is then $\E_\mu\langle U_m-\E_\mu U_m,U_m\rangle=v_m$.
Hence, by the Cauchy--Schwarz inequality,
\begin{equation}\label{eq:bk-series-cs}
 \begin{aligned}
 v_m^2
 &=\bigl(\E_\mu\langle U_{m+1},\nabla U_m\rangle\bigr)^2\\
 &\leq\E_\mu\|U_{m+1}\|_{\mathrm{HS}}^2\,\E_\mu\|\nabla U_m\|_{\mathrm{HS}}^2\\
 &\leq(v_{m+1}+c_{m+1})\bigl(v_{m-1}-a\E_\mu\|U_m\|_{\mathrm{HS}}^2\bigr)\\
 &\leq(v_{m+1}+c_{m+1})(v_{m-1}-av_m),
 \end{aligned}
\end{equation}
where the second inequality follows from \eqref{eq:bk-tensor-curvature} with $m-1$, and the last inequality holds since $\E_\mu\|U_m\|_{\mathrm{HS}}^2\geq v_m$.
Now assume $v_{m-1},v_m>0$ and let $r_m=v_m/v_{m-1}$.
Since $v_m^2>0$ and $v_{m+1}+c_{m+1}\geq0$, \eqref{eq:bk-series-cs} forces $v_{m-1}-av_m>0$, that is, $r_m<1/a$.
Dividing \eqref{eq:bk-series-cs} by $v_m(v_{m-1}-av_m)>0$ gives
\[
 \frac{r_m}{1-ar_m}
 =\frac{v_m}{v_{m-1}-av_m}
 \leq\frac{v_{m+1}+c_{m+1}}{v_m}
 =r_{m+1}+\frac{c_{m+1}}{v_m}.
\]
Rearranging and using $1/(1-x)\geq1+x$ for $0\leq x<1$ with $x=ar_m$, we get
\begin{equation}\label{eq:bk-series-recursion}
 r_{m+1}
 \geq\frac{r_m}{1-ar_m}-\frac{c_{m+1}}{v_m}
 \geq r_m+ar_m^2-\frac{c_{m+1}}{v_m}.
\end{equation}

Using this recursion, we prove \eqref{eq:bk-series-scalar} by contradiction: if $v_1$ were larger, then the ratios $r_m$ would tend to infinity, contradicting $r_m<1/a$.
Let $S=\sum_{m\geq2}c_m/s^{m-1}$.
If $S=\infty$, there is nothing to prove, so assume $S<\infty$ and $v_1>s+S$.
We show by induction that for every $m\geq1$, we have $v_m>s^m$ and
\begin{equation}\label{eq:bk-series-induction}
 r_m\geq v_1+(m-1)as^2-\sum_{j=2}^{m}\frac{c_j}{s^{j-1}}.
\end{equation}
The right-hand side of \eqref{eq:bk-series-induction} is at least $v_1+(m-1)as^2-S\ge v_1-S>s$, so \eqref{eq:bk-series-induction} implies $r_m>s$.
For $m=1$, both claims hold since $r_1=v_1/v_0=v_1>s$.
If both claims hold for some $m\geq1$, then $v_m>0$ and $v_{m-1}=v_m/r_m>0$, so from \eqref{eq:bk-series-recursion},
\[
 r_{m+1}
 \geq r_m+ar_m^2-\frac{c_{m+1}}{v_m}
 \geq r_m+as^2-\frac{c_{m+1}}{s^m}
 \geq v_1+mas^2-\sum_{j=2}^{m+1}\frac{c_j}{s^{j-1}},
\]
and hence $v_{m+1}=r_{m+1}v_m>s\cdot s^m=s^{m+1}$.
This completes the induction.
Since the right-hand side of \eqref{eq:bk-series-induction} tends to infinity as $m\to\infty$, so does $r_m$, contradicting $r_m<1/a$.
This proves \eqref{eq:bk-series-scalar}, and hence the desired bound \eqref{eq:cumulant-energy-series}.

Finally, we prove the last assertion.
Under \eqref{eq:cumulant-geometric}, taking $s=10R^2$ in \eqref{eq:cumulant-energy-series} gives
\[
 \|f\|_{H^{-1}(\mu)}^2
 \leq10R^2+\sum_{d\geq1}\frac{R^{2d}}{(10R^2)^{d-1}}
 =10R^2+R^2\sum_{k\geq0}10^{-k}
 \lesssim R^2.
\]
Since this holds for every centered $f\in L^2(\mu)$ of unit variance, \cref{lem:dual-poincare} gives $\CP(\mu)\lesssim R^2$.
\end{proof}

In the rest of this section, we explain how this proof strategy can be interpreted, and in fact motivated, from the Bayesian estimation viewpoint of the Poincar\'e inequality.

\subsection{The low-degree MMSE approach to the KLS conjecture: the Schramm--Wein method}\label{sec:SW_cumulants}
In what follows, let us focus on the usual by now Gaussian additive model, where the prior is $X\sim\mu$, and the statistician observes $Y_t=\sqrt tX+G.$ Recall also the definition of the MMSE of any label $f$, $\mmse_t(f),$ of the model, \eqref{eq:dfn_mmse_PI}.

Based on \cref{cor:poincare-mmse}, the following corollary immediately holds.
\begin{corollary}
    The following are equivalent. 
    \begin{enumerate}[label=(\roman*)]
        \item The KLS conjecture: for some universal $C>0$, for any isotropic log-concave measure, $\CP(\mu) \leq C.$ 
         
        \item For some universal constants $c,C>0,$ for any isotropic log-concave measure $\mu$ and any centered $f \in L^2(\mu)$ with unit variance it holds for $t=C,$ \[\mmse_t(f) \geq c.\]
    \end{enumerate}
\end{corollary}

There are of course multiple ways to approach the KLS conjecture, even from this Bayesian information-theoretic angle. In fact, \cref{sec:balanced} employs some of the natural approaches, which fall short of proving the KLS conjecture but give new direct proofs of the Poincar\'e constant bounds of \citet{KLS1995} and \citet{LeeVempala}.

Another reasonable way to do this is via first studying the easier $D$-degree MMSE which first restricts the estimators to $D$-degree polynomials \cite{SchrammWein2022}. Specifically, for an integer $D\geq0$, define
\[
 \mmse_{t,\leq D}(f)
 :=\inf_{q:\deg q\leq D}\E\bigl(f(X)-q(Y_t)\bigr)^2.
\]
For a log-concave prior $\mu$, the moment generating function of $Y_t$ is finite in a
neighborhood of the origin, so polynomials are dense in its $L^2$ space.
Consequently $\mmse_{t,\leq D}(f)\to\mmse_t(f)$ as $D\to\infty$.
Thus if one is able to prove a lower bound uniform in $D$ on $ \mmse_{t,\leq D}(f)$ it will lead to the desired $\mmse_t(f)$ information-theoretic lower bound.

One might wonder at this point why the $D$-degree MMSE would be easier to bound. 
The main reason is that over the last decade there has been a tremendous amount of work aiming to bound the $D$-degree MMSE in the literature of computational-statistical trade-offs \cite{SchrammWein2022,SohnWein2025,MaoWeinZhang2023,LuoGao2024,EvenGiraudVerzelen2024,MontanariWein2025,Semerjian2024,WeinSurvey25,CGGV2026}.
The motivation there is that $D$-degree estimators with $D=O(\log n)$ might be a tractable proxy for polynomial-time computation \cite{HKPPRSS2017,HopkinsSteurer2017,HopkinsThesis18}.
Independently though of the motivation, which is not directly relevant to this work, 
one might try multiple methods to lower bound $ \mmse_{t,\leq D}(f)$ for GAMs with any isotropic log-concave prior $\mu.$

Perhaps surprisingly, the perhaps most relevant tool in this community for such a lower bound is a seminal approach by Schramm and Wein which is fundamentally related with the cumulant tensor $\kappa_d^\mu(f)$ used in the proofs of the KLS conjecture. To be more precise, the cumulant bound of \cite[Theorem~2.2 and Claim~2.14]{SchrammWein2022},
written in this notation, gives for any prior $\mu,$
\begin{equation}\label{eq:cumulant-low-degree}
1-\mmse_{t,\leq D}(f)
 \leq\sum_{d=1}^D\frac{t^d}{d!}
             \|\kappa_d^\mu(f)\|_{\mathrm{HS}}^2.
\end{equation}

The resemblance between \eqref{eq:cumulant-low-degree} and \eqref{eq:cumulant-energy-series} is striking, and we argue that this at least somewhat makes the use of cumulants as a justifiable approach for the KLS conjecture. We remark though that the bounds are different. \eqref{eq:cumulant-energy-series} is a stronger bound per term, as it contains a factor $d!$ less compared to \eqref{eq:cumulant-low-degree}. On the other hand, \eqref{eq:cumulant-low-degree} holds for any prior $\mu$ and in particular does not require log-concavity. On top of it, its proof is a smart application of Jensen's inequality, on top of routine linear algebraic calculations.

One can nevertheless apply the all-order cumulant bounds  \eqref{eq:cumulant-geometric} on the Schramm--Wein bound, to obtain
\begin{equation}\label{eq:cumulant-low-degree-SW22}
1-\mmse_{t,\leq D}(f)
 \leq  \sum_{d=1}^D d!(tR^2)^d
\end{equation}
with a universal $R$, uniformly over the relevant priors $\mu$ and labels $f$. In particular, for constant $t=\Theta(1)$ the upper bound diverges as $D$ grows, making at least the direct application of the Schramm--Wein bound insufficient to obtain the desired uniform lower bound in $D$ of $\mmse_{t,\leq D}(f)$. That being said, in the field of computational-statistical thresholds it is known that the cumulant bound by Schramm--Wein is often non-tight. For these reasons, tighter methods have been recently obtained, one of which is proposed by \citet{SohnWein2025}.  

\subsection{The low-degree MMSE approach to the KLS conjecture: the Sohn--Wein method}\label{sec:sohn1}

The method by \citet[Proposition~1.3]{SohnWein2025} to lower bound the $D$-degree MMSE is as follows. Their main observation is that if for any GAM with observation $Y_t = \sqrt{t}X+G$ we define $\Xi_{\mu,t,D}$ as the set of all square-integrable random variables $\eta(X,G)$ such that $\eta(X,G)-f(X)$ is in the orthogonal complement of all polynomials in $\mathbb{R}_{\leq D}[Y]$, i.e., for all polynomials $q(Y)$ of degree at most $D$ it holds,
\[\E[\eta(X,G) q(Y_t)]=\E[f(X)q(Y_t)],\]then
\begin{equation}\label{eq:cumulant-low-degree-SW25}
1-\mmse_{t,\leq D}(f)
 \le\inf_{\eta \in \Xi_{\mu,t,D}}  \E[\eta^2(X,G)].
\end{equation}
In fact, equality holds in \eqref{eq:cumulant-low-degree-SW25} by the same orthogonal projection argument as in the proof of \cref{prop:sohn-wein}, with the conditional expectation replaced by the optimal degree-$D$ polynomial estimator.
Depending on the GAM, researchers have been constructing appropriate ``certificates'' $\eta$ with small $\E[\eta^2(X,G)]$ leading to surprisingly tight bounds on the $D$-degree MMSE in multiple settings \cite{SohnWein2025,CMSW25,Li26,FuSohn26}.

While the motivation in this work is to focus on finite $D$-degree polynomials, one might immediately generalize their argument to the case $D=\infty$ to capture the information-theoretic MMSE. We present here this easy extension as a separate proposition for completeness.

\begin{proposition}\label{prop:sohn-wein}
    Fix any prior $\mu$, any $t>0$ and any $f\in L^2(\mu)$ with $\E_\mu f^2=1$, and define $\Xi_{\mu,t,\infty}$ as the set of all square-integrable random variables $\eta(X,G)$ such that for every $q:\R^n\to\R$ with $\E[q(Y_t)^2]<\infty$,
\begin{equation}\label{eq:sohn_wein}
 \E[\eta(X,G)q(Y_t)]=\E[f(X)q(Y_t)].
\end{equation}
Then
\begin{equation}\label{eq:cumulant-low-degree_2}
1-\mmse_{t}(f)
 =\inf_{\eta \in \Xi_{\mu,t,\infty}}  \E[\eta^2(X,G)].
\end{equation}
\end{proposition}

\begin{proof}
Let $m(y)=\E_{\mu_{t,y}}f$ be the posterior mean, so that $m(Y_t)=\E[f(X)\mid Y_t]$.
By \cref{lem:bayes-risk} with $\Phi(z)=z^2$,
\[
 \mmse_t(f)
 =\var_\mu(f)-\var_{\mu K_t}(m)
 =1-\E[m(Y_t)^2],
\]
where the last equality holds since $\E[m(Y_t)]=\E[f(X)]$ and $\E[f(X)^2]=1$.
Hence it suffices to show that the infimum in \eqref{eq:cumulant-low-degree_2} equals $\E[m(Y_t)^2]$.
For every $\eta\in\Xi_{\mu,t,\infty}$, taking $q=m$ in \eqref{eq:sohn_wein} and using the tower property gives $\E[m(Y_t)\eta(X,G)]=\E[m(Y_t)f(X)]=\E[m(Y_t)^2]$, so
\[
 \begin{aligned}
 \E[\eta(X,G)^2]
 &=\E[m(Y_t)^2]+2\E\bigl[m(Y_t)\bigl(\eta(X,G)-m(Y_t)\bigr)\bigr]+\E\bigl[\bigl(\eta(X,G)-m(Y_t)\bigr)^2\bigr]\\
 &=\E[m(Y_t)^2]+\E\bigl[\bigl(\eta(X,G)-m(Y_t)\bigr)^2\bigr]
 \geq\E[m(Y_t)^2].
 \end{aligned}
\]
Equality holds for $\eta(X,G)=m(Y_t)$, which belongs to $\Xi_{\mu,t,\infty}$ since $\E[m(Y_t)q(Y_t)]=\E[f(X)q(Y_t)]$ for every $q$ by the tower property.
\end{proof}

\subsection{Constructing small Sohn--Wein certificates from \texorpdfstring{$\nabla(-\cL)^{-1}f$}{the Poisson equation}}\label{sec:sohn2}
In light of the above Sohn--Wein method, a natural now approach to solve the KLS conjecture would be to find for any isotropic log-concave measure a certificate with small norm. We now want to argue that $U_f$ leads to a very natural such certificate. In what follows, we call the random variables $\eta(X,G)$ satisfying \eqref{eq:sohn_wein} as Sohn--Wein certificates.

Let $\mu$ be regular and let $f\in L^2(\mu)$ be centered with $\E_\mu f^2=1$.
By \eqref{eq:poisson-weak} and \eqref{eq:poisson}, for every $h\in H^1(\mu)$, the field $U=U_f$ satisfies
\begin{equation}\label{eq:cumulant-poisson-ibp}
 \E_\mu[f(X)h(X)]=\E_\mu\langle U(X),\nabla h(X)\rangle.
\end{equation}
Moreover, $U_f$ has the smallest $L^2(\mu)$-norm among all vector fields $U\in L^2(\mu)$ satisfying \eqref{eq:cumulant-poisson-ibp} for every $h\in H^1(\mu)$.
Indeed, for such $U$ and every $h\in H^1(\mu)$ with $\E_\mu\|\nabla h\|_2^2\leq1$, the Cauchy--Schwarz inequality gives
\[
 \E_\mu[fh]=\E_\mu\langle U,\nabla h\rangle\leq\bigl(\E_\mu\|U\|_2^2\bigr)^{1/2}.
\]
Taking the supremum over $h$ and applying \cref{lem:poisson-dual-norm} gives $\E_\mu\|U\|_2^2\geq\|f\|_{H^{-1}(\mu)}^2=\E_\mu\|U_f\|_2^2$.

Now, interestingly any $U$ satisfying \eqref{eq:cumulant-poisson-ibp} allows us to create a Sohn--Wein certificate with the same $L^2(\mu)$-norm (up to a factor of \(\sqrt{t}\)). In particular, $U_f$ leads to the minimum-norm Sohn--Wein certificate among this class.
Indeed, define
\[
 \eta_U(X,G):=\sqrt{t}\langle G,U(X)\rangle,
\]which satisfies $\E[\eta_U(X,G)^2]=t\E[\|U(X)\|^2_2].$ We claim $\eta_U$ is a Sohn--Wein certificate.

Indeed, for any $q$ let $h(x)=\E_G[q(\sqrt{t}x+G)]$.
Applying Stein's lemma \cite{Stein1981} to $g(z)=q(\sqrt tx+z)$ gives
\[
 \nabla h(x)
 =\sqrt t\,\E_G[\nabla q(\sqrt tx+G)]
 =\sqrt t\,\E_G[G\,q(\sqrt tx+G)].
\]
Since $G$ is independent of $X$, we obtain
\begin{align*}
 \E[f(X)q(Y_t)]
 &=\E[f(X)h(X)]\\
 &=\E\langle U(X),\nabla h(X)\rangle\\
 &=\sqrt t\,\E[\langle G,U(X)\rangle q(Y_t)]\\
 &=\E[\eta_U(X,G)q(Y_t)],
\end{align*}
where the second equality follows from \eqref{eq:cumulant-poisson-ibp}.
Since this holds for every $q$, the random variable $\eta_U(X,G)$ satisfies \eqref{eq:sohn_wein}, that is, it is a Sohn--Wein certificate.

Using that, by \cref{prop:sohn-wein}, for any $U$ satisfying \eqref{eq:cumulant-poisson-ibp},
\[1-\mmse_{t}(f) \leq t\E[\|U(X)\|^2_2].\]In particular, optimizing over such $U$'s gives that $U_f$ satisfies \[1-t\E[\|U_f(X)\|^2_2] \leq \mmse_{t}(f).\]

We can improve this lower bound by combining $\eta_{U_f}(X,G)$ with the trivial Sohn--Wein certificate $f(X)$.
Let $b=t\E[\|U_f(X)\|_2^2]$.
Since \eqref{eq:sohn_wein} is preserved under affine combinations, for every $\lambda\in\R$ the random variable $\lambda f(X)+(1-\lambda)\eta_{U_f}(X,G)$ is a Sohn--Wein certificate.
Since $G$ is centered and independent of $X$, we have $\E[f(X)\eta_{U_f}(X,G)]=\sqrt t\,\E[f(X)\langle\E[G],U_f(X)\rangle]=0$, so
\[
 \E\bigl[\bigl(\lambda f(X)+(1-\lambda)\eta_{U_f}(X,G)\bigr)^2\bigr]
 =\lambda^2\E[f(X)^2]+(1-\lambda)^2\E[\eta_{U_f}(X,G)^2]
 =\lambda^2+(1-\lambda)^2b,
\]
where the last equality holds since $\E[f(X)^2]=1$ and $\E[\eta_{U_f}(X,G)^2]=t\E[\|U_f(X)\|_2^2]=b$.
The right-hand side is minimized at $\lambda=b/(1+b)$, where it equals
\[
 \frac{b^2}{(1+b)^2}+\frac{b}{(1+b)^2}
 =\frac{b}{1+b}.
\]
Hence \cref{prop:sohn-wein} gives $1-\mmse_t(f)\leq b/(1+b)$, that is,
\begin{equation}\label{eq:poisson_SW}
 \mmse_t(f)\geq\frac{1}{1+t\E[\|U_f(X)\|_2^2]}.
\end{equation}
Therefore, to lower bound the MMSE, it remains to upper bound $\E[\|U_f(X)\|_2^2]$.

\subsection{Concluding the proof of the KLS conjecture}
We now combine the lower bound \eqref{eq:poisson_SW} with the analytic input \cref{prop:cumulant-dual-norm} to conclude the proof of the KLS conjecture.
Let $\mu$ be an isotropic log-concave measure, which we may assume to be regular (see e.g., \cite[Section~2]{BizeulKlartagLehec2026}), and let $f\in L^2(\mu)$ be centered with $\E_\mu f^2=1$.
By \eqref{eq:poisson_SW}, for every $t,s>0$,
\[
 \mmse_t(f)
 \geq\frac{1}{1+t\E[\|U_f(X)\|_2^2]}
 =\frac{1}{1+t\|f\|_{H^{-1}(\mu)}^2}
 \geq\biggl(1+ts+t\sum_{d\geq1}\frac{\|\kappa_d^\mu(f)\|_{\mathrm{HS}}^2}{(d!)^2s^{d-1}}\biggr)^{-1},
\]
where the equality follows from \cref{lem:poisson-dual-norm}, and the last inequality follows from \cref{prop:cumulant-dual-norm}.
Unlike the Schramm--Wein bound \eqref{eq:cumulant-low-degree-SW22}, whose right-hand side diverges as $D\to\infty$, this lower bound stays bounded away from zero under the cumulant bound \eqref{eq:cumulant-geometric}.
Indeed, taking $t=1$ and $s=10R^2$, \eqref{eq:cumulant-geometric} gives
\[
 \sum_{d\geq1}\frac{\|\kappa_d^\mu(f)\|_{\mathrm{HS}}^2}{(d!)^2(10R^2)^{d-1}}
 \leq\sum_{d\geq1}\frac{R^{2d}}{(10R^2)^{d-1}}
 =R^2\sum_{k\geq0}10^{-k}
 =\frac{10}{9}R^2,
\]
and hence
\[
 \mmse_1(f)
 \geq\frac{1}{1+10R^2+\frac{10}{9}R^2}
 =\frac{9}{9+100R^2}.
\]
Since this holds for every centered $f\in L^2(\mu)$ with unit variance, and $\mmse_1(c+\sigma f)=\sigma^2\mmse_1(f)$ for all constants $c,\sigma\in\R$, every $f\in\cA_0^\R$ satisfies $\mmse_1(f)\geq\frac{9}{9+100R^2}\var_\mu(f)$.
Hence the second part of \cref{cor:poincare-mmse} with $T=1$ gives $\CP(\mu)\lesssim1$, since $R$ is a universal constant.

\section{Proof of the main equivalence and direct corollaries}\label{sec:equivalence}

We start with introducing some necessary notation.

\paragraph{Posterior distribution} Let $\mu$ be a probability measure on $\R^n$, and let $Y_t$ be the GAM per \cref{dfn:GAM} for prior $\mu$ and SNR $t$. An important object is the so-called posterior of the GAM with prior $\mu$ which we denote by $\mu_{t,y}=\Law(X\mid Y_t=y)$ for any fixed \(Y_t = y\).
Bayes' theorem gives
\begin{equation}\label{eq:posterior}
 \mu_{t,y}(\mathrm{d}x)
 =\frac{e^{-\|y-\sqrt t\,x\|_2^2/2}\,\mu(\mathrm{d}x)}{\int e^{-\|y-\sqrt t\,x'\|_2^2/2}\,\mu(\mathrm{d}x')}.
\end{equation}

\begin{lemma}[Posterior log-concavity]\label{lem:posterior-log-concavity}
If $\mu$ is log-concave, then for every $t>0$ and $y\in\R^n$, the posterior $\mu_{t,y}$ is $t$-strongly log-concave, that is, $\mu_{t,y}(\mathrm{d}x)=e^{-W(x)}\mathrm{d}x$ with $\nabla^2W\succeq tI_n$.
\end{lemma}
\begin{proof}
Write $\mu(\mathrm{d}x)=e^{-V(x)}\mathrm{d}x$ with $V$ convex.
By \eqref{eq:posterior}, $\mu_{t,y}$ has density proportional to $e^{-W(x)}$ with $W(x)=V(x)+t\|x\|_2^2/2-\sqrt t\langle y,x\rangle$, and $\nabla^2W=\nabla^2V+tI_n\succeq tI_n$.
\end{proof}

\paragraph{Planted vs null detection} Recall from \cref{sec:bayesian-estimation} the kernel $K_{t,x}=\cN(\sqrt t\,x,I_n)$, the law $\zeta K_t$ of the observation $Y_t$ when \(X \sim \zeta\), and the planted model $P_t=\Law(X,Y_t)$. Similar to the above sections, it is useful to adopt a hypothesis testing viewpoint. For prior $\mu$, we compare the planted model with the ``null'' (or ``pure noise'') model $\mu\otimes\gamma_n=\Law(X,Y_0)$, under which the observation is independent of the signal. The likelihood ratio (or Radon--Nikodym derivative) of the observation $Y_t$ relative to the pure noise $Y_0$ is
\begin{equation}\label{eq:likelihood}
 L_t(y):=\frac{\mathrm{d}(\mu K_t)}{\mathrm{d}\gamma_n}(y)
 =\int\exp\!\left(\sqrt t\langle y,x\rangle-\frac t2\|x\|_2^2\right)\mu(\mathrm{d}x).
\end{equation}Notice also that the posterior satisfies for any $y,$ $\mu_{t,y}(\mathrm{d}x) =\frac{\exp(\sqrt t\langle y,x\rangle-t\|x\|_2^2/2)}{L_t(y)}\,\mu(\mathrm{d}x).$

By Bayes' theorem, also referred to in the statistical physics literature as the Nishimori identity \cite{Nishimori2001,Iba1999}, the planted model can be sampled in either order: draw $X\sim\mu$ and then $Y_t\sim K_{t,X}$, or draw $Y_t\sim\mu K_t$ and then $X\sim\mu_{t,Y_t}$. That is,
\[
 P_t(\mathrm{d}x,\mathrm{d}y)=\mu(\mathrm{d}x)\,K_{t,x}(\mathrm{d}y)=(\mu K_t)(\mathrm{d}y)\,\mu_{t,y}(\mathrm{d}x).
\]
Integrating out $y$ shows that averaging the posterior over the observation recovers the prior,
\begin{equation}\label{eq:posterior-average}
 \E_{y\sim Y_t}\mu_{t,y}=\mu.
\end{equation}

\paragraph{Bayes risk decomposition} We will use the following standard decomposition of the Bayes risk \cite{BanerjeeGuoWang2005}.

\begin{lemma}[Bayes risk decomposition]\label{lem:bayes-risk}
Let $J$ be an open interval, let $\Phi:J\to\R$ be smooth and convex, let $f$ be a $J$-valued function, and let $t>0$.
Then the posterior mean $m(y)=\E_{\mu_{t,y}}f$ attains the infimum in the $\Phi$-optimal risk \(
 \cR_\Phi(t,f)=\inf_h\E B_\Phi(f(X),h(Y_t))\), and
\begin{equation}\label{eq:bayes-risk}
 \cR_\Phi(t,f)
 =\E_{y\sim Y_t}\Ent_{\mu_{t,y}}^\Phi(f)
 =\Ent_\mu^\Phi(f)-\Ent_{\mu K_t}^\Phi(m).
\end{equation}
In particular, for $\Phi(z)=z^2$ on $J=\R$, we have $\mmse_t(f)=\E\var(f(X)\mid Y_t)$.
\end{lemma}
\begin{proof}
Since $J$ is an interval, $m(y)\in J$.
For every $y$ and every $a\in J$, expanding the Bregman divergence gives
\[
 \E_{\mu_{t,y}}B_\Phi(f,a)
 =\E_{\mu_{t,y}}\Phi(f)-\Phi(a)-\Phi'(a)\bigl(m(y)-a\bigr)
 =\Ent_{\mu_{t,y}}^\Phi(f)+B_\Phi(m(y),a).
\]
Since $\mu_{t,y}$ is the conditional law of $X$ given $Y_t=y$ and $B_\Phi\geq0$ by convexity, taking $a=h(y)$ and averaging over $y\sim Y_t$ gives, for every measurable $h$ with values in $J$,
\[
 \E B_\Phi(f(X),h(Y_t))
 =\E_{y\sim Y_t}\bigl[\Ent_{\mu_{t,y}}^\Phi(f)+B_\Phi(m(y),h(y))\bigr]
 \geq\E_{y\sim Y_t}\Ent_{\mu_{t,y}}^\Phi(f),
\]
with equality for $h=m$.
Taking the infimum over $h$ shows that $h=m$ attains the infimum, and proves the first equality in \eqref{eq:bayes-risk}.
The second equality is the entropy decomposition (\cref{lem:entropy-decomposition}) for the mixture $\mu=\E_{y\sim Y_t}\mu_{t,y}$ in \eqref{eq:posterior-average}.
For $\Phi(z)=z^2$, we have $B_\Phi(z,a)=(z-a)^2$ and $\Ent_{\mu_{t,y}}^\Phi(f)=\var(f(X)\mid Y_t=y)$, which gives the last claim.
\end{proof}
The proof only uses that $\mu_{t,y}$ is the conditional law of $X$ given $Y_t=y$, so \cref{lem:bayes-risk} holds for the output $Y$ of any channel, with the posterior $\Law(X\mid Y=y)$ in place of $\mu_{t,y}$ and $Y$ in place of $Y_t$.

\subsection{The main equivalence: proof of \texorpdfstring{\cref{thm:general-equivalence}}{Theorem \getrefnumber{thm:general-equivalence}}}

\begin{proof}
Throughout the proof, $f\in\cA_0^J$, and $m(y)=\E_{\mu_{t,y}}f$ is the posterior mean. We start with a standard observation that the posterior mean is the optimal estimator achieving the $\Phi$-optimal risk.
\cref{lem:bayes-risk} gives
\begin{equation}\label{eq:phi-residual}
 \cR_\Phi(t,f)
 =\E_{y\sim Y_t}\Ent_{\mu_{t,y}}^\Phi(f)
 =\Ent_\mu^\Phi(f)-\Ent_{\mu K_t}^\Phi(m).
\end{equation}

By \cref{lem:posterior-log-concavity}, the posterior $\mu_{t,y}$ is $t$-strongly log-concave,
which implies by Bakry--\'Emery (\cref{fact:bakry-emery}) that the posterior $\mu_{t,y}$ satisfies the $\Phi$-Sobolev inequality with constant \(1/t\), 
\[
 \Ent_{\mu_{t,y}}^\Phi(f)
 \leq\frac1{2t}\E_{\mu_{t,y}}[\Phi''(f)\|\nabla f\|_2^2].
\]
Averaging over \(y \sim Y_t\) and using \eqref{eq:posterior-average} and \eqref{eq:phi-residual} therefore gives the following key inequality,
\begin{equation}\label{eq:key_step_main_proof}
 \cR_\Phi(t,f)\leq\frac1{2t}\E_\mu[\Phi''(f)\|\nabla f\|_2^2].   
\end{equation}

\paragraph{The ``impossibility of estimation $\Rightarrow$ $\Phi$-Sobolev inequality'' direction.} The above inequality \eqref{eq:key_step_main_proof} allows us to prove directly this direction. If \eqref{eq:phi-risk} holds for every $t>0$, then \eqref{eq:key_step_main_proof} yields \[\Ent_\mu^\Phi(f)\le (1+Ct) \cR_\Phi(t,f)\leq\frac{1+Ct}{2t}\E_\mu[\Phi''(f)\|\nabla f\|_2^2].\] Letting \(t\to+\infty\) yields the $\Phi$-Sobolev inequality \eqref{eq:phi-functional}.

We can also directly prove the last statement of the theorem. If $\cR_\Phi(T,f)\geq b\Ent_\mu^\Phi(f)$ for every $f\in\cA_0^J$, then \eqref{eq:key_step_main_proof} gives  \[\Ent_\mu^\Phi(f)\le \frac{1}{b} \cR_\Phi(T,f)\leq\frac1{2bT}\E_\mu[\Phi''(f)\|\nabla f\|_2^2].\] 

\paragraph{The ``$\Phi$-Sobolev inequality $\Rightarrow$ impossibility of estimation'' direction.}For the converse direction, we deduce \eqref{eq:phi-risk} by using the contraction property of $\Phi$-divergences under the \(\Phi\)-Sobolev inequality (\cref{fact:divergence-contraction}).
Recall from \cref{def:phi-divergence} that for a convex $\Psi$ with $\Psi(1)=0$, the $\Psi$-divergence is $D_\Psi(\nu\Vert\mu)=\E_\mu\Psi(\mathrm{d}\nu/\mathrm{d}\mu)$.

Fix $f$.
We will consider the probability measure with density proportional to $f$, so we first reduce to $f>0$.
For $c>-\min f$, replacing $f$ by $f+c$, $\Phi$ by $\Phi(\cdot-c)$, and $J$ by $J+c$ changes neither the assumptions nor the two sides of \eqref{eq:phi-functional} and \eqref{eq:phi-risk}, so we may assume that $f>0$.

Now, consider the convex function \(\Psi(r)=\Phi(r\E_\mu f)-\Phi(\E_\mu f)\) for \(r\in J/\E_\mu f\), which satisfies $\Psi(1)=0$.
 
 We first prove that  \(\mu\) satisfies the \(\Psi\)-Sobolev inequality with constant $C$. Indeed, for any $g\in\cA_0^{J/\E_\mu f}$, we have $g\E_\mu f\in\cA_0^J$, $\Psi''(g)=(\E_\mu f)^2\Phi''(g\E_\mu f)$, and  $\Ent_\mu^\Psi(g)=\Ent_\mu^\Phi(g\E_\mu f)$. 
Since \(\mu\) satisfies the \(\Phi\)-Sobolev inequality with constant $C$, applying \eqref{eq:phi-functional} to $g\E_\mu f$ shows for any \(g\),
\[
 \Ent_\mu^\Psi(g)=\Ent_\mu^\Phi(g\E_\mu f)\leq \frac{C}{2} \E_\mu\bigl[(\E_\mu f)^2\Phi''(g\E_\mu f) \|\nabla g\|_2^2 \bigr] =\frac C2\E_\mu\bigl[\Psi''(g)\|\nabla g\|_2^2\bigr].
\]
Hence, $\mu$ satisfies the $\Psi$-Sobolev inequality with constant $C$ as we wanted.

Next, define the probability measure $\nu$ satisfying
\(
 \frac{\mathrm{d}\nu}{\mathrm{d}\mu}=\frac f{\E_\mu f}\). Notice for our choice of $\Psi,$ \begin{equation}\label{eq:RHS_KL} D_\Psi(\nu\Vert\mu)=\E_\mu\Phi(f)-\Phi(\E_\mu f)=\Ent_\mu^\Phi(f).\end{equation}
Since the posterior $\mu_{t,y}$ is the conditional law of $X$ given $Y_t=y$, the output likelihood ratio (\cref{lem:output-density}) shows that the output law has density
\[
 \frac{\mathrm{d}(\nu K_t)}{\mathrm{d}(\mu K_t)}(y)=\E_{\mu_{t,y}}\frac{\mathrm{d}\nu}{\mathrm{d}\mu}=\frac{m(y)}{\E_\mu f} = \frac{m(y)}{\E_{y'\sim Y_t}m(y')}
\]
for $\mu K_t$-almost every $y$, where the last equality uses \eqref{eq:posterior-average}.
This gives \begin{equation}\label{eq:LHS_KL}D_\Psi(\nu K_t\Vert\mu K_t)=\E_{y\sim Y_t}\Phi(m(y))-\Phi(\E_\mu f)=\Ent_{\mu K_t}^\Phi(m).\end{equation}
Hence \cref{fact:divergence-contraction} gives
\[\Ent_{\mu K_t}^\Phi(m)=D_\Psi(\nu K_t\Vert\mu K_t)\leq\frac{Ct}{1+Ct}D_\Psi(\nu\Vert\mu)=\frac{Ct}{1+Ct}\Ent_\mu^\Phi(f).\]Combining this inequality with \eqref{eq:phi-residual} gives \[\cR_\Phi(t,f)=\Ent_\mu^\Phi(f)-\Ent_{\mu K_t}^\Phi(m)\geq\Ent_\mu^\Phi(f)/(1+Ct). \]
\end{proof}

\subsection{Specialization to Poincar\'e inequalities}\label{sec:poincare-mmse}

\paragraph{Additional background} We first need some additional background on both Bayesian estimation and Poincar\'e inequalities.
For $f\in L^2(\mu)$, note that the minimum mean-square error defined in \eqref{eq:dfn_mmse_PI} satisfies
\[
 \mmse_t(f):=\inf_h\E\bigl(f(X)-h(Y_t)\bigr)^2
 =\E\var(f(X)\mid Y_t).
\]
The infimum is over measurable estimators $h:\R^n\to\R$, and the second equality is the Bayes risk decomposition (\cref{lem:bayes-risk}) with $\Phi(z)=z^2$, by which the posterior mean \(h(Y_t) = \E[f(X) \mid Y_t]\) attains the infimum.

Now, we highlight that the Poincar\'e inequality for log-concave measures is closely tied to isoperimetry.
For a Borel set $A$ and $r>0$, let $A_r=\{x:\operatorname{dist}(x,A)\leq r\}$.
The boundary measure of $A$ is (see, e.g., \cite{BobkovHoudre1997,Milman2009})
\[
 \mu^+(A)=\liminf_{r\searrow0}\frac{\mu(A_r)-\mu(A)}r,
\]
and the so-called Cheeger or isoperimetric constant of $\mu$ is
\[
 h(\mu)=\inf_{A}\frac{\mu^+(A)}{\min\{\mu(A),1-\mu(A)\}}.
\]
For log-concave $\mu$, the infimum is known to be achieved by sets of mass $1/2$ \cite{Milman2009}, that is, 
\begin{equation}\label{eq:half-mass}
 h(\mu)=2\inf_{A:\,\mu(A)=1/2}\mu^+(A).
\end{equation}
Notably, Cheeger's inequality directly bounds the Poincar\'e constant with isoperimetry via $\CP(\mu)\leq4/h(\mu)^2$ \cite{Cheeger1970}, \cite[Theorem~3.1 and Remark~3.6]{BobkovHoudre1997}, and for log-concave $\mu$ the two are comparable up to universal multiplicative factors \cite{Milman2009}.

\paragraph{Proof of \cref{cor:poincare-mmse}} \Cref{cor:poincare-mmse} is the case $\Phi(z)=(z-1)^2$ of \cref{thm:general-equivalence}.
It states that the Poincar\'e inequality is equivalent to a lower bound on the MMSE of every observable at every SNR, and that a bound at a single SNR already implies it.
By \eqref{eq:half-mass} and Cheeger's inequality, it moreover suffices to bound the MMSE of indicators of sets of mass $1/2$.

\begin{proof}
Take $J=\R$ and $\Phi(z)=(z-1)^2$ in \cref{thm:general-equivalence}.
Then $\Ent_\mu^\Phi(f)=\var_\mu(f)$, and the Bregman divergence is the squared error,
\[
 B_\Phi(z,a)=(z-1)^2-(a-1)^2-2(a-1)(z-a)=(z-a)^2,
\]
so $\cR_\Phi(t,f)=\mmse_t(f)$.
Moreover, $\Phi''=2$, so the $\Phi$-Sobolev inequality \eqref{eq:phi-functional} with constant $C$ is
\[
 \var_\mu(f)\leq\frac C2\E_\mu\bigl[2\|\nabla f\|_2^2\bigr]=C\E_\mu\|\nabla f\|_2^2,
\]
which is the Poincar\'e inequality \eqref{eq:Poincare_ineq} with constant $C$ for $f\in\cA_0^\R$.
Since both inequalities are tested on $\cA_0^\R$, $\mu$ satisfies the $\Phi$-Sobolev inequality with constant $C$ if and only if $\CP(\mu)\leq C$.
The equivalence and the one-SNR implication (that is, all but the last statement of the corollary) now follow from \cref{thm:general-equivalence}.
This also follows from \citet[Theorem~1.1]{KlartagOrdentlich}.

For the last statement, fix $A$ with $\mu(A)=1/2$.
By \cref{lem:posterior-log-concavity}, the posterior $\mu_{T,y}$ is $T$-strongly log-concave, so it must hold $\CP(\mu_{T,y})\leq1/T$ by Bakry--\'Emery (\cref{fact:bakry-emery}).
Since \(h(\eta) \asymp 1/\sqrt{\CP(\eta)}\) for log-concave \(\eta\) \cite[Theorem~1.5]{Milman2009}, we have $h(\mu_{T,y}) \gtrsim\sqrt T$, and therefore it must hold for the event $A,$
\[
 \mu_{T,y}^+(A)\gtrsim\sqrt T\min\{\mu_{T,y}(A),1-\mu_{T,y}(A)\}\geq\sqrt T\,\mu_{T,y}(A)(1-\mu_{T,y}(A)).
\]
By \eqref{eq:posterior-average}, $\mu(B)=\E_{y\sim Y_T}\mu_{T,y}(B)$ for every Borel set $B$.
Applying Fatou's lemma (\cref{fact:fatou}) gives
\begin{align*}
 \mu^+(A)
 &=\liminf_{r\searrow0}\E_{y\sim Y_T}\frac{\mu_{T,y}(A_r)-\mu_{T,y}(A)}{r}
 \geq\E_{y\sim Y_T}\mu_{T,y}^+(A)\\
 &\gtrsim\sqrt T\,\E_{y\sim Y_T}\bigl[\mu_{T,y}(A)(1-\mu_{T,y}(A))\bigr]
 =\sqrt T\,\E\var(\mathbf1_A(X)\mid Y_T)\\
 &=\sqrt T\,\mmse_T(\mathbf1_A).
\end{align*}
Therefore, if $\mmse_T(\mathbf1_A)\geq b$, we have $ \mu^+(A) \gtrsim b \sqrt{T}$, and since $A$ with $\mu(A)=1/2$ was arbitrary, \eqref{eq:half-mass} gives $h(\mu)\gtrsim b\sqrt T$. Finally, Cheeger's inequality gives $\CP(\mu)\lesssim1/(b^2T)$.
\end{proof}

\subsection{Specialization to log-Sobolev inequalities}
Let $U\in\{0,1\}$ be a \emph{randomized binary label}, per \cref{dfn:random_label}, that is in words a binary label generated from $X$ independently of the Gaussian noise. 

We now prove \cref{cor:lsi-information}, our equivalence between log-Sobolev inequalities and impossibility of estimation in terms of mutual information error.

\begin{proof}[Proof of \cref{cor:lsi-information}] 
We first reduce the lower bounds on the mutual information of labels in (ii) and in the last statement to lower bounds on the relative entropy lost by priors.
Fix $t>0$.
By \cref{lem:label-contraction} with $\pi=\mu$, $K=K_t$ and $b=1/(1+St)$, every randomized binary label $U$ of $X$ satisfies \eqref{eq:info-retention} if and only if every $\zeta\ll\mu$ with bounded density $\mathrm{d}\zeta/\mathrm{d}\mu$ satisfies
\begin{equation}\label{eq:entropy-retention}
 \KL(\zeta\Vert\mu)-\KL(\zeta K_t\Vert\mu K_t)\geq\frac{\KL(\zeta\Vert\mu)}{1+St}.
\end{equation}
In the same way, the assumption of the last statement implies that every such $\zeta$ satisfies \eqref{eq:entropy-retention} with $T$ and $b$ in place of $t$ and $1/(1+St)$.
Hence it suffices to show that (i) holds if and only if \eqref{eq:entropy-retention} holds for every $t>0$, and that \eqref{eq:entropy-retention} with $T$ and $b$ in place of $t$ and $1/(1+St)$ implies $\LS(\mu)\leq1/(bT)$.

Next, we write the log-Sobolev inequality as a $\Phi$-Sobolev inequality, for $\Phi(z)=z\log z$ on $J=\R_{>0}$.
Here $\Ent_\mu^\Phi=\Ent_\mu$ and $\Phi''(z)=1/z$.
For $g\in\cA_0^{\R_{>0}}$, since $\nabla(g^2)=2g\nabla g$, the $\Phi$-Sobolev inequality \eqref{eq:phi-functional} with constant $S$ for $f=g^2$ becomes
\[
 \Ent_\mu(g^2)\leq\frac S2\E_\mu\frac{\|\nabla(g^2)\|_2^2}{g^2}=\frac S2\E_\mu\frac{\|2g\nabla g\|_2^2}{g^2}=2S\E_\mu\|\nabla g\|_2^2,
\]
which is the log-Sobolev inequality \eqref{eq:lsi-convention} with constant $S$ for $g$.
The map $g\mapsto g^2$ is a bijection of $\cA_0^{\R_{>0}}$ onto itself, with inverse $f\mapsto\sqrt f$, since for $f=c+h\in\cA_0^{\R_{>0}}$, the function $\sqrt f$ is smooth and positive, and $\sqrt f-\sqrt c$ vanishes outside the support of $h$.
Hence $\mu$ satisfies the $\Phi$-Sobolev inequality with constant $S$ if and only if $\LS(\mu)\leq S$.

We now show that (i) implies \eqref{eq:entropy-retention} for every $t>0$.
By the substitution $f=g^2$, (i) is the $\Phi$-Sobolev inequality with constant $S$, and $\Phi(1)=0$, so the contraction of $\Phi$-divergences (\cref{fact:divergence-contraction}) with $\Psi=\Phi$ applies.
Since $D_\Phi$ is the relative entropy, it gives, for every $\zeta\ll\mu$ with bounded density,
\[
 \KL(\zeta\Vert\mu)-\KL(\zeta K_t\Vert\mu K_t)
 \geq\KL(\zeta\Vert\mu)-\frac{St}{1+St}\KL(\zeta\Vert\mu)
 =\frac{\KL(\zeta\Vert\mu)}{1+St}.
\]

Finally, we show the converse and the last statement, by writing the $\Phi$-optimal risk as a loss of relative entropy.
Fix $t>0$ and $f\in\cA_0^{\R_{>0}}$, let $\zeta$ be the prior with $\mathrm{d}\zeta/\mathrm{d}\mu=f/\E_\mu f$, and let $m(y)=\E_{\mu_{t,y}}f$ be the posterior mean.
Since $\mathrm{d}\zeta/\mathrm{d}\mu=f/\E_\mu f$, the entropy of $f$ is
\begin{align*}
 \Ent_\mu(f)
 &=\E_\mu[f\log f]-\E_\mu f\log\E_\mu f
 =\E_\mu\Bigl[f\log\frac f{\E_\mu f}\Bigr]\\
 &=\E_\mu f\cdot\E_\zeta\log\frac{\mathrm{d}\zeta}{\mathrm{d}\mu}
 =\E_\mu f\cdot\KL(\zeta\Vert\mu).
\end{align*}
By the output likelihood ratio (\cref{lem:output-density}), $\mathrm{d}(\zeta K_t)/\mathrm{d}(\mu K_t)=m/\E_{y\sim Y_t}m(y)$, where \eqref{eq:posterior-average} gives
\[
 \E_{y\sim Y_t}m(y)=\E_{y\sim Y_t}\E_{x\sim\mu_{t,y}}f(x)=\E_{x\sim\mu}f(x).
\]
Hence, in the same way,
\[
 \Ent_{\mu K_t}(m)
 =\E_{y\sim Y_t}\Bigl[m(y)\log\frac{m(y)}{\E_{y'\sim Y_t}m(y')}\Bigr]
 =\E_\mu f\cdot\E_{\zeta K_t}\log\frac{\mathrm{d}(\zeta K_t)}{\mathrm{d}(\mu K_t)}
 =\E_\mu f\cdot\KL(\zeta K_t\Vert\mu K_t),
\]
and the Bayes risk decomposition (\cref{lem:bayes-risk}) gives
\begin{equation}\label{eq:risk-entropy-loss}
 \cR_\Phi(t,f)=\Ent_\mu(f)-\Ent_{\mu K_t}(m)=\E_\mu f\cdot\bigl(\KL(\zeta\Vert\mu)-\KL(\zeta K_t\Vert\mu K_t)\bigr).
\end{equation}
Since $f$ is bounded, so is $\mathrm{d}\zeta/\mathrm{d}\mu$.
Hence, if \eqref{eq:entropy-retention} holds, then \eqref{eq:risk-entropy-loss} and the identity $\Ent_\mu(f)=\E_\mu f\cdot\KL(\zeta\Vert\mu)$ give
\[
 \cR_\Phi(t,f)
 \geq\frac{\E_\mu f\cdot\KL(\zeta\Vert\mu)}{1+St}
 =\frac{\Ent_\mu(f)}{1+St}.
\]
If this holds for every $t>0$ and every $f\in\cA_0^{\R_{>0}}$, then \cref{thm:general-equivalence} shows that $\mu$ satisfies the $\Phi$-Sobolev inequality with constant $S$, which is (i) by the substitution $f=g^2$.
In the same way, \eqref{eq:entropy-retention} with $T$ and $b$ in place of $t$ and $1/(1+St)$ gives $\cR_\Phi(T,f)\geq b\Ent_\mu(f)$ for every $f\in\cA_0^{\R_{>0}}$, so the last statement of \cref{thm:general-equivalence} gives the $\Phi$-Sobolev inequality with constant $1/(bT)$, that is, $\LS(\mu)\leq1/(bT)$.
\end{proof}

\section{Proofs of the Poincar\'e inequality bounds}\label{sec:balanced}

By the last statement of \cref{cor:poincare-mmse}, we know that an MMSE lower bound for balanced indicators at a single SNR yields a Poincar\'e inequality. We now show how to leverage this connection to obtain information-theoretic proofs of two key results from the Poincar\'e inequalities literature.

\subsection{Connecting MMSE with mutual information: proof of \texorpdfstring{\cref{lem:pinsker-mmse}}{Lemma \getrefnumber{lem:pinsker-mmse}}}\label{sec:pinsker-mmse}
\begin{proof}[Proof of \cref{lem:pinsker-mmse}]
Fix any SNR value $t>0.$ Fix also a Borel set $A$, and let its label function $\mathbf1_A(X)$ and its mass $p=\mu(A)$.
A key object to control the MMSE of recovering $\mathbf1_A(X)$ is its mutual information with the observation $Y_t$ which equals
\[
 I(\mathbf1_A(X);Y_t)
 =\E_{y\sim Y_t}\KL\bigl(\Law(\mathbf1_A(X)\mid Y_t=y)\Vert\Law(\mathbf1_A(X))\bigr)
 =\E_{y\sim Y_t}\KL\bigl(\Bern(\mu_{t,y}(A))\Vert\Bern(p)\bigr),
\]
where $\Bern(a)$ denotes the Bernoulli law with mean $a$.
In particular, Pinsker's inequality (\cref{fact:pinsker}) gives
\[
 I(\mathbf1_A(X);Y_t)
 \geq2\E_{y\sim Y_t}\bigl(\mu_{t,y}(A)-p\bigr)^2.
\]
But now comparing with the MMSE, which equals the expected variance of the posterior by the Bayes risk decomposition (\cref{lem:bayes-risk}) with $\Phi(z)=z^2$, we obtain
\[
 \begin{aligned}
 \mmse_t(\mathbf1_A)
 &=\E_{y\sim Y_t}\bigl[\mu_{t,y}(A)(1-\mu_{t,y}(A))\bigr]
 =\var_\mu(\mathbf1_A)-\E_{y\sim Y_t}\bigl(\mu_{t,y}(A)-p\bigr)^2\\
 &\geq\var_\mu(\mathbf1_A)-\frac12I(\mathbf1_A(X);Y_t),
 \end{aligned}
\]
where the second equality uses \eqref{eq:posterior-average}.
\end{proof}
Note that \cref{lem:pinsker-mmse} holds for any probability measure $\mu$; in particular, no log-concavity of $\mu$ is assumed.

From this point on, the MMSE lower bounds, and therefore the Poincar\'e inequalities, are derived by directly bounding the mutual information $I(\mathbf1_A(X);Y_t)$. To do this we adopt a ``detection to estimation'' strategy. Specifically, we compare the planted model $P_t$ with the ``null'' model $\mu\otimes\gamma_n$, under which the observation $Y_0\sim\gamma_n$ is pure noise. Note that as $Y_0$ is independent from $X$, $I(\mathbf1_A(X);Y_0)=0$. Naturally, if the observation $Y_t$ is sufficiently close in law to the pure noise $Y_0$, then $I(\mathbf1_A(X);Y_t)$ should also be small. But there is a plethora of tools in theoretical statistics and information theory to prove such ``impossibility of detection/hypothesis testing'' between $Y_t$ and $Y_0$. We prove that via this strategy, \begin{itemize}
    \item  a relatively naive such bound obtains the classical bound of \citet{KLS1995} (\cref{sec:kls}), and,
    
    \item a slightly more advanced ``truncated'' bound obtains the bound of \citet{LeeVempala} (\cref{sec:output-label,sec:truncation}).
\end{itemize}

For the rest of the section, let $\Sigma=\cov(X)$ be the covariance of the random vector $X \sim \mu.$
Finally, the bounds of \cref{thm:intro-kls-lv} only involve $\Sigma$ and $\CP(\mu)$ for a log-concave $\mu$, and translating $\mu$ changes none of $\Sigma$, $\CP(\mu)$ and the log-concavity of $\mu$.
We therefore assume from now on that $\mu$ is centered, i.e., $\E X=0$.

\subsection{Obtaining the \texorpdfstring{\citet{KLS1995}}{Kannan--Lov\'asz--Simonovits} bound by data processing}\label{sec:kls}
We first compute the mutual information between the signal and the observation, together with the relative entropy between the observation and the pure noise.
\begin{lemma}[Information and the null model]\label{lem:mi-null}
Let $\mu$ be a probability measure on $\R^n$ with $\E\|X\|_2^2<\infty$.
Then for every $t>0$,
\[
 I(X;Y_t)+\KL(Y_t\Vert\gamma_n)=\frac t2\E\|X\|_2^2.
\]
\end{lemma}
\begin{proof}
For each input $x$, $K_{t,x}=\cN(\sqrt t\,x,I_n)$ has relative entropy $\KL(K_{t,x}\Vert\gamma_n)=t\|x\|_2^2/2$ with respect to $\gamma_n$.
Hence the golden formula (\cref{lem:golden-formula}) with the reference law $\gamma_n$ gives
\[
 I(X;Y_t)+\KL(Y_t\Vert\gamma_n)=\E\KL(K_{t,X}\Vert\gamma_n)=\frac t2\E\|X\|_2^2.
\]
\end{proof}
Since $\mu$ is centered, $\E\|X\|_2^2=\tr(\Sigma)$, and \cref{lem:mi-null} allows us to obtain the bound from \citet{KLS1995}.

\begin{proposition}\label{prop:kls}
Every log-concave $\mu$ satisfies $\CP(\mu)\lesssim\tr(\Sigma)$.
\end{proposition}
\begin{proof}
Let $T=1/(2\tr(\Sigma))$, and let $A$ satisfy $\mu(A)=1/2$.
By the data processing inequality (\cref{lem:data-processing}),
\[
 I(\mathbf1_A(X);Y_T)
 \leq I(X;Y_T)
 \leq I(X;Y_T)+\KL(Y_T\Vert\gamma_n)
 =\frac T2\E\|X\|_2^2
 =\frac T2\tr(\Sigma)
 =\frac14,
\]
where the second inequality holds since $\KL(Y_T\Vert\gamma_n)\geq0$, the first equality follows from \cref{lem:mi-null}, and the second equality holds since $\mu$ is centered.
Hence \cref{lem:pinsker-mmse} gives $\mmse_T(\mathbf1_A)\geq1/4-1/8=1/8$.
The last statement of \cref{cor:poincare-mmse} then gives $\CP(\mu)\lesssim1/T\lesssim\tr(\Sigma)$.
\end{proof}
As claimed, for isotropic $\mu$, this is the bound $\CP(\mu)\lesssim n$ obtained by \citet{KLS1995}.

\subsection{An improved bound on the mutual information by Ornstein--Uhlenbeck contraction}\label{sec:output-label}

A better bound on $I(\mathbf1_A(X);Y_t)$ than the bound $I(\mathbf1_A(X);Y_t)\leq I(X;Y_t)\leq t\tr(\Sigma)/2$ from data processing (\cref{lem:data-processing}) and \cref{lem:mi-null} can be obtained in terms of the relative entropy between $Y_t$ and $\gamma_n$ at a slightly inflated SNR value. In particular, as long as $Y_t$ is close to the pure noise $Y_0$ in the relative entropy sense, the mutual information is small.
\begin{lemma}\label{lem:label-information}
Let $\mu$ be any probability measure on $\R^n$, let $t>0$ and $0<q\leq1$, and let $U$ be a random variable such that $U-X-Y_t$ is a Markov chain.
Then
\[
 I(U;Y_{qt})\leq q\bigl(I(U;Y_t)+\KL(Y_t\Vert\gamma_n)\bigr).
\]
In particular, if $U\in\{0,1\}$, for instance $U=\mathbf1_A(X)$ for a Borel set $A$, then
\[
 I(U;Y_{qt})\leq q\bigl(\log2+\KL(Y_t\Vert\gamma_n)\bigr).
\]
\end{lemma}
\begin{proof}
First, let us make some easy observations of how one can inflate the SNR of a Gaussian additive model.
Since $U-X-Y_t$ is a Markov chain and $G=Y_t-\sqrt t\,X$, the noise $G$ is conditionally independent of $U$ given $X$, and since $G$ is independent of $X$, it is independent of $(U,X)$.
Let $G'\sim\gamma_n$ be independent of $(U,X,G)$, and let $W=\sqrt q\,Y_t+\sqrt{1-q}\,G'$.
Then \begin{equation}\label{eq:W_OU}
W=\sqrt{qt}\,X+\sqrt q\,G+\sqrt{1-q}\,G'=\sqrt{qt}\,X + G'',\end{equation} where \(G''\) is standard Gaussian independent of $(U,X)$, so $(U,X,W)$ has the law of $(U,X,Y_{qt})$.

Now, for a probability measure $\nu$ on $\R^n$, let $\nu O_q$ be the law of $\sqrt q\,V+\sqrt{1-q}\,G'$ with $V\sim\nu$ independent of $G'$, i.e., \(O_q\) is the Ornstein--Uhlenbeck (OU) semigroup at time $\frac12\log(1/q)$.
Therefore, the Gaussian law is invariant under this semigroup, i.e., $\gamma_nO_q=\gamma_n$, and the entropy decay along this semigroup (\cref{fact:ou-decay}) gives, for any probability measure $\nu$ on $\R^n$, \[\KL(\nu O_q\Vert\gamma_n)\leq q\KL(\nu\Vert\gamma_n).\] 
For each $u$, let $\nu_u$ be the conditional law of $Y_t$ given $U=u$.
Since $G'$ is independent of $(U,Y_t)$, the measure $\nu_uO_q$ is the conditional law of $W$ given $U=u$.
Applying the golden formula (\cref{lem:golden-formula}) with the reference law $\gamma_n$ to $U$ and $W$ (defined in \eqref{eq:W_OU}), and to $U$ and $Y_t$, gives
\[
 I(U;W)+\KL(W\Vert\gamma_n)=\E\KL(\nu_UO_q\Vert\gamma_n)\leq q\E\KL(\nu_U\Vert\gamma_n)=q\bigl(I(U;Y_t)+\KL(Y_t\Vert\gamma_n)\bigr).
\]
Since $\KL(W\Vert\gamma_n)\geq0$ and $(U,W)$ has the law of $(U,Y_{qt})$, the first claim follows.
If $U\in\{0,1\}$, then $I(U;Y_t)\leq H(U)\leq\log2$ by \cref{lem:entropy-bound}, which gives the second claim.
\end{proof}

\begin{remark}\label{rem:label-information}
\cref{lem:label-information} offers an alternative route to proving \cref{prop:kls}, which uses \cref{lem:label-information} in place of the data processing step $I(\mathbf1_A(X);Y_T)\leq I(X;Y_T)$.
Indeed, for every Borel set $A$, combining \cref{lem:pinsker-mmse,lem:label-information} gives
\begin{equation}\label{eq:label-mmse}
 \mmse_{qt}(\mathbf1_A)
 \geq\var_\mu(\mathbf1_A)-\frac12I(\mathbf1_A(X);Y_{qt})
 \geq\var_\mu(\mathbf1_A)-\frac q2\bigl(\log2+\KL(Y_t\Vert\gamma_n)\bigr).
\end{equation}
Now for \(A\) such that $\mu(A)=1/2$, $\var_\mu(\mathbf1_A)=1/4$, and let $t=1/\tr(\Sigma)$.
Since $I(X;Y_t)\geq0$ and $\E\|X\|_2^2=\tr(\Sigma)$, \cref{lem:mi-null} gives $\KL(Y_t\Vert\gamma_n)\leq t\tr(\Sigma)/2=1/2$.
Taking $q=1/5$ in \eqref{eq:label-mmse} then gives
\[
 \mmse_{qt}(\mathbf1_A)\geq\frac14-\frac1{10}\Bigl(\log2+\frac12\Bigr)\geq\frac1{10},
\]
so the last statement of \cref{cor:poincare-mmse} with $T=qt$ gives $\CP(\mu)\lesssim1/T\lesssim\tr(\Sigma)$.
More generally, the same argument with $q=1/5$ shows that $\CP(\mu)\lesssim1/t$ for every $t>0$ with $\KL(Y_t\Vert\gamma_n)\leq1$.
Hence, to improve \cref{prop:kls}, it suffices to show that the observation $Y_t$ stays close to the pure noise $Y_0$ at a larger SNR $t$ than the bound $\KL(Y_t\Vert\gamma_n)\leq t\tr(\Sigma)/2$ allows.
\Cref{sec:truncation} does so for a truncation of $\mu$, which yields the bound of \citet{LeeVempala}.
\end{remark}

\subsection{Obtaining the \texorpdfstring{\citet{LeeVempala}}{Lee--Vempala} bound via a truncated \texorpdfstring{$\chi^2$}{chi-square}-bound}\label{sec:truncation}
Towards bounding the relative entropy between $Y_t$ and $\gamma_n$, we start with an application of Jensen's inequality (\cref{lem:kl-chi2}), which gives \[\KL(Y_t\Vert\gamma_n)\leq\log(1+\chi^2(Y_t\Vert\gamma_n)).\] Now the $\chi^2$-divergence can be expressed as an integral of the Euclidean inner product between two independent copies of the signal, see e.g.\ \cite[Proposition~2.5]{KuniskyWeinBandeira}.

\begin{lemma}[Replica identity]\label{lem:replica}
Let $\mu$ be any probability measure on $\R^n$.
For every $t>0$ and $x\in\R^n$,
\[
 \E_{Y\sim K_{t,x}}L_t(Y)=\E e^{t\langle x,X\rangle}.
\]
Consequently, if $X'$ is an independent copy of $X$, then \[\chi^2(Y_t\Vert\gamma_n)=\E e^{t\langle X,X'\rangle}-1,\] where both sides may be infinite.
\end{lemma}
\begin{proof}
Let $G\sim\gamma_n$ be independent of $X$, so that $\sqrt t\,x+G\sim K_{t,x}$.
By \eqref{eq:likelihood} and Tonelli's theorem (\cref{fact:tonelli}),
\begin{align*}
 \E_{Y\sim K_{t,x}}L_t(Y)
 &=\E_G\E_X e^{\sqrt t\langle\sqrt t\,x+G,X\rangle-t\|X\|_2^2/2}\\
 &=\E_X\!\left[e^{t\langle x,X\rangle-t\|X\|_2^2/2}
   \E_G e^{\sqrt t\langle G,X\rangle}\right]
 =\E_X e^{t\langle x,X\rangle},
\end{align*}
where the last step uses the Gaussian moment generating function $\E_Ge^{\sqrt t\langle G,X\rangle}=e^{t\|X\|_2^2/2}$.
Since $L_t=\mathrm{d}(\mu K_t)/\mathrm{d}\gamma_n$ and $\mu K_t=\int K_{t,x}\,\mu(\mathrm{d}x)$,
\[
 1+\chi^2(Y_t\Vert\gamma_n)=\int L_t^2\,\mathrm{d}\gamma_n=\E_{y\sim Y_t}L_t(y)
 =\int\E_{Y\sim K_{t,x}}L_t(Y)\,\mu(\mathrm{d}x)=\E e^{t\langle X,X'\rangle}.
\]
\end{proof}

\begin{remark}One might attempt to directly bound $\chi^2(Y_t\Vert\gamma_n)$ for any log-concave measure $\mu.$ We remark though that $\chi^2(Y_t\Vert\gamma_n)$ can be in fact infinite at every SNR, even for log-concave $\mu$. 
For example, let $X$ have independent coordinates, each with density proportional to $e^{-\sqrt{1+x^2}}$ on $\R$. So, up to an additive constant, the potential $V(x)=\sum_i\sqrt{1+x_i^2}$ is smooth and convex.
If $X'$ is an independent copy of $X$, then $\E e^{t\langle X,X'\rangle}=(\E e^{tX_1X_1'})^n$.
Since $\sqrt{1+x^2}\leq1+|x|$, the conditional expectation $\E[e^{tX_1X_1'}\mid X_1]$ is infinite whenever $|X_1|\geq1/t$, so $\E e^{t\langle X,X'\rangle}=\infty$ for every $t>0$, and \cref{lem:replica} gives $\chi^2(Y_t\Vert\gamma_n)=\infty$.
\end{remark}
To overcome this issue, we apply \cref{lem:label-information} to a truncation of $\mu$, in the spirit of the conditional second moment method, see \cite[Section~1.2]{KuniskyWeinBandeira} and the references there.

\paragraph{The conditioning event} As mentioned above, a standard idea would be to condition the $\chi^2$ calculation on a high probability event. But, how to choose the right conditioning event? 

Recall $1+\chi^2(Y_t\Vert\gamma_n)=\E e^{t\langle X,X'\rangle}$. Conditionally on $X=x$, notice that the exponent of the integrand $t\langle x,X'\rangle$ is a centered random variable with variance $t^2x^\tran\Sigma x$. In particular, $\E e^{t\langle x,X'\rangle}$ is close to $1$ when $t^2x^\tran\Sigma x$ is small enough.
Since $\E[X^\tran\Sigma X]=\tr(\Sigma\,\E[XX^\tran])=\tr(\Sigma^2)$, this suggests that, after integrating on the high probability event that the $x$'s with $t^2x^\tran\Sigma x $ of order at most $\tr(\Sigma^2)$, for SNR $t \ll \tr(\Sigma^2)^{-1/2}$, the truncated integrand $\E e^{t\langle X,X'\rangle}$ should be bounded.

We therefore restrict $X$ to the ellipsoid
\[
 \cE=\{x\in\R^n:x^\tran\Sigma x\leq10\tr(\Sigma^2)\},
\]
and let $\mu_\cE$ be the ``conditional'' probability measure with $\mathrm{d}\mu_\cE/\mathrm{d}\mu=\mathbf1_\cE/\mu(\cE)$.

We now turn to bounding $\chi^2(\mu_\cE K_{t}\Vert\gamma_n).$ An elementary lemma we will use is the following.
\begin{lemma}\label{lem:scalar-mgf}
There is a universal $c>0$ such that every log-concave $\xi$ with $\E\xi=0$ and $\var(\xi)=\sigma^2$ satisfies, whenever $|\lambda|\sigma\leq c$,
\[
 \E e^{\lambda\xi}\leq1+\frac{2\lambda^2\sigma^2}{c^2}.
\]
\end{lemma}
\begin{proof}
By \cite[Lemma~5.7]{LovaszVempala2007} $\xi$ is subexponential with parameter of order $\sigma$.
Hence, since $\E\xi=0$, there is a universal constant $c>0$ such that $\E e^{\lambda\xi}\leq e^{\lambda^2\sigma^2/c^2}$ whenever $|\lambda|\sigma\leq c$.
The claim then follows from the elementary inequality $e^z\leq1+2z$ for $z\in[0,1]$.
\end{proof}

We are now in a position to prove the conditional $\chi^2$ bound.
\begin{lemma}\label{lem:truncation}
Let $t_*=c/(10\sqrt{\tr(\Sigma^2)})$, where $c$ is the constant of \cref{lem:scalar-mgf}.
Then \begin{itemize}
    \item(``$\cE^c$ is rare'') \[\mu(\cE^c)\leq 1/10.\]

    \item(``Conditional $\chi^2$ is bounded'') \[\chi^2(\mu_\cE K_{t_*}\Vert\gamma_n)\leq 1/5.\]

\end{itemize}
\end{lemma}
\begin{proof}

The first part follows directly by Markov's inequality since $\mu(\cE^c)\leq\E[X^\tran\Sigma X]/(10\tr(\Sigma^2))=1/10$.

Now for the second part, let $X'$ be an independent copy of $X \sim \mu$.
By \cref{lem:replica} with $\mu_\cE$ in place of $\mu$, and since the integrand is nonnegative,
\begin{align*}
 1+\chi^2(\mu_\cE K_{t_*}\Vert\gamma_n)
 &=\E_{x,x'\sim\mu_\cE}e^{t_*\langle x,x'\rangle}
 =\frac{\E_{X,X'\sim\mu}\bigl[\mathbf1_\cE(X)\mathbf1_\cE(X')e^{t_*\langle X,X'\rangle}\bigr]}{\mu(\cE)^2}\\
 &\leq\frac{\E_{X\sim\mu}\bigl[\mathbf1_\cE(X)\,\E_{X'\sim\mu}e^{t_*\langle X,X'\rangle}\bigr]}{\mu(\cE)^2}.
\end{align*}
For any fixed $x\in\cE$, the variable $\langle x,X'\rangle$ is centered with variance $x^\tran\Sigma x\leq10\tr(\Sigma^2)$, and is log-concave by the Pr\'ekopa--Leindler inequality (see \cref{fact:prekopa}).
Hence, $t_*(x^\tran\Sigma x)^{1/2}\leq c/\sqrt{10}\leq c$, and therefore \cref{lem:scalar-mgf} gives \[\E_{X'\sim\mu}e^{t_*\langle x,X'\rangle}\leq1+2t_*^2x^\tran\Sigma x/c^2.\]
Hence,
\begin{align*}
 1+\chi^2(\mu_\cE K_{t_*}\Vert\gamma_n)
 &\leq\frac{\E_{X\sim\mu}\bigl[\mathbf1_\cE(X)\bigl(1+2t_*^2X^\tran\Sigma X/c^2\bigr)\bigr]}{\mu(\cE)^2}
 \leq\frac{\mu(\cE)+2t_*^2\tr(\Sigma^2)/c^2}{\mu(\cE)^2}\\
 &=\frac1{\mu(\cE)}+\frac1{50\mu(\cE)^2}
 \leq\frac{10}9+\frac2{81}
 \leq\frac65,
\end{align*}
where the equality uses $t_*^2\tr(\Sigma^2)=c^2/100$, and the next inequality uses $\mu(\cE)\geq9/10$.
\end{proof}

Then by a simple application of the law of total expectation, the above lemma implies the desired conclusion.
\begin{theorem}\label{thm:lee-vempala}
Every log-concave $\mu$ with $\Sigma=\cov(X)$ satisfies $\CP(\mu)\lesssim\sqrt{\tr(\Sigma^2)}$. In particular, for isotropic log-concave $\mu,$ it holds $\CP(\mu)\lesssim n^{1/2}$, which is proved by \citet{LeeVempala}.
\end{theorem}

\begin{proof}
For $t_*$ defined in \cref{lem:truncation}, let $s_*=t_*/10$, and fix an arbitrary Borel set $A$ with $\mu(A)=1/2$.

Since $\mu(\cE^c)\leq1/10$, both $\mu(A\cap\cE)$ and $\mu(A^c\cap\cE)$ lie in $[2/5,1/2]$, so $\mu_\cE(A)\in[2/5,5/9]$ and \[\var_{\mu_\cE}(\mathbf1_A)=\mu_\cE(A)(1-\mu_\cE(A))\geq\frac25\cdot\frac49\geq\frac1{10}.\]
By \cref{lem:truncation} and \cref{lem:kl-chi2}, \[\KL(\mu_\cE K_{t_*}\Vert\gamma_n)\leq\chi^2(\mu_\cE K_{t_*}\Vert\gamma_n)\leq\frac15.\]
Applying \cref{lem:pinsker-mmse,lem:label-information} with $\mu_\cE$ in place of $\mu$ (note that since $\cE$ is convex, $\mu_\cE$ is a  log-concave measure), $t=t_*$ and $q=1/10$ shows that, when $X\sim\mu_\cE$, every estimator of $\mathbf1_A(X)$ from $Y_{s_*}$ has mean-square error at least
\(
 \frac1{10}-\frac1{20}\Bigl(\log2+\frac15\Bigr)\geq\frac1{20}.
\)

Combining the above, for every measurable $h$,
\begin{align*}
 \E\bigl(\mathbf1_A(X)-h(Y_{s_*})\bigr)^2
 &=\mu(\cE)\,\E\bigl[\bigl(\mathbf1_A(X)-h(Y_{s_*})\bigr)^2\bigm|X\in\cE\bigr]+\mu(\cE^c)\,\E\bigl[\bigl(\mathbf1_A(X)-h(Y_{s_*})\bigr)^2\bigm|X\notin\cE\bigr]\\
 &\geq\mu(\cE)\cdot\frac1{20}.
\end{align*}
Taking the infimum over $h$ gives \[\mmse_{s_*}(\mathbf1_A)\geq\mu(\cE)\cdot\frac1{20}\geq\frac9{10}\cdot\frac1{20}\geq\frac1{50}.\]
Hence, the last statement of \cref{cor:poincare-mmse} with $T=s_*$ gives $\CP(\mu)\lesssim1/s_*\lesssim \sqrt{\tr(\Sigma^2)}$. This completes the proof of \cref{thm:lee-vempala}.
\end{proof}

\begin{remark}
One could argue in an alternative way to finish the above argument by using the stability of the Poincar\'e constant under total variation perturbations by Milman \cite[Theorems~1.5 and~5.5]{Milman2009}. Indeed, by directly combining \cref{lem:truncation} with \cref{lem:label-information,lem:pinsker-mmse} and the last statement of \cref{cor:poincare-mmse} we have $\CP(\mu_\cE)\lesssim\sqrt{\tr(\Sigma^2)}$.
Since $\TV(\mu,\mu_\cE)=\mu(\cE^c)\leq1/10$ and the aforementioned property that the Poincar\'e constant of log-concave measures is stable under such perturbations \cite[Theorems~1.5 and~5.5]{Milman2009}, this Poincar\'e constant bound transfers to $\mu$. Above, we found a way to argue directly about it by lower bounding the MMSE under $\mu$ with the law of total expectation.
\end{remark}
\section{Proofs of log-Sobolev inequality bounds via the Gaussian null model}\label{sec:info-output}
In this section, we prove \cref{thm:output-information,cor:lsi-quarter}.
Since \cref{thm:output-information} reduces \cref{cor:lsi-quarter} to a bound on $\chi^2(Y_t\Vert\gamma_n)$, we first prove this bound and deduce the corollary, and then prove the theorem.

\subsection{Proof of \texorpdfstring{\cref{cor:lsi-quarter}}{Corollary \getrefnumber{cor:lsi-quarter}}}\label{sec:proof-lsi-quarter}
By \cref{thm:output-information}, to obtain a log-Sobolev inequality it suffices to bound $\chi^2(Y_t\Vert\gamma_n)$, which measures how far $Y_t$ is from the pure noise $Y_0$.
By the replica identity (\cref{lem:replica}), $1+\chi^2(Y_t\Vert\gamma_n)=\E e^{t\langle X,X'\rangle}$, where $X'$ is an independent copy of $X$. In particular, subgaussianity implies \[1+\chi^2(Y_t\Vert\gamma_n) \leq \E e^{t^2\|X\|_2^2/2},\] so it remains to bound the moment generating function of $\|X\|_2^2$. A simple calculation gives the following lemma.
\begin{lemma}\label{lem:normalized-moments}
Let $G\sim\gamma_n$ be independent of $X$. For any \(0 \le s < 1/2\),
\begin{equation}\label{eq:norm-moment}
 \E e^{s\|X\|_2^2}=\E_G\E_Xe^{\sqrt{2s}\langle G,X\rangle}
 \leq(1-2s)^{-n/2}.
\end{equation}
Moreover, $\cov(X)\preceq I_n$ and $\E\|X\|_2^2\leq n$.
\end{lemma}
\begin{proof}
We linearize the square with an independent Gaussian vector, so that \eqref{eq:normalized} applies to the linear function $\langle G,X\rangle$.
For fixed $x \in \R^n$, the Gaussian moment generating function gives
\[
 \E_G e^{\sqrt{2s}\langle G,x\rangle}=e^{s\|x\|_2^2}.
\]
By Tonelli's theorem (\cref{fact:tonelli}), \(\E e^{s \|X\|_2^2} = \E_X \E_G e^{\sqrt{2s}\langle G,X\rangle} = \E_G \E_X e^{\sqrt{2s}\langle G,X\rangle} \).
Conditional on $G$, \eqref{eq:normalized} gives \(\E_G\E_X e^{\sqrt{2s}\langle G,X\rangle} \le \E_G e^{s\|G\|_2^2} \).
Since \(\|G\|_2^2 \sim \chi^2_n\), by the moment generating function of the chi-squared distribution, for \(0 \le s < 1/2\),
\[
 \E_Ge^{s\|G\|_2^2} =(1-2s)^{-n/2}.
\]

For the second moments, we expand \eqref{eq:normalized} to second order in $u$.
For $v\in\R^n$, let $Z=\langle v,X\rangle$.
Applying \eqref{eq:normalized} with $u=\lambda v$ and $u=-\lambda v$ gives
\[
 \E\cosh(\lambda Z)
 =\frac{\E e^{\lambda Z}+\E e^{-\lambda Z}}2
 \leq e^{\lambda^2\|v\|_2^2/2}.
\]
Since $\cosh z\geq1+z^2/2$, it follows that, for $\lambda\ne0$,
\[
 \E Z^2
 \leq\frac{2}{\lambda^2}\left(e^{\lambda^2\|v\|_2^2/2}-1\right).
\]
Letting $\lambda\to0$ and using $\E Z=0$ yields
\[
 v^\tran\cov(X)v=\E Z^2\leq\|v\|_2^2.
\]
This proves $\cov(X)\preceq I_n$.
Taking the trace and using $\E X=0$ yields $\E\|X\|_2^2=\tr\cov(X)\leq n$.
\end{proof}

Combining \cref{lem:replica,lem:normalized-moments} bounds the comparison cost.
\begin{proposition}\label{lem:output-chi2}
For every $0<t\leq1/\sqrt2$, $\log(1+\chi^2(Y_t\Vert\gamma_n))\leq nt^2$.
\end{proposition}
\begin{proof}
Let $X'$ be an independent copy of $X$.
By \cref{lem:replica},
\[
 1+\chi^2(Y_t\Vert\gamma_n)=\E e^{t\langle X,X'\rangle}\leq\E e^{t^2\|X\|_2^2/2}\leq(1-t^2)^{-n/2},
\]
where the first inequality follows from \eqref{eq:normalized} applied to $X'$ with $u=tX$, conditionally on $X$, and the second inequality follows from \cref{lem:normalized-moments} with $s=t^2/2$.
Since $\log(1/(1-t^2))\leq2t^2$ for $t^2\leq1/2$, this gives $\log(1+\chi^2(Y_t\Vert\gamma_n))\leq nt^2$.
\end{proof}

We now prove \cref{cor:lsi-quarter}.
\begin{proof}[Proof of \cref{cor:lsi-quarter}]
Let $0<t\leq1/\sqrt2$.
By \cref{lem:output-chi2}, $\log(1+\chi^2(Y_t\Vert\gamma_n))\leq nt^2$, and \cref{thm:output-information} gives
\[
 \LS(\mu)\lesssim\frac1t\bigl(1+nt^2\bigr)=\frac1t+nt.
\]
Choosing $t=n^{-1/2}$, which balances the two terms and is at most $1/\sqrt2$ since $n\geq2$, gives $\LS(\mu)\lesssim n^{1/2}$.
\end{proof}

\subsection{Proof of \texorpdfstring{\cref{thm:output-information}}{Theorem \getrefnumber{thm:output-information}} given a key KL-contraction lemma}\label{sec:proof-output-information}
The proof of \cref{thm:output-information} relies on the following relative entropy contraction, which we prove in \cref{sec:contraction}.
Here and below, we say that $\zeta\ll\pi$ has finite entropy, or is finite-entropy, if $\KL(\zeta\Vert\pi)<\infty$.
\begin{lemma}[Relative entropy contraction]\label{lem:restricted-contraction}
Let $\pi\ll\eta$ be probability measures on $\R^n$ such that $\eta$ is absolutely continuous, $\LS(\eta)<\infty$, and
\[
 \Delta:=\log\bigl(1+\chi^2(\pi\Vert\eta)\bigr)<\infty.
\]
Let $0<a\leq1/\bigl(100\LS(\eta)(1+\Delta)\bigr)$. Then every finite-entropy $\zeta\ll\pi$ satisfies
\[
 \KL(\zeta\Vert\pi)-\KL(\zeta K_a\Vert\pi K_a)
 \geq\frac{\KL(\zeta\Vert\pi)}{10^3}.
\]
\end{lemma}

\begin{remark}
For log-concave $\pi$, the transference principle of \citet[Theorem~1.1]{BartheMilman} gives $\LS(\pi)\lesssim\LS(\eta)(1+\Delta)$, and our equivalence (\cref{cor:lsi-information}) then implies \cref{lem:restricted-contraction} up to universal constants.
However, our proof follows a different route via information theory and avoids the equivalence between concentration and isoperimetry used there.
It also yields a stronger statement, in the sense that the relative entropy contraction of \cref{lem:restricted-contraction} holds for every law $\pi$, while \cite{BartheMilman} requires $\pi$ to be log-concave.
\end{remark}

We now prove \cref{thm:output-information}, assuming \cref{lem:restricted-contraction}.

\begin{proof}[Proof of \cref{thm:output-information}]We first show that every finite-entropy $\zeta\ll\mu$ satisfies
\begin{equation}\label{eq:output-kl}
 \KL(\zeta\Vert\mu)-\KL(\zeta K_s\Vert\mu K_s)\geq\frac{\KL(\zeta\Vert\mu)}{10^3}.
\end{equation}
We choose the SNR $a$ of the second channel so that $at/(1+a)=s$, that is, $a=s/(t-s)$.
Since
\[
 a=\frac s{t-s}=\frac1{1000(1+\Delta_t)-1}\leq\frac1{100(1+\Delta_t)},
\]
\cref{lem:restricted-contraction} with $(\pi,\eta)=(\mu K_t,\gamma_n)$ and $\LS(\gamma_n)=1$ applies to the second channel $K_a$, and shows that every finite-entropy $\rho\ll\mu K_t$ satisfies
\[
 \KL(\rho\Vert\mu K_t)-\KL(\rho K_a\Vert\mu K_tK_a)\geq\frac{\KL(\rho\Vert\mu K_t)}{10^3}.
\]
We apply this to the law $\rho=\zeta K_t$ of $Y_t$ when $X\sim\zeta$, which has finite entropy because $\KL(\zeta K_t\Vert\mu K_t)\leq\KL(\zeta\Vert\mu)$ by data processing (\cref{lem:data-processing}).
For $r>0$, let
\[
 R_\zeta(r)=\KL(\zeta\Vert\mu)-\KL(\zeta K_r\Vert\mu K_r)
\]
be the relative entropy lost in the channel $K_r$, which is nonnegative by data processing (\cref{lem:data-processing}).

It remains to identify the two channels in a row with the channel $K_s$.
For every $x$, if $Y\sim K_{t,x}$ and $G'\sim\gamma_n$ is independent of $Y$, then $\sqrt a\,Y+G'\sim\cN(\sqrt{at}\,x,(1+a)I_n)$, and $at/(1+a)=s$.
Hence $K_{t,x}K_a$ is the image of $K_{s,x}$ under the bijection $y\mapsto\sqrt{1+a}\,y$, and so are $\zeta K_tK_a$ and $\mu K_tK_a$ of $\zeta K_s$ and $\mu K_s$.
Relative entropy is invariant under this bijection (\cref{lem:data-processing}), so $\KL(\zeta K_tK_a\Vert\mu K_tK_a)=\KL(\zeta K_s\Vert\mu K_s)$.
Therefore, the loss at the SNR $s$ is the loss at the SNR $t$ plus the loss in the second channel, and
\begin{align*}
 R_\zeta(s)
 &=R_\zeta(t)+\KL(\zeta K_t\Vert\mu K_t)-\KL(\zeta K_tK_a\Vert\mu K_tK_a)
 \geq R_\zeta(t)+\frac{\KL(\zeta K_t\Vert\mu K_t)}{10^3}\\
 &\geq\frac{R_\zeta(t)+\KL(\zeta K_t\Vert\mu K_t)}{10^3}
 =\frac{\KL(\zeta\Vert\mu)}{10^3},
\end{align*}
where the second inequality uses $R_\zeta(t)\geq0$, and the last step is the definition of $R_\zeta(t)$.
This proves \eqref{eq:output-kl}.
By \cref{lem:label-contraction} with $\pi=\mu$, $K=K_s$ and $b=10^{-3}$, \eqref{eq:output-kl} implies
\[
 I(U;X\mid Y_s)\geq\frac{I(U;X)}{10^3}
\]
for every randomized binary label $U$ of $X$.

Finally, for the last part, apply the one-SNR implication in \cref{cor:lsi-information} to \eqref{eq:output-entropy} at $T=s$, which gives $\LS(\mu)\leq10^3/s\lesssim(1+\Delta_t)/t$.
\end{proof}

\subsection{Proof of \texorpdfstring{\cref{lem:restricted-contraction}}{Lemma \getrefnumber{lem:restricted-contraction}}}\label{sec:contraction}

By the following lemma, it suffices to prove \cref{lem:restricted-contraction} for finite-entropy $\zeta\ll\pi$ supported on a set of $\pi$-mass at most $1/2$.

\begin{lemma}[Uniform entropy loss]\label{lem:uniform-loss}
Let $K$ be a channel with input law $\pi$.
Suppose every finite-entropy $\zeta\ll\pi$ supported on a set of $\pi$-mass at most $1/2$ satisfies
\begin{equation}\label{eq:small-set-contraction}
 \KL(\zeta K\Vert\pi K)\leq\frac1{10}\KL(\zeta\Vert\pi).
\end{equation}
Then every finite-entropy $\zeta\ll\pi$ satisfies 
\begin{equation}\label{eq:uniform-loss}
 \KL(\zeta\Vert\pi)-\KL(\zeta K\Vert\pi K)
 \geq\frac{\KL(\zeta\Vert\pi)}{10^3}.
\end{equation}
\end{lemma}
We prove \cref{lem:uniform-loss} in \cref{app:channel}, and now prove \cref{lem:restricted-contraction} assuming it.

\begin{proof}[Proof of \cref{lem:restricted-contraction}]
Given \cref{lem:uniform-loss}, we assume from now on that $\zeta$ is supported on a set of $\pi$-mass at most $1/2$ and aim to prove \eqref{eq:small-set-contraction}. We first compare $\zeta$ with the reference law $\eta$, which is the only step where the comparison cost enters.
By the chain rule $\mathrm{d}\zeta/\mathrm{d}\eta=(\mathrm{d}\zeta/\mathrm{d}\pi)(\mathrm{d}\pi/\mathrm{d}\eta)$ and entropy duality (\cref{lem:entropy-variational}) under $\pi$ with $g=\log(\mathrm{d}\pi/\mathrm{d}\eta)$,
\[
 \KL(\zeta\Vert\eta)
 =\KL(\zeta\Vert\pi)+\E_\zeta\log\frac{\mathrm{d}\pi}{\mathrm{d}\eta}
 \leq2\KL(\zeta\Vert\pi)+\log\E_\pi\frac{\mathrm{d}\pi}{\mathrm{d}\eta}
 =2\KL(\zeta\Vert\pi)+\Delta,
\]
where the last step uses $\E_\pi[\mathrm{d}\pi/\mathrm{d}\eta]=1+\chi^2(\pi\Vert\eta)=e^\Delta$.
Thus $\zeta$ is close to $\eta$ whenever it is close to $\pi$, up to the additive cost $\Delta$.
Taking $\zeta=\pi$ gives $\KL(\pi\Vert\eta)\leq\Delta$.

Next, we turn these bounds into a transport bound between $\zeta$ and $\pi$, passing through $\eta$.
By the triangle inequality for $W_2$ (\cref{fact:w2-triangle}) and $(u+v)^2\leq2u^2+2v^2$,
\[
 W_2(\zeta,\pi)^2
 \leq2W_2(\zeta,\eta)^2+2W_2(\pi,\eta)^2
 \leq4\LS(\eta)\bigl(\KL(\zeta\Vert\eta)+\KL(\pi\Vert\eta)\bigr)
 \leq8\LS(\eta)\bigl(\KL(\zeta\Vert\pi)+\Delta\bigr),
\]
where the second inequality follows from the Otto--Villani inequality (\cref{fact:otto-villani}) applied to $\zeta$ and $\pi$, and the last inequality follows from the two bounds of the previous step.

Finally, we push this transport bound through the channel.
The channel sends two inputs $z_1,z_2\in\R^n$ to two Gaussian laws with the same covariance, whose relative entropy is
\begin{equation}\label{eq:gaussian-shift}
 \KL(K_{a,z_1}\Vert K_{a,z_2})
 =\KL\bigl(\cN(\sqrt a\,z_1,I_n)\Vert\cN(\sqrt a\,z_2,I_n)\bigr)
 =\frac a2\|z_1-z_2\|_2^2.
\end{equation}
For any coupling of $Z_1\sim\zeta$ and $Z_2\sim\pi$, the output laws $\zeta K_a$ and $\pi K_a$ are the averages of $K_{a,Z_1}$ and $K_{a,Z_2}$ over the coupling, so joint convexity of KL (\cref{lem:joint-convexity}) and \eqref{eq:gaussian-shift} give
\[
 \KL(\zeta K_a\Vert\pi K_a)
 \leq\E\KL(K_{a,Z_1}\Vert K_{a,Z_2})
 =\frac a2\E\|Z_1-Z_2\|_2^2.
\]
Taking the infimum over couplings gives
\[
 \KL(\zeta K_a\Vert\pi K_a)
 \leq\frac a2W_2(\zeta,\pi)^2
 \leq4a\LS(\eta)\bigl(\KL(\zeta\Vert\pi)+\Delta\bigr).
\]
It remains to absorb the additive cost $\Delta$, and this is where we use that $\zeta$ is supported on the set $A$ with $\pi(A)\leq1/2$.
Let $\pi_A$ be the probability measure with $\mathrm{d}\pi_A/\mathrm{d}\pi=\mathbf1_A/\pi(A)$.
Since $\zeta$ is supported on $A$, we have $\mathrm{d}\zeta/\mathrm{d}\pi=(\mathrm{d}\zeta/\mathrm{d}\pi_A)/\pi(A)$ $\zeta$-almost surely, so
\[
 \KL(\zeta\Vert\pi)=\KL(\zeta\Vert\pi_A)+\log\frac1{\pi(A)}\geq\log2.
\]
Hence $\Delta\leq(\Delta/\log2)\KL(\zeta\Vert\pi)\leq2\Delta\KL(\zeta\Vert\pi)$, and thus
\begin{align*}
 \KL(\zeta K_a\Vert\pi K_a)
 &\leq4a\LS(\eta)\bigl(\KL(\zeta\Vert\pi)+2\Delta\KL(\zeta\Vert\pi)\bigr)\\
 &\leq8a\LS(\eta)(1+\Delta)\KL(\zeta\Vert\pi)
 \leq\frac{1}{10}\KL(\zeta\Vert\pi),
\end{align*}
where the last step uses $100a\LS(\eta)(1+\Delta)\leq1$.
\end{proof}

\section{Proofs of log-Sobolev inequality bounds via the null-posterior model}\label{sec:info-reference}
In this section, we prove \cref{prop:reference-lsi,thm:reference-information,cor:lsi-sixth}.

\subsection{Proof of \texorpdfstring{\cref{cor:lsi-sixth}}{Corollary \getrefnumber{cor:lsi-sixth}}}\label{sec:proof-lsi-sixth}
We prove the following bound on the divergence between $\mu$ and $\nu_t$ in \cref{sec:ratio}.
\begin{proposition}\label{prop:likelihood-L2}
For $0<t\leq1/2$,
\begin{equation*}
 \log\bigl(1+\chi^2(\mu\Vert\nu_t)\bigr)\lesssim nt^3.
\end{equation*}
\end{proposition}
We now prove \cref{cor:lsi-sixth}.
\begin{proof}[Proof of \cref{cor:lsi-sixth}]
Let $0<t\leq1/2$.
By \cref{prop:likelihood-L2}, $\log(1+\chi^2(\mu\Vert\nu_t))\lesssim nt^3$, and \cref{thm:reference-information} gives
\[
 \LS(\mu)\lesssim\frac1t\bigl(1+nt^3\bigr)=\frac1t+nt^2.
\]
Choosing $t=\tfrac12n^{-1/3}$, which balances the two terms, gives $\LS(\mu)\lesssim n^{1/3}$.
\end{proof}

\subsection{Proof of \texorpdfstring{\cref{thm:reference-information}}{Theorem \getrefnumber{thm:reference-information}} given the log-Sobolev inequality of the null prior}\label{sec:reference-information}
The proof is an application of \cref{lem:restricted-contraction} with $(\pi,\eta)=(\mu,\nu_t)$, given the log-Sobolev inequality $\LS(\nu_t)\leq2/t$ of \cref{prop:reference-lsi}, which we prove in \cref{sec:reference-lsi}.

\begin{proof}[Proof of \cref{thm:reference-information}]
We first show that the channel $K_s$ loses a fixed fraction of the relative entropy of every prior.
By \cref{prop:reference-lsi}, $\LS(\nu_t)\leq2/t$.
Moreover, $\nu_t$ has a strictly positive density because each posterior does, so $\mu\ll\nu_t$.
Since
\[
 100s\LS(\nu_t)(1+\Delta'_t)
 \leq100\cdot\frac t{1000(1+\Delta'_t)}\cdot\frac2t\cdot(1+\Delta'_t)
 =\frac15
 <1,
\]
\cref{lem:restricted-contraction} with $(\pi,\eta)=(\mu,\nu_t)$ and $a=s$ shows that every finite-entropy $\zeta\ll\mu$ satisfies
\begin{equation}\label{eq:reference-kl}
 \KL(\zeta\Vert\mu)-\KL(\zeta K_s\Vert\mu K_s)\geq\frac{\KL(\zeta\Vert\mu)}{10^3}.
\end{equation}
By \cref{lem:label-contraction} with $\pi=\mu$, $K=K_s$ and $b=10^{-3}$, \eqref{eq:reference-kl} implies
\[
 I(U;X\mid Y_s)\geq\frac{I(U;X)}{10^3}
\]
for every randomized binary label $U$ of $X$.

Finally, the last statement of \cref{cor:lsi-information} with $T=s$, applied to \eqref{eq:reference-entropy}, gives \[\LS(\mu)\leq10^3/s\lesssim(1+\Delta'_t)/t.\]
\end{proof}

\subsection{Proof of \texorpdfstring{\cref{prop:reference-lsi}}{Proposition \getrefnumber{prop:reference-lsi}}}\label{sec:reference-lsi}
To prove \cref{prop:reference-lsi}, we need to know how the posterior $\mu_{t,y}$ changes with $y$.

We will use the following posterior differentiation identities, see also \cite[Section~2.3]{ElAlaouiMontanari}.
By \eqref{eq:posterior}, the posterior has density $e^{\sqrt t\langle y,x\rangle-t\|x\|_2^2/2}/L_t(y)$ w.r.t.\ $\mu$, so $y$ enters only through the linear term $\sqrt t\langle y,x\rangle$ in the exponent and through the normalizing constant $L_t(y)$.
Differentiating the exponent in $y$ brings down the factor $\sqrt t\,x$, and differentiating the normalizing constant subtracts its posterior mean, so the gradient in $y$ of a posterior average $\E_{\mu_{t,y}}f$ is $\sqrt t$ times the posterior covariance of $f$ and $X$. Specifically, the following lemma holds.

\begin{lemma}[Posterior differentiation]\label{lem:posterior-formulas}
For every $y\in\R^n$,
\[
 \nabla^2\log L_t(y)=t\cov_{\mu_{t,y}}(X)\succeq0.
\]
Moreover, every bounded measurable $f:\R^n\to\R$ satisfies
\[
 \nabla_y\E_{\mu_{t,y}}f=\sqrt t\cov_{\mu_{t,y}}(f,X).
\]
\end{lemma}
\begin{proof}
Both identities follow by differentiating the definition of $\mu_{t,y}$ (see \eqref{eq:posterior}) w.r.t.\ $y$.
Let $W_y(x)=\sqrt t\langle y,x\rangle-t\|x\|_2^2/2$.
Since $W_y(x)\leq\|y\|_2^2-t\|x\|_2^2/4$, \cref{fact:differentiation} allows us to differentiate under the integral in \eqref{eq:likelihood}, which gives
\begin{align*}
 \nabla L_t&=\sqrt t\,L_t\E_{\mu_{t,y}}X,\\
 \nabla^2L_t&=tL_t\E_{\mu_{t,y}}[XX^\tran].
\end{align*}
The quotient rule gives
\[
 \nabla^2\log L_t
 =\frac{\nabla^2L_t}{L_t}-\frac{\nabla L_t\nabla L_t^\tran}{L_t^2}
 =t\E_{\mu_{t,y}}[XX^\tran]-t\,\E_{\mu_{t,y}}X\,(\E_{\mu_{t,y}}X)^\tran
 =t\cov_{\mu_{t,y}}(X).
\]
For the second identity, \eqref{eq:posterior} gives $\E_{\mu_{t,y}}f=\int fe^{W_y}\,\mathrm{d}\mu/L_t(y)$.
Differentiating the numerator under the integral gives $\nabla_y\int fe^{W_y}\,\mathrm{d}\mu=\sqrt t\int fxe^{W_y}\,\mu(\mathrm{d}x)=\sqrt t\,L_t(y)\E_{\mu_{t,y}}[fX]$, so the quotient rule gives
\[
 \nabla_y\E_{\mu_{t,y}}f
 =\sqrt t\,\E_{\mu_{t,y}}[fX]-\E_{\mu_{t,y}}f\,\frac{\nabla L_t(y)}{L_t(y)}
 =\sqrt t\bigl(\E_{\mu_{t,y}}[fX]-\E_{\mu_{t,y}}f\,\E_{\mu_{t,y}}X\bigr)
 =\sqrt t\cov_{\mu_{t,y}}(f,X).
\]
\end{proof}

By \cref{lem:posterior-log-concavity}, for each fixed $t,y$ the posterior $\mu_{t,y}$ is $t$-strongly log-concave, so $\LS(\mu_{t,y})\leq1/t$ by Bakry--\'Emery (\cref{fact:bakry-emery}).
In particular, this leads to the following mean estimate for posteriors.
It bounds the distance between the means of $\mu_{t,y}$ and of any $\theta\ll\mu_{t,y}$ by the relative entropy $\KL(\theta\Vert\mu_{t,y})$.

\begin{lemma}\label{lem:posterior-mean}
For every probability measure $\theta\ll\mu_{t,y}$ with finite relative entropy,
\[
 \|\E_\theta X-\E_{\mu_{t,y}}X\|_2^2
 \leq\frac2t\KL(\theta\Vert\mu_{t,y}).
\]
\end{lemma}
\begin{proof}
Write $d=\E_\theta X-\E_{\mu_{t,y}}X$ for the shift of the mean.
We first turn the LSI of the posterior into a subgaussian bound for linear functions.
Since $\LS(\mu_{t,y})\leq1/t$ and for any fixed vector $v$, $x\mapsto\langle v,x\rangle$ is $\|v\|_2$-Lipschitz, Herbst's argument (\cref{lem:herbst}) gives, for every $v\in\R^n$,
\begin{equation}\label{eq:posterior-herbst}
 \log\E_{\mu_{t,y}}e^{\langle v,X-\E_{\mu_{t,y}}X\rangle}
 \leq\frac{\|v\|_2^2}{2t}.
\end{equation}

Next, we turn this bound into a bound on the component $\langle v,d\rangle$ of the shift.
Entropy duality (\cref{lem:entropy-variational}) under $\mu_{t,y}$ with $g(x)=\langle v,x-\E_{\mu_{t,y}}X\rangle$ and \eqref{eq:posterior-herbst} give
\[
 \langle v,d\rangle
 =\E_\theta g
 \leq\KL(\theta\Vert\mu_{t,y})+\log\E_{\mu_{t,y}}e^{g}
 \leq\KL(\theta\Vert\mu_{t,y})+\frac{\|v\|_2^2}{2t}.
\]

Finally, we choose $v$ to optimize this bound.
Taking $v=td$ gives
\[
 t\|d\|_2^2\leq\KL(\theta\Vert\mu_{t,y})+\frac t2\|d\|_2^2,
\]
and rearranging gives $\|d\|_2^2\leq\frac2t\KL(\theta\Vert\mu_{t,y})$, which is the claim.
\end{proof}

We now explain the strategy of the proof of \cref{prop:reference-lsi}.
By \eqref{eq:reference-law}, a sample $X\sim\nu_t$ can be generated by first drawing $Y'\sim \gamma_n$ and then $X\sim\mu_{t,Y'}$.
For a label $U$ of $X$, the chain rule (\cref{lem:chain-rule}) splits the information $I(U;X)$ into the information $I(U;Y')$ carried by $Y'$ and the information $I(U;X\mid Y')$ that remains in $X$ once $Y'$ is known.
The remaining information is controlled by the log-Sobolev inequality of the $t$-strongly log-concave posteriors $\mu_{t,y}$.
The information carried by $Y'$ is controlled by the log-Sobolev inequality of the Gaussian law $\gamma_n$, together with the fact that the posteriors $\mu_{t,y}$ vary slowly with $y$, at a rate controlled by \cref{lem:posterior-formulas,lem:posterior-mean}.

We first bound the information carried by $Y'$.
\begin{lemma}\label{lem:mixing-information}
Let $t>0$, let $Y'\sim\gamma_n$, and let $X$ have conditional law $\mu_{t,Y'}$ given $Y'$, so that $X\sim\nu_t$ by \eqref{eq:reference-law}.
Let $U\in\{0,1\}$ be a random variable such that $U-X-Y'$ is a Markov chain.
Then
\begin{equation}\label{eq:mixing-information}
 I(U;Y')\leq I(U;X\mid Y').
\end{equation}
\end{lemma}
\begin{proof}
We first write both sides of \eqref{eq:mixing-information} in terms of the classes of $U$.
In this form, \eqref{eq:mixing-information} becomes a bound on the entropy under $\gamma_n$ of the class weights, which we then prove with the log-Sobolev inequality of the Gaussian law $\gamma_n$.
If $U$ is almost surely constant, both sides vanish, so we may assume that $\P(U=0)$ and $\P(U=1)$ are positive.
For $u\in\{0,1\}$, let $g_u(y)=\P(U=u\mid Y'=y)$ be the weight of the class $u$ given $Y'=y$, and let $\theta_{u,y}=\Law(X\mid U=u,Y'=y)$ be this class.
Since $X\sim\mu_{t,y}$ given $Y'=y$, \cref{lem:label-entropy} gives the first identity below, and since mutual information can be written as an average of relative entropies (\cref{fact:mi-average}), we obtain the second one:
\begin{align*}
 I(U;Y')&=\sum_{u\in\{0,1\}}\Ent_{\gamma_n}(g_u),\\
 I(U;X\mid Y')&=\E_{y\sim\gamma_n}\sum_{u\in\{0,1\}}g_u(y)\KL(\theta_{u,y}\Vert\mu_{t,y}).
\end{align*}
Therefore, it suffices to show that each $u\in\{0,1\}$ satisfies
\begin{equation}\label{eq:class-goal}
 \Ent_{\gamma_n}(g_u)\leq\E_{y\sim\gamma_n}\bigl[g_u(y)\KL(\theta_{u,y}\Vert\mu_{t,y})\bigr].
\end{equation}

The Gaussian log-Sobolev inequality (\cref{fact:gaussian-lsi}) bounds $\Ent_{\gamma_n}(g_u)$ by $\frac12\E_{\gamma_n}[g_u\|\nabla\log g_u\|_2^2]$, so it suffices to show that every $y$ satisfies $\|\nabla\log g_u(y)\|_2^2\leq2\KL(\theta_{u,y}\Vert\mu_{t,y})$.
Let $p_u(x)=\P(U=u\mid X=x)$.
Since $U-X-Y'$ is a Markov chain and $X\sim\mu_{t,y}$ given $Y'=y$, Bayes' theorem gives
\begin{equation}\label{eq:class-weight}
 g_u(y)=\E_{\mu_{t,y}}p_u,
\end{equation}
and
\begin{equation}\label{eq:class-posterior}
 \frac{\mathrm{d}\theta_{u,y}}{\mathrm{d}\mu_{t,y}}=\frac{p_u}{g_u(y)}.
\end{equation}
Applying \cref{lem:posterior-formulas} with $f=p_u$ gives
\[
 \nabla g_u(y)
 =\sqrt t\cov_{\mu_{t,y}}(p_u,X)
 =\sqrt t\bigl(\E_{\mu_{t,y}}[p_uX]-g_u(y)\E_{\mu_{t,y}}X\bigr)
 =\sqrt t\,g_u(y)\bigl(\E_{\theta_{u,y}}X-\E_{\mu_{t,y}}X\bigr).
\]
Dividing by $g_u(y)$ gives
\[
 \nabla\log g_u(y)=\sqrt t\bigl(\E_{\theta_{u,y}}X-\E_{\mu_{t,y}}X\bigr).
\]
Hence \cref{lem:posterior-mean} gives
\begin{equation}\label{eq:class-score}
 \|\nabla\log g_u(y)\|_2^2
 =t\|\E_{\theta_{u,y}}X-\E_{\mu_{t,y}}X\|_2^2
 \leq t\cdot\frac2t\KL(\theta_{u,y}\Vert\mu_{t,y})
 =2\KL(\theta_{u,y}\Vert\mu_{t,y}).
\end{equation}

Finally, the Gaussian log-Sobolev inequality (\cref{fact:gaussian-lsi}) and \eqref{eq:class-score} give
\[
 \Ent_{\gamma_n}(g_u)
 \leq\frac12\E_{y\sim\gamma_n}\bigl[g_u(y)\|\nabla\log g_u(y)\|_2^2\bigr]
 \leq\E_{y\sim\gamma_n}\bigl[g_u(y)\KL(\theta_{u,y}\Vert\mu_{t,y})\bigr],
\]
which is \eqref{eq:class-goal}.
\end{proof}

Now we are almost ready to prove \cref{prop:reference-lsi}.  For the log-Sobolev inequality of \(\nu_t\), we will use the equivalence in \cref{cor:lsi-information} to turn the information inequality \eqref{eq:null-information} into a log-Sobolev inequality.
However, this requires that \(\nu_t\) is log-concave, as follows. 
\begin{lemma}[Log-concavity of the null prior]\label{lem:null-log-concavity}
For every $t>0$, the null prior $\nu_t$ is log-concave.
\end{lemma}
We prove \cref{lem:null-log-concavity} in \cref{app:null-law}.

We now prove \cref{prop:reference-lsi}.
Recall that we generate $Y'\sim\gamma_n$, then $X\sim\mu_{t,Y'}$, so that $X\sim\nu_t$ by \eqref{eq:reference-law}, and finally $Y_s=\sqrt sX+G$ at a fixed SNR $s>0$, with $G\sim\gamma_n$ independent of $(Y',X)$.
Fix a randomized binary label $U$ of $X$.
We prove \eqref{eq:null-information} through the chain
\begin{equation}\label{eq:null-chain}
 I(U;X\mid Y_s)
 \geq I(U;X\mid Y',Y_s)
 \geq\frac t{t+s}I(U;X\mid Y')
 \geq\frac t{2(t+s)}I(U;X).
\end{equation}
\begin{proof}[Proof of \cref{prop:reference-lsi}]
Note that given $X$, the variables $U$, $Y'$ and $G$ are independent.

The first inequality of \eqref{eq:null-chain} follows from data processing in \cref{lem:chain-rule}, for the Markov chain $U-X-(Y_s,Y')$. 

For the second inequality of \eqref{eq:null-chain}, given $Y'=y$, the pair $(X,Y_s)$ is the GAM with prior $\mu_{t,y}$ and SNR $s$, and $U$ is a randomized binary label of $X$.
The prior $\mu_{t,y}$ is $t$-strongly log-concave by \cref{lem:posterior-log-concavity}, so $\LS(\mu_{t,y})\leq1/t$ by Bakry--\'Emery (\cref{fact:bakry-emery}), and \cref{cor:lsi-information} with $\mu_{t,y}$ in place of $\mu$ and $S=1/t$ gives
\[
 I(U;X\mid Y_s,Y'=y)
 \geq\frac{I(U;X\mid Y'=y)}{1+s/t}
 =\frac t{t+s}I(U;X\mid Y'=y).
\]
Averaging over $Y'\sim\gamma_n$ (\cref{lem:mi-conditioning}) gives the second inequality.

For the third inequality of \eqref{eq:null-chain}, the chain rule (\cref{lem:chain-rule}) for the Markov chain $U-X-Y'$ and \cref{lem:mixing-information} give
\[
 I(U;X)=I(U;Y')+I(U;X\mid Y')\leq2I(U;X\mid Y').
\]
Combining the three inequalities of \eqref{eq:null-chain} proves \eqref{eq:null-information}.

Finally, we deduce the log-Sobolev inequality for $\nu_t$.
By \cref{lem:null-log-concavity}, $\nu_t$ is log-concave.
Since \eqref{eq:null-information} holds for every randomized binary label $U$ of $X\sim\nu_t$, the last part of \cref{cor:lsi-information}, applied with $\nu_t$ in place of $\mu$, $T=s$, and $b=t/(2(t+s))$, gives
\[
 \LS(\nu_t)\leq\frac1{bs}=\frac{2(t+s)}{ts}=\frac2t+\frac2s.
\]
Since $s>0$ is arbitrary, letting $s\to\infty$ gives $\LS(\nu_t)\leq2/t$.
\end{proof}

\subsection{Proof of \texorpdfstring{\cref{prop:likelihood-L2}}{Proposition \getrefnumber{prop:likelihood-L2}}}\label{sec:ratio}
We use the LSI of $\nu_t$ to estimate the divergence between $\mu$ and $\nu_t$.
Define the log-likelihood ratio
\begin{equation*}
 F_t(x):=\log\frac{\mathrm{d}\mu}{\mathrm{d}\nu_t}(x).
\end{equation*}
By definition,
\begin{equation}\label{eq:F-normalization}
 \E_\mu e^{-F_t}=\int\mathrm{d}\nu_t=1,
\end{equation}
and therefore, the $\chi^2$-divergence can be seen as the mean of the exponential function on $F_t$ under $\nu_t$, i.e.,
\begin{equation}\label{eq:chi2-F}
 1+\chi^2(\mu\Vert\nu_t)
 =\E_{\nu_t}\Bigl(\frac{\mathrm{d}\mu}{\mathrm{d}\nu_t}\Bigr)^2
 =\E_{\nu_t}e^{2F_t}.
\end{equation}
If $\LS(\eta)<\infty$, then Herbst's argument (\cref{lem:herbst}) bounds $\E_\eta e^{\lambda f}$ in terms of the gradient of $f$.
So we first bound the gradient of $F_t$ in \cref{lem:F-bounds}, and then run Herbst's argument for $F_t$ to prove \cref{prop:likelihood-L2}.
Heuristically, \cref{lem:F-bounds} gives $\|\nabla F_t\|_2^2\lesssim t^4n$ for typical $x$, since $\E\|X\|_2^2\leq n$ by \cref{lem:normalized-moments}, and with $\LS(\nu_t)\leq2/t$, Herbst's argument then gives an exponent of order $\LS(\nu_t)\,t^4n\lesssim nt^3$.
The next lemma supplies a few technical bounds on the behavior of \(F_t\) that we need.
\begin{lemma}\label{lem:F-bounds}
For every $t>0$ and $x\in\R^n$,
\begin{equation}\label{eq:F-upper}
 F_t(x)\leq\frac{t^2}{2}\|x\|_2^2,
\end{equation}
\begin{equation}\label{eq:F-lower}
 F_t(x)\geq-2t^2n-2t^2\sqrt n\|x\|_2,
\end{equation}
and
\begin{equation}\label{eq:F-gradient}
 \|\nabla F_t(x)\|_2^2\lesssim t^4(n+\|x\|_2^2).
\end{equation}
Moreover, $F_t$ is finite, smooth, and convex, and $0<\mathrm{d}\nu_t/\mathrm{d}\mu\leq e^{tn/2}$.
\end{lemma}
\begin{proof}
We first express the likelihood ratio $\mathrm{d}\nu_t/\mathrm{d}\mu$ through the channel.
By the definition of $\nu_t$ and Bayes' theorem in \eqref{eq:posterior}, which gives $\mathrm{d}\mu_{t,y}/\mathrm{d}\mu(x)=e^{\sqrt t\langle y,x\rangle-t\|x\|_2^2/2}/L_t(y)$,
\begin{equation}\label{eq:w-formulas}
 \begin{aligned}
 e^{-F_t(x)}=\frac{\mathrm{d}\nu_t}{\mathrm{d}\mu}(x)
 &=\int\frac{e^{\sqrt t\langle y,x\rangle-t\|x\|_2^2/2}}{L_t(y)}\,\gamma_n(\mathrm{d}y)\\
 &=\int\frac1{L_t(y)}\,K_{t,x}(\mathrm{d}y)
 =\E_{Y\sim K_{t,x}}\frac1{L_t(Y)},
 \end{aligned}
\end{equation}
where the second equality uses $K_{t,x}(\mathrm{d}y)=e^{\sqrt t\langle y,x\rangle-t\|x\|_2^2/2}\gamma_n(\mathrm{d}y)$.
In words, $e^{-F_t(x)}$ is the average of the reciprocal likelihood ratio $1/L_t=\mathrm{d}\gamma_n/\mathrm{d}(\mu K_t)$ over an observation $Y$ generated from the input $x$.

We next check that $F_t$ is finite and smooth.
By Jensen's inequality in \eqref{eq:likelihood}, $\log L_t(y)\geq\sqrt t\langle y,\E X\rangle-t\E\|X\|_2^2/2\geq-tn/2$, where the last inequality uses $\E X=0$ and \cref{lem:normalized-moments}.
Hence $0<\mathrm{d}\nu_t/\mathrm{d}\mu\leq e^{tn/2}$.
Since $1/L_t\leq e^{tn/2}$, \cref{fact:differentiation} allows us to differentiate under the integral in \eqref{eq:w-formulas}, so $F_t$ is smooth.

For the quadratic upper bound \eqref{eq:F-upper}, Jensen's inequality gives
\[
 F_t(x)=-\log\E_{Y\sim K_{t,x}}\frac1{L_t(Y)}
 \leq\log\E_{Y\sim K_{t,x}}L_t(Y)
 =\log\E_{X\sim\mu}e^{t\langle x,X\rangle}
 \leq\frac{t^2}{2}\|x\|_2^2,
\]
where the second equality follows from \cref{lem:replica}, and the last inequality follows from \eqref{eq:normalized} with $u=tx$.

For the gradient bound \eqref{eq:F-gradient} and the lower bound \eqref{eq:F-lower} we prove and leverage that $F_t$ is convex.
The rough idea is that a convex function lies above its tangent planes, so if it is bounded above by a quadratic and is not too negative, then its gradient is of the same order as the gradient of the quadratic.

We now show that $F_t$ is convex, using the Pr\'ekopa--Leindler inequality.
By \eqref{eq:w-formulas} and $K_{t,x}(\mathrm{d}y)=(2\pi)^{-n/2}e^{-\|y-\sqrt t\,x\|_2^2/2}\,\mathrm{d}y$, we have $e^{-F_t(x)}=(2\pi)^{-n/2}\int e^{\psi(x,y)}\,\mathrm{d}y$, where
\[
 \psi(x,y)=-\frac{\|y-\sqrt t\,x\|_2^2}2-\log L_t(y).
\]
The function $\psi$ is concave on $\R^n\times\R^n$, since the first term is a concave quadratic function of the linear map $(x,y)\mapsto y-\sqrt t\,x$, and the second term is concave in $y$ because $\nabla^2\log L_t\succeq0$ by \cref{lem:posterior-formulas}.
Hence $e^\psi$ is log-concave on $\R^n\times\R^n$, and the Pr\'ekopa--Leindler inequality (\cref{fact:prekopa}) shows that its marginal $e^{-F_t}$ is log-concave.
Since $e^{-F_t}$ is positive and finite, $F_t$ is convex.

We now combine convexity with the quadratic upper bound \eqref{eq:F-upper}, and first bound $F_t(0)$ from below.
Convexity and \eqref{eq:F-upper} imply
\[
 F_t(x)\leq\frac{F_t(0)+F_t(2x)}2
 \leq\frac{F_t(0)}2+t^2\|x\|_2^2.
\]
By \eqref{eq:F-normalization} and Jensen's inequality,
\[
 1 = \E_\mu e^{-F_t} \ge \E_{X\sim\mu} \exp \left( -\frac{F_t(0)}{2} - t^2 \|X\|_2^2 \right)\geq\exp\!\left(-\frac{F_t(0)}2-t^2\E\|X\|_2^2\right),
\]
so $F_t(0)\geq-2t^2 \E\|X\|_2^2 \ge -2t^2n$ by \cref{lem:normalized-moments}.

Next, we bound $\nabla F_t(0)$, which gives the lower bound \eqref{eq:F-lower}.
Since \(F_t\) lies above its tangent plane at $0$, \eqref{eq:F-upper} gives $F_t(0)+\langle\nabla F_t(0),z\rangle \le F_t(z) \leq t^2\|z\|_2^2/2$ for every $z\in\R^n$.
Taking $z=\nabla F_t(0)/t^2$ gives $\|\nabla F_t(0)\|_2^2/(2t^2)\leq-F_t(0)\leq2t^2n$, that is, $\|\nabla F_t(0)\|_2\leq2t^2\sqrt n$, and therefore
\begin{equation*}
 F_t(x)\geq F_t(0) + \langle \nabla F_t(0), x \rangle \ge -2t^2n-2t^2\sqrt n\|x\|_2.
\end{equation*}
Finally, we bound the gradient at $x$ to get \eqref{eq:F-gradient}.
Let $r=\|x\|_2+\sqrt n$, and let \(u\) be a unit vector.
By convexity,
\[
 \langle\nabla F_t(x),u\rangle\leq\frac{F_t(x+ru)-F_t(x)}r \le \frac{t^2 \|x+ru\|_2^2/2 + 2t^2n + 2t^2 \sqrt{n}\|x\|_2}{r} \le 4t^2r,
\]
where the second inequality follows from the upper bound \eqref{eq:F-upper} at $x+ru$ and the lower bound \eqref{eq:F-lower} at $x$, and the last inequality uses $\|x+ru\|_2\leq2r$ and $n+\sqrt n\|x\|_2\leq r^2$.
Taking the supremum over $u$ proves \(\|\nabla F_t(x)\|_2 \le 4t^2r\), and thus
\[
 \|\nabla F_t(x)\|_2^2\leq16t^4\bigl(\|x\|_2+\sqrt n\bigr)^2\lesssim t^4(n+\|x\|_2^2).
\]
\end{proof}

We now run Herbst's argument for $F_t$.
Since $F_t$ is not globally Lipschitz, we use the gradient bound \eqref{eq:F-gradient} instead, which grows with $\|x\|_2$, and control the resulting moments by \eqref{eq:normalized}.

\begin{proof}[Proof of \cref{prop:likelihood-L2}]
As in Herbst's argument, we interpolate between $\E_{\nu_t}e^{F_t}$ and $\E_{\nu_t}e^{2F_t}$, and bound the derivative along the way by the LSI of $\nu_t$.
For \(q \in [1, 2]\), define
\[
 \Lambda_t(q):=\log\int e^{qF_t}\,\mathrm{d}\nu_t
          =\log\int e^{(q-1)F_t}\,\mathrm{d}\mu.
\]
By \eqref{eq:F-normalization} and \eqref{eq:chi2-F}, $\Lambda_t(1)=\log 1=0$ and $\Lambda_t(2)=\log(1+\chi^2(\mu\Vert\nu_t))$.

Set $S=\LS(\nu_t)$, $Z_q=e^{\Lambda_t(q)}=\int e^{qF_t}\,\mathrm{d}\nu_t$.
We first check that the integrals of $e^{qF_t}$, $|F_t|e^{qF_t}$ and $e^{qF_t}\|\nabla F_t\|_2^2$ under $\nu_t$ are finite for $q\in[1,2]$, so that we can differentiate $Z_q$ and apply the LSI to $e^{qF_t/2}$.
By \cref{lem:F-bounds} and \(t \le 1/2\), we have $|F_t(x)|\lesssim n+\|x\|_2^2$, and these three functions are upper bounded by $O(e^{\|x\|_2^2/4}(n+\|x\|_2^2))$,
which is $\mu$-integrable by \cref{lem:normalized-moments}.
They are also $\nu_t$-integrable because $\mathrm{d}\nu_t/\mathrm{d}\mu\leq e^{tn/2}$ by \cref{lem:F-bounds}.
So \(Z_q\) is differentiable on \(q \in [1,2]\), and by \cref{rem:test-functions} we can apply the LSI for \(e^{q F_t/2}\) on \(\nu_t\).

Now, direct computations give $Z_q'=\int F_t e^{qF_t}\,\mathrm{d}\nu_t$, $\Lambda_t'(q)=Z_q'/Z_q$, and
\begin{align*}
 \Ent_{\nu_t}(e^{qF_t})
 &=q\int F_t e^{qF_t}\,\mathrm{d}\nu_t-Z_q\log Z_q
 =Z_q\bigl(q\Lambda_t'(q)-\Lambda_t(q)\bigr),\\
 \int\|\nabla e^{qF_t/2}\|_2^2\,\mathrm{d}\nu_t
 &=\int\left\|\frac q2e^{qF_t/2}\nabla F_t\right\|_2^2\,\mathrm{d}\nu_t
 =\frac{q^2}{4}\int e^{qF_t}\|\nabla F_t\|_2^2\,\mathrm{d}\nu_t.
\end{align*}
Hence, applying the LSI for \(e^{q F_t/2}\) on \(\nu_t\) gives
\begin{align*}
\left( \frac{\Lambda_t(q)}{q} \right)' &= \frac{q\Lambda_t'(q) - \Lambda_t(q)}{q^2} = \frac{1}{q^2Z_q} \Ent_{\nu_t}(e^{qF_t}) \le \frac{2S}{q^2Z_q} \int\|\nabla e^{qF_t/2}\|_2^2\,\mathrm{d}\nu_t = \frac{S}{2Z_q} \int e^{qF_t}\|\nabla F_t\|_2^2\,\mathrm{d}\nu_t.
\end{align*}
It remains to bound the right-hand side, which is an average of $\|\nabla F_t\|_2^2$ under a tilt of $\mu$. Hence,
\begin{equation*}
 \frac{\int e^{qF_t}\|\nabla F_t\|_2^2\,\mathrm{d}\nu_t}{Z_q} =
     \frac{\int e^{(q-1)F_t}\|\nabla F_t\|_2^2\,\mathrm{d}\mu}
      {\int e^{(q-1)F_t}\,\mathrm{d}\mu} = \E_{\mu_q} \|\nabla F_t\|_2^2,
\end{equation*}
where \(\mu_q(\mathrm{d} x) = \frac{e^{(q-1)F_t}}{Z_q}  \mu(\mathrm{d} x)\) is a probability measure.
By the gradient bound \eqref{eq:F-gradient}, it suffices to show that this tilt does not increase the second moment of $X$ much.

Since $|F_t(x)|\lesssim n+\|x\|_2^2$ by \eqref{eq:F-upper} and \eqref{eq:F-lower}, $F_t$ is $\mu$-integrable by \cref{lem:normalized-moments}.
Jensen's inequality and \eqref{eq:F-normalization} give $\E_\mu F_t\geq0$, so
\[
 Z_q=\E_\mu e^{(q-1)F_t}\geq e^{(q-1)\E_\mu F_t}\geq1.
\]
By \eqref{eq:F-gradient}, we have \(\E_{\mu_q}\|\nabla F_t\|_2^2 \lesssim t^4(n + \E_{X\sim \mu_q}\|X\|_2^2)\).
Again by Jensen's inequality,
\begin{align*}
\E_{X\sim \mu_q}\|X\|_2^2 \le 8\log \E_{X\sim\mu_q} e^{\|X\|_2^2/8} = 8\log \frac{\int e^{(q-1)F_t+\|x\|_2^2/8}\,\mathrm{d}\mu}
      {\int e^{(q-1)F_t}\,\mathrm{d}\mu} \le 8\log \E_{X\sim\mu} e^{\|X\|_2^2/4} \lesssim n,
\end{align*}
where the second inequality is because $Z_q\geq1$ and \((q-1)F_t(x) \le {t^2}\|x\|_2^2/2 \le \|x\|_2^2/8\) by \eqref{eq:F-upper}, and the last inequality is from \cref{lem:normalized-moments}.
Therefore,
\begin{equation*}
\left( \frac{\Lambda_t(q)}{q} \right)' \le \frac{S}{2} \E_{\mu_q} \|\nabla F_t\|_2^2 \lesssim Snt^4.
\end{equation*}

Integrating from $q=1$ to $q=2$ gives
\[
{\Lambda_t(2)} = 2 \left( \int_1^2 \left( \frac{\Lambda_t(q)}{q} \right)' \mathrm{d} q + {\Lambda_t(1)} \right) \lesssim Snt^4.
\]
By \cref{prop:reference-lsi}, $S=\LS(\nu_t)\leq2/t$, and thus
\[
 \log\bigl(1+\chi^2(\mu\Vert\nu_t)\bigr)=\Lambda_t(2)\lesssim Snt^4\lesssim nt^3.
\]
\end{proof}

\section{Improving the log-Sobolev inequality bound to order \texorpdfstring{$n^{1/4}$}{n to the power 1/4}}\label{sec:improvement}
The perhaps surprising success of the measure $\nu_t$ in the preceding section leads to the natural question of whether one can leverage it even further and obtain an even better bound on the log-Sobolev constant, perhaps leveraging tools outside the Bayesian estimation toolkit. In this section, we answer this question affirmatively and obtain $\LS(\mu)\lesssim n^{1/4}$.

Throughout this section, $\mu$ is as in \cref{sec:lsi-applications}, so in particular it satisfies \eqref{eq:normalized}.

To start this approach, instead of our equivalence \cref{cor:lsi-information}, we use a result of Milman, which shows that Gaussian concentration behavior implies an LSI for log-concave measures.
With $A_r$ as in \cref{sec:poincare-mmse}, define the concentration function \(\alpha_\mu: \R_+ \to [0,1/2]\) of a measure $\mu$ as
\[
 \alpha_\mu(r)=\sup_{A: \mu(A) = 1/2}\mu(A_r^c).
\]

\begin{theorem}[{\cite{Milman2010}, \cite[Theorem 23]{BizeulLSI}}]\label{thm:milman}
If $\mu$ is log-concave and, for some $B>0$ and all $r>0$,
\[
 \alpha_\mu(r)\leq e^{-r^2/B},
\]
then $\LS(\mu)\lesssim B$.
\end{theorem}
It therefore suffices to prove $\alpha_\mu(r)\leq e^{-r^2/B}$ for all $r>0$ with $B\lesssim n^{1/4}$.
For small $r$, we use a known two-level concentration bound for log-concave measures due to \citet{BizeulSubspace}.
For large $r$, we transfer concentration to $\mu$ from the key measure $\nu_t$, whose LSI constant is at most $2/t$, at the cost of the $\chi^2$-divergence between $\mu$ and $\nu_t$ we have bounded in \cref{sec:info-reference}.
Yet, unlike in \cref{sec:info-reference}, where we had to choose a fixed $t$ to leverage the $\chi^2$ bound, here we may choose a different $t$ for each $r$, which gives a tail of order $e^{-r^3/\sqrt n}$, and is the key idea why we can improve the log-Sobolev bound to $n^{1/4}$.

\subsection{Small \texorpdfstring{$r$}{r}: two-level concentration}\label{sec:external}
We will need the following concentration bound for log-concave measures.
Let $\CP^*(n)$ be the supremum of $\CP(\eta)$ over isotropic log-concave probability measures $\eta$ on $\R^n$.

\begin{theorem}[{\cite[Corollary~1.2]{BizeulSubspace}}]\label{thm:two-level}
There are universal $C_0\geq1$ and $c_0>0$ such that every log-concave $\mu$ on $\R^n$, $n\geq2$, with covariance matrix $\Sigma$ satisfies for all \(r>0\),
\begin{equation*}
 \alpha_\mu(r)\leq C_0\exp\left(-c_0\min\left\{\frac{r}{\|\Sigma\|_{\mathrm{op}}^{1/2}},\frac{r^2}{\CP^*(n)\log n\,\|\Sigma\|_{\mathrm{op}}}\right\}\right).
\end{equation*}
\end{theorem}

Recent preprints \cite{SongZhang2026,BizeulKlartagLehec2026,BalasubramanianKasiviswanathan2026} give proofs of the KLS conjecture, that is, $\CP^*(n)\lesssim1$.
However, for our purpose, the earlier subpolynomial bound of \citet{Chen2021} suffices.

\begin{theorem}[{\cite{Chen2021}}]\label{thm:chen}
There is a universal $C$ such that $\CP^*(n)\leq n^{C\sqrt{\log\log n/\log n}}$.
\end{theorem}

\subsection{Large \texorpdfstring{$r$}{r}: transferring concentration from \texorpdfstring{$\nu_t$}{the null prior}}\label{sec:cubic}
Now, for large $r$ we combine \cref{prop:reference-lsi,prop:likelihood-L2} to get the following bound.
\begin{proposition}\label{prop:cubic}
There is a universal constant $c>0$ such that for every $r>0$,
\begin{equation*}
 \alpha_\mu(r)\leq2\exp\!\left[-c
   \min\left\{\frac{r^3}{\sqrt n},r^2\right\}\right].
\end{equation*}
\end{proposition}

\begin{proof}
Fix $r>0$ and a measurable set $A$ with $\mu(A)=1/2$, and define $f_A(x)=\operatorname{dist}(x,A)$, which is $1$-Lipschitz.
Let $X,X' \sim \mu$ be independent.
Since $A_r^c=\{x:f_A(x)>r\}$ and $f_A(X')=0$ on $\{X'\in A\}$, independence implies
\[
 \frac12\mu(A_r^c)
 =\P\{f_A(X)>r,\ X'\in A\}
 \leq\P\{f_A(X)-f_A(X')\geq r\}.
\]
For every $\lambda>0$, by Markov's inequality,
\[
 \begin{aligned}
 \P\{f_A(X)-f_A(X')\geq r\}
 &=\P\{e^{\lambda(f_A(X)-f_A(X'))}\geq e^{\lambda r}\}\\
 &\leq e^{-\lambda r}\E e^{\lambda(f_A(X)-f_A(X'))}.
 \end{aligned}
\]
Therefore, it suffices to upper bound the MGF \(\E e^{\lambda(f_A(X)-f_A(X'))}\).

Fix $0<t\leq1/2$.
For every $s\in\R$, changing measure and applying Cauchy--Schwarz gives
\begin{align*}
 \E_\mu e^{s(f_A-\E_{\nu_t}f_A)}
 &=\E_{\nu_t}\!\left[\frac{\mathrm{d}\mu}{\mathrm{d}\nu_t}e^{s(f_A-\E_{\nu_t}f_A)}\right]\leq\bigl(1+\chi^2(\mu\Vert\nu_t)\bigr)^{1/2}
       \left(\E_{\nu_t}e^{2s(f_A-\E_{\nu_t}f_A)}\right)^{1/2}.
\end{align*}
By \cref{prop:likelihood-L2}, there is a universal constant $C\geq1$ such that $1+\chi^2(\mu\Vert\nu_t)\leq e^{Cnt^3}$.
And from Herbst's argument (\cref{lem:herbst}) and \cref{prop:reference-lsi},
\[
 \log\E_{\nu_t}e^{2s(f_A-\E_{\nu_t}f_A)}
 \leq\frac{(2s)^2}{2}\LS(\nu_t)
 \leq\frac{4s^2}{t}.
\]
Therefore, for every $s\in\R$,
\begin{equation}\label{eq:herbst-concentration}
 \E_\mu e^{s(f_A-\E_{\nu_t}f_A)}
 \leq\exp\!\left(\frac{Cnt^3}2+\frac{2s^2}{t}\right).
\end{equation}
Applying \cref{eq:herbst-concentration} at $s=\lambda$ and $s=-\lambda$ to the independent pair \(X\) and \(X'\) yields
\begin{align*}
 \E e^{\lambda(f_A(X)-f_A(X'))}
 &=\E e^{\lambda(f_A(X)-\E_{\nu_t}f_A)}\E e^{-\lambda(f_A(X')-\E_{\nu_t}f_A)}\leq\exp\!\left(Cnt^3+\frac{4\lambda^2}{t}\right).
\end{align*}
Thus, taking $\lambda=tr/10$,
\[
 \P\{f_A(X)-f_A(X')\geq r\}
 \leq\exp\!\left(Cnt^3+\frac{4\lambda^2}{t}-\lambda r\right)
 \leq\exp\!\left(Cnt^3-\frac{tr^2}{20}\right).
\]

Set $t=\tau\min\{1,r/\sqrt n\}$ with $\tau=(100C)^{-1/2}\leq1/2$.
Then $t^2=\tau^2\min\{1,r^2/n\}$, so
\[
 Cnt^3=Cn\tau^2\,t\min\Bigl\{1,\frac{r^2}n\Bigr\}=\frac{t\min\{n,r^2\}}{100}\leq\frac{tr^2}{100},
\]
and the exponent is at most
\[
 -\frac{tr^2}{20}+\frac{tr^2}{100}
 \leq-\frac{tr^2}{50}
 =-\frac\tau{50}\min\Bigl\{r^2,\frac{r^3}{\sqrt n}\Bigr\}.
\]
Therefore,
\(
     \alpha_\mu(r) \le 2 \exp \left( -\frac{\tau}{50} \min \left\{ \frac{r^3}{\sqrt{n}}, r^2 \right\} \right),
\)
which is the claim with $c=\tau/50$.
\end{proof}

\subsection{Proof of the \texorpdfstring{$n^{1/4}$}{n to the power 1/4} bound}\label{sec:reduction}
\begin{proof}[Proof of \cref{thm:subgaussian-lsi}]
Let
\[
 M_n=\max\{1,n^{1/4},\CP^*(n)\log n\}.
\]
For $r\leq M_n$, \cref{thm:two-level} gives
\begin{align*}
 \alpha_\mu(r)
 &\leq C_0\exp\left(-c_0\min\left\{\frac{r}{\|\Sigma\|_{\mathrm{op}}^{1/2}},\frac{r^2}{\CP^*(n)\log n\,\|\Sigma\|_{\mathrm{op}}}\right\}\right)\\
 &\leq C_0\exp\left(-c_0\min\left\{r,\frac{r^2}{\CP^*(n)\log n}\right\}\right)
 \leq C_0e^{-c_0r^2/M_n},
\end{align*}
where $\Sigma$ is the covariance matrix of $\mu$, the second inequality holds since $\|\Sigma\|_{\mathrm{op}}\leq1$ by \cref{lem:normalized-moments}, and the last inequality holds since $r\leq M_n$ and $\CP^*(n)\log n\leq M_n$.
For $r\geq M_n$, \cref{prop:cubic} gives
\[
 \alpha_\mu(r)\leq2\exp\left(-c\min\left\{\frac{r^3}{\sqrt n},r^2\right\}\right)
 \leq2e^{-cr^2/M_n}.
\]
The last step uses $rM_n\geq M_n^2\geq\sqrt n$ and $M_n\geq1$.
Therefore, $\alpha_\mu(r)\leq e^{-\Omega(r^2/M_n)}$ for all $r>0$.
\Cref{thm:milman} now gives
\[
 \LS(\mu)\lesssim M_n\lesssim n^{1/4}.
\]
The last inequality holds because $\CP^*(n)\log n\lesssim n^{1/4}$ by \cref{thm:chen}.
\end{proof}

\appendix
\section{Background facts}\label{app:background}
This appendix recalls the standard definitions and facts from analysis, information theory and functional inequalities used in the main text, for readers less familiar with the background.

\subsection{Analysis}
We first recall the smooth functions that make up the test classes of the functional inequalities.
\begin{definition}[Smooth functions and compact support]\label{def:smooth}
A function $g:\R^n\to\R$ is \emph{smooth} if it has continuous partial derivatives of all orders.
Its \emph{support} $\supp\,g$ is the closure of $\{x:g(x)\neq0\}$, and $g$ has \emph{compact support} if $\supp\,g$ is bounded, and hence compact.
\end{definition}

We also recall absolute continuity and densities, which we use to define divergences.
\begin{definition}[Absolute continuity and densities {\cite[Chapter~3]{Folland}}]\label{def:density}
A measure $\nu$ is \emph{absolutely continuous} w.r.t.\ a measure $\mu$, written $\nu\ll\mu$, if $\mu(A)=0$ implies $\nu(A)=0$ for every measurable $A$.
If $\nu\ll\mu$ for probability measures $\mu$ and $\nu$, then $\nu(\mathrm{d}x)=g(x)\,\mu(\mathrm{d}x)$ for a measurable $g\geq0$, which is unique $\mu$-almost everywhere and is called the \emph{density} $\mathrm{d}\nu/\mathrm{d}\mu$.
A measure on $\R^n$ is \emph{absolutely continuous} if it is absolutely continuous w.r.t.\ the Lebesgue measure.
\end{definition}

We also use the following standard results of measure theory, stated on a measure space $(\Omega,\mathcal F,\nu)$.
The first one exchanges limits and integrals for nondecreasing sequences of nonnegative functions.
\begin{fact}[Monotone convergence {\cite[Chapter~2]{Folland}}]\label{fact:monotone}
Let $f_k:\Omega\to[0,\infty]$, $k\geq1$, and $f:\Omega\to[0,\infty]$ be measurable functions such that $f_1\leq f_2\leq\cdots$ and $f_k\to f$ $\nu$-almost everywhere.
Then
\[
 \lim_{k\to\infty}\int f_k\,\mathrm{d}\nu=\int f\,\mathrm{d}\nu.
\]
\end{fact}

Without monotonicity, the integral of the limit inferior of nonnegative functions is still at most the limit inferior of their integrals.
\begin{fact}[Fatou's lemma {\cite[Chapter~2]{Folland}}]\label{fact:fatou}
Let $f_k:\Omega\to[0,\infty]$, $k\geq1$, be measurable functions.
Then
\[
 \int\liminf_{k\to\infty}f_k\,\mathrm{d}\nu\leq\liminf_{k\to\infty}\int f_k\,\mathrm{d}\nu.
\]
\end{fact}

For sequences bounded by a fixed integrable function, limits and integrals can again be exchanged.
\begin{fact}[Dominated convergence {\cite[Chapter~2]{Folland}}]\label{fact:dominated}
Let $f_k$, $k\geq1$, $f$ and $g$ be measurable functions on $\Omega$ such that $f_k\to f$ $\nu$-almost everywhere, $\int|g|\,\mathrm{d}\nu<\infty$, and $|f_k|\leq|g|$ $\nu$-almost everywhere for every $k$.
Then $f$ is $\nu$-integrable and
\[
 \lim_{k\to\infty}\int f_k\,\mathrm{d}\nu=\int f\,\mathrm{d}\nu.
\]
In particular, if $\nu$ is a probability measure, this holds whenever $\sup_k\sup_\Omega|f_k|<\infty$ (bounded convergence).
\end{fact}

For nonnegative functions on a product space, the order of integration can be exchanged.
\begin{fact}[Tonelli's theorem {\cite[Chapter~2]{Folland}}]\label{fact:tonelli}
Let $(\Omega_1,\mathcal F_1,\nu_1)$ and $(\Omega_2,\mathcal F_2,\nu_2)$ be $\sigma$-finite measure spaces, and let $F:\Omega_1\times\Omega_2\to[0,\infty]$ be measurable w.r.t.\ the product $\sigma$-field.
Then $x\mapsto\int F(x,y)\,\nu_2(\mathrm{d}y)$ and $y\mapsto\int F(x,y)\,\nu_1(\mathrm{d}x)$ are measurable, and
\[
 \int\Bigl(\int F(x,y)\,\nu_2(\mathrm{d}y)\Bigr)\nu_1(\mathrm{d}x)
 =\int F\,\mathrm{d}(\nu_1\otimes\nu_2)
 =\int\Bigl(\int F(x,y)\,\nu_1(\mathrm{d}x)\Bigr)\nu_2(\mathrm{d}y).
\]
\end{fact}

Combining dominated convergence with the mean value theorem allows us to differentiate under the integral.
\begin{fact}[Differentiation under the integral {\cite[Chapter~2]{Folland}}]\label{fact:differentiation}
Let $O\subseteq\R^m$ be open, and let $g:\Omega\times O\to\R$ be such that $g(\cdot,y)$ is $\nu$-integrable for every $y\in O$ and $g(x,\cdot)$ is differentiable on $O$ for every $x\in\Omega$.
Suppose that $\|\nabla_yg(x,y)\|_2\leq G(x)$ for all $x\in\Omega$ and $y\in O$, for a $\nu$-integrable function $G$.
Then $y\mapsto\int g(x,y)\,\nu(\mathrm{d}x)$ is differentiable on $O$, and
\[
 \nabla_y\int g(x,y)\,\nu(\mathrm{d}x)=\int\nabla_yg(x,y)\,\nu(\mathrm{d}x).
\]
\end{fact}

\subsection{Information theory}
For random variables $U$, $X$ and $Y$, we write $P_X$ for the law of $X$, $P_{UX}$ for the joint law of $(U,X)$, and $P_{X\mid U}$ and $P_{UX\mid Y}$ for conditional laws.

We first recall the divergences between probability measures that we use.
\begin{definition}[Relative entropy]\label{def:kl}
For probability measures $\zeta\ll\eta$, the \emph{relative entropy}, or Kullback--Leibler divergence, is
\[
 \KL(\zeta\Vert\eta)=\int\log\frac{\mathrm{d}\zeta}{\mathrm{d}\eta}\,\mathrm{d}\zeta\in[0,\infty],
\]
and $\KL(\zeta\Vert\eta)=\infty$ for other pairs of probability measures.
It is nonnegative by Jensen's inequality.
\end{definition}

\begin{definition}[$\chi^2$-divergence]\label{def:chi2}
For probability measures $\zeta\ll\eta$, the \emph{$\chi^2$-divergence} is
\[
 \chi^2(\zeta\Vert\eta)=\int\Bigl(\frac{\mathrm{d}\zeta}{\mathrm{d}\eta}-1\Bigr)^2\mathrm{d}\eta=\E_\eta\Bigl(\frac{\mathrm{d}\zeta}{\mathrm{d}\eta}\Bigr)^2-1.
\]
\end{definition}

\begin{definition}[$\Phi$-divergence]\label{def:phi-divergence}
For a convex $\Phi:J\to\R$ and probability measures $\nu\ll\mu$ such that $\mathrm{d}\nu/\mathrm{d}\mu$ takes values in $J$, the \emph{$\Phi$-divergence} is
\[
 D_\Phi(\nu\Vert\mu)=\E_\mu\Phi\Bigl(\frac{\mathrm{d}\nu}{\mathrm{d}\mu}\Bigr).
\]
For $\Phi(z)=(z-1)^2$ and $\Phi(z)=z\log z$, it is the $\chi^2$-divergence and the relative entropy.
\end{definition}

The following comparisons relate the relative entropy to the $\chi^2$-divergence and to the total variation distance.
\begin{lemma}[Relative entropy and $\chi^2$-divergence]\label{lem:kl-chi2}
For probability measures $\zeta\ll\eta$,
\[
 \KL(\zeta\Vert\eta)\leq\log\bigl(1+\chi^2(\zeta\Vert\eta)\bigr)\leq\chi^2(\zeta\Vert\eta).
\]
\end{lemma}
\begin{proof}
Jensen's inequality for the concave logarithm gives
\[
 \KL(\zeta\Vert\eta)=\E_\zeta\log\frac{\mathrm{d}\zeta}{\mathrm{d}\eta}\leq\log\E_\zeta\frac{\mathrm{d}\zeta}{\mathrm{d}\eta}=\log\E_\eta\Bigl(\frac{\mathrm{d}\zeta}{\mathrm{d}\eta}\Bigr)^2=\log\bigl(1+\chi^2(\zeta\Vert\eta)\bigr),
\]
and $\log(1+z)\leq z$.
\end{proof}

\begin{fact}[Pinsker's inequality {\cite[Theorem~7.10]{PolyanskiyWuBook}}]\label{fact:pinsker}
For probability measures $\zeta$ and $\eta$ on the same space, with total variation distance $\TV(\zeta,\eta)=\sup_A|\zeta(A)-\eta(A)|$,
\[
 \KL(\zeta\Vert\eta)\geq2\TV(\zeta,\eta)^2.
\]
In particular, for $a,b\in[0,1]$, the total variation distance between the Bernoulli laws $\Bern(a)$ and $\Bern(b)$ with means $a$ and $b$ is $|a-b|$, and hence
\[
 \KL\bigl(\Bern(a)\Vert\Bern(b)\bigr)\geq2(a-b)^2.
\]
\end{fact}

\begin{fact}[{\cite[Theorem~9]{SasonVerdu2016}}]\label{fact:sason-verdu}
If $\zeta\ll\eta$ are probability measures with $\mathrm{d}\zeta/\mathrm{d}\eta\leq M$ for some $M>1$, then
\[
 \chi^2(\zeta\Vert\eta)\leq\frac{(M-1)^2}{M\log M-M+1}\KL(\zeta\Vert\eta).
\]
\end{fact}

The relative entropy also gives the following variational upper bound on expectations.
\begin{lemma}[Entropy duality]\label{lem:entropy-variational}
Let $\theta\ll\eta$ be probability measures, and let $g$ be a measurable function with $\E_\eta e^g<\infty$.
Then
\begin{equation*}
 \E_\theta g\leq\KL(\theta\Vert\eta)+\log\E_\eta e^g.
\end{equation*}
\end{lemma}
\begin{proof}
For bounded $g$, set $Z=\E_\eta e^g$ and $\mathrm{d}\eta_g=e^g\mathrm{d}\eta/Z$.
Nonnegativity of relative entropy gives
\[
 0\leq\KL(\theta\Vert\eta_g)
 =\KL(\theta\Vert\eta)-\E_\theta g+\log Z.
\]
If the relative entropy is infinite the claim is immediate.
Otherwise, apply the bounded case first to $g_+\wedge M$.
Since $e^{g_+}\leq1+e^g$, monotone convergence (\cref{fact:monotone}) proves $\E_\theta g_+<\infty$.
If $\E_\theta g_-=\infty$, the desired upper bound is automatic.
In the remaining case use $g_M=\max\{-M,\min\{g,M\}\}$.
Then $|g_M|\leq|g|$ and $e^{g_M}\leq1+e^g$, so dominated convergence (\cref{fact:dominated}) on both sides proves the result.
\end{proof}

Next, we recall entropy and mutual information.
\begin{definition}[Shannon entropy]\label{def:entropy}
For a random variable $U$ taking finitely many values, the \emph{entropy} is $H(U)=-\sum_u\P(U=u)\log\P(U=u)$, and for another random variable $Y$, the \emph{conditional entropy} $H(U\mid Y)$ is the expectation over $Y$ of the entropy of $P_{U\mid Y}$.
The \emph{binary entropy} is $\Hb(z)=-z\log z-(1-z)\log(1-z)$ for $z\in[0,1]$, with $0\log0=0$, so that $H(U)=\Hb(\P(U=1))$ for $U\in\{0,1\}$.
\end{definition}

\begin{definition}[Mutual information]\label{def:mutual-information}
For random variables $U$ and $X$, the \emph{mutual information} is
\[
 I(U;X)=\KL(P_{UX}\Vert P_U\otimes P_X).
\]
For a third random variable $Y$, the \emph{conditional mutual information} is
\[
 I(U;X\mid Y)=\E_Y\KL\bigl(P_{UX\mid Y}\Vert P_{U\mid Y}\otimes P_{X\mid Y}\bigr).
\]
\end{definition}

Mutual information is an average of relative entropies, and conditioning on an additional variable averages it.
\begin{fact}[Mutual information as an average relative entropy {\cite[Chapter~3]{PolyanskiyWuBook}}]\label{fact:mi-average}
For random variables $U$, $X$ and $Y$,
\[
 I(U;X)=\E_U\KL(P_{X\mid U}\Vert P_X),
\]
and
\[
 I(U;X\mid Y)=\E_{(U,Y)}\KL(P_{X\mid U,Y}\Vert P_{X\mid Y}).
\]
Indeed, $P_{UX}$ has density $\mathrm{d}P_{X\mid U=u}/\mathrm{d}P_X(x)$ at $(u,x)$ w.r.t.\ $P_U\otimes P_X$, and in the same way, $P_{UX\mid Y=y}$ has density $\mathrm{d}P_{X\mid U=u,Y=y}/\mathrm{d}P_{X\mid Y=y}(x)$ at $(u,x)$ w.r.t.\ $P_{U\mid Y=y}\otimes P_{X\mid Y=y}$.
\end{fact}

\begin{lemma}[Conditional mutual information as an average]\label{lem:mi-conditioning}
Let $U$, $X$, $Y$ and $Z$ be random variables.
For $P_Y$-almost every $y$, write $I(U;X\mid Y=y)$ for the mutual information of $U$ and $X$, and $I(U;X\mid Z,Y=y)$ for their conditional mutual information given $Z$, both under the conditional law $P_{UXZ\mid Y=y}$.
Then
\[
 I(U;X\mid Y)=\E_{y\sim P_Y}I(U;X\mid Y=y),
\]
and
\[
 I(U;X\mid Y,Z)=\E_{y\sim P_Y}I(U;X\mid Z,Y=y).
\]
\end{lemma}
\begin{proof}
The first identity is \cref{def:mutual-information}.
For the second, under $P_{UXZ\mid Y=y}$, the conditional law of $(U,X)$ given $Z=z$ is $P_{UX\mid Y=y,Z=z}$ for $P_{Z\mid Y=y}$-almost every $z$, and likewise for $U$ and for $X$ alone.
So \cref{def:mutual-information} under $P_{UXZ\mid Y=y}$ gives
\[
 I(U;X\mid Z,Y=y)
 =\E_{z\sim P_{Z\mid Y=y}}\KL\bigl(P_{UX\mid Y=y,Z=z}\Vert P_{U\mid Y=y,Z=z}\otimes P_{X\mid Y=y,Z=z}\bigr).
\]
By the tower property, averaging over $y\sim P_Y$ and then over $z\sim P_{Z\mid Y=y}$ is averaging over $(y,z)\sim P_{YZ}$, and the result is $I(U;X\mid Y,Z)$ by \cref{def:mutual-information}.
\end{proof}

For a label taking finitely many values, \cref{fact:mi-average} gives the following formula in terms of the conditional probabilities of its values.
\begin{lemma}\label{lem:label-entropy}
Let $U$ take finitely many values, let $Y$ be a random variable, and let $g_u(y)=\P(U=u\mid Y=y)$.
Then
\[
 I(U;Y)=\sum_u\Ent_{P_Y}(g_u).
\]
\end{lemma}
\begin{proof}
If $\P(U=u)=0$, then $g_u=0$ almost surely, and $u$ contributes nothing to either side, so let $\P(U=u)>0$.
Bayes' theorem gives $\mathrm{d}P_{Y\mid U=u}/\mathrm{d}P_Y=g_u/\P(U=u)$, where $\P(U=u)=\E_{P_Y}g_u$.
Hence
\[
 \P(U=u)\KL(P_{Y\mid U=u}\Vert P_Y)
 =\E_{P_Y}\Bigl[g_u\log\frac{g_u}{\E_{P_Y}g_u}\Bigr]
 =\E_{P_Y}[g_u\log g_u]-\E_{P_Y}g_u\log\E_{P_Y}g_u
 =\Ent_{P_Y}(g_u).
\]
Summing over $u$ and using \cref{fact:mi-average} proves the claim.
\end{proof}

In particular, a label taking $k$ values carries at most $\log k$ of information.
\begin{lemma}\label{lem:entropy-bound}
Let $U$ take $k$ values, and let $Y$ be a random variable.
Then
\[
 I(U;Y)=H(U)-H(U\mid Y)\leq H(U)\leq\log k.
\]
\end{lemma}
\begin{proof}
Let $g_u(y)=\P(U=u\mid Y=y)$.
By \cref{def:entropy}, $\sum_u\E_{P_Y}[g_u\log g_u]=-H(U\mid Y)$ and $\sum_u\E_{P_Y}g_u\log\E_{P_Y}g_u=\sum_u\P(U=u)\log\P(U=u)=-H(U)$.
So \cref{lem:label-entropy} gives $I(U;Y)=H(U)-H(U\mid Y)$, which is at most $H(U)$ since $H(U\mid Y)\geq0$.
Finally, $H(U)=\E\log(1/\P(U))\leq\log\E[1/\P(U)]\leq\log k$ by Jensen's inequality.
\end{proof}

Next, we recall Markov chains, which we use for a label $U$ and an observation $Y$ of the same input $X$.
\begin{definition}[Markov chain]\label{def:markov}
For random variables $U$, $X$ and $Y$, we say that $U-X-Y$ is a \emph{Markov chain} if $U$ and $Y$ are conditionally independent given $X$, that is, $P_{UY\mid X}=P_{U\mid X}\otimes P_{Y\mid X}$.
\end{definition}

Finally, we consider Markov kernels, which we also call channels, and first compute the likelihood ratio between two output laws.
\begin{lemma}[Output likelihood ratio]\label{lem:output-density}
Let $\zeta\ll\eta$ be probability measures, let $K$ be a Markov kernel, let $X\sim\eta$, and let $Y$ have conditional law $K(X,\cdot)$ given $X$.
Then $\zeta K\ll\eta K$, and with $g=\mathrm{d}\zeta/\mathrm{d}\eta$, almost surely,
\[
 \frac{\mathrm{d}(\zeta K)}{\mathrm{d}(\eta K)}(Y)=\E[g(X)\mid Y].
\]
\end{lemma}
\begin{proof}
Let $h$ be a nonnegative measurable function with $h(Y)=\E[g(X)\mid Y]$.
For every measurable set $B$, since $\P(Y\in B\mid X)=K(X,B)$, the tower property gives
\[
 (\zeta K)(B)
 =\int K(x,B)\,g(x)\,\eta(\mathrm{d}x)
 =\E[g(X)\mathbf1_B(Y)]
 =\E[h(Y)\mathbf1_B(Y)]
 =\int_Bh\,\mathrm{d}(\eta K).
\]
Hence $\zeta K\ll\eta K$ with density $h$.
\end{proof}

Divergences and mutual information do not increase through a channel.
\begin{lemma}[Data processing]\label{lem:data-processing}
Let $\zeta\ll\eta$ be probability measures, let $K$ be a Markov kernel, and let $\Phi$ be convex on an interval containing the values of $\mathrm{d}\zeta/\mathrm{d}\eta$.
Then $D_\Phi(\zeta K\Vert\eta K)\leq D_\Phi(\zeta\Vert\eta)$, with equality if $K$ is given by a measurable bijection with measurable inverse.
In particular, this holds for the relative entropy and the $\chi^2$-divergence, and $I(U;Y)\leq I(U;X)$ whenever $U-X-Y$ is a Markov chain.
\end{lemma}
\begin{proof}
Let $X\sim\eta$, let $Y$ have conditional law $K(X,\cdot)$ given $X$, and let $g=\mathrm{d}\zeta/\mathrm{d}\eta$.
By the output likelihood ratio (\cref{lem:output-density}), the density $\mathrm{d}(\zeta K)/\mathrm{d}(\eta K)$ evaluated at $Y$ is $\E[g(X)\mid Y]$, so Jensen's inequality for conditional expectations gives
\[
 D_\Phi(\zeta K\Vert\eta K)=\E\Phi\bigl(\E[g(X)\mid Y]\bigr)\leq\E\Phi(g(X))=D_\Phi(\zeta\Vert\eta).
\]
For a bijection, applying this to $K$ and to its inverse gives equality.
For a Markov chain $U-X-Y$, let $K$ be the conditional law of $Y$ given $X$, so that $P_{Y\mid U}=P_{X\mid U}K$ and $P_Y=P_XK$.
Since mutual information can be written as an average of relative entropies (\cref{fact:mi-average}), the inequality for the relative entropy gives $I(U;Y)=\E_U\KL(P_{X\mid U}K\Vert P_XK)\leq\E_U\KL(P_{X\mid U}\Vert P_X)=I(U;X)$.
\end{proof}

Applied to mixtures, data processing gives the joint convexity of relative entropy.
\begin{lemma}[Joint convexity of relative entropy]\label{lem:joint-convexity}
If $\zeta=\int\zeta_s\,\rho(\mathrm{d}s)$ and $\eta=\int\eta_s\,\rho(\mathrm{d}s)$ for a probability measure $\rho$ and probability measures $\zeta_s\ll\eta_s$, then
\[
 \KL(\zeta\Vert\eta)\leq\int\KL(\zeta_s\Vert\eta_s)\,\rho(\mathrm{d}s).
\]
\end{lemma}
\begin{proof}
Let $\bar\zeta(\mathrm{d}s,\mathrm{d}x)=\rho(\mathrm{d}s)\zeta_s(\mathrm{d}x)$ and $\bar\eta(\mathrm{d}s,\mathrm{d}x)=\rho(\mathrm{d}s)\eta_s(\mathrm{d}x)$.
Then $\mathrm{d}\bar\zeta/\mathrm{d}\bar\eta(s,x)=\mathrm{d}\zeta_s/\mathrm{d}\eta_s(x)$, so $\KL(\bar\zeta\Vert\bar\eta)=\int\KL(\zeta_s\Vert\eta_s)\,\rho(\mathrm{d}s)$.
The map $(s,x)\mapsto x$ sends $\bar\zeta$ to $\zeta$ and $\bar\eta$ to $\eta$, so data processing (\cref{lem:data-processing}) gives the claim.
\end{proof}

The chain rule splits mutual information along an additional variable.
\begin{lemma}[Chain rule]\label{lem:chain-rule}
For random variables $U$, $X$ and $Y$,
\[
 I(U;(X,Y))=I(U;Y)+I(U;X\mid Y).
\]
In particular, if $U-X-Y$ is a Markov chain and $I(U;X)<\infty$, then $I(U;X\mid Y)=I(U;X)-I(U;Y)$.
Moreover, if $U-X-(Y,Z)$ is a Markov chain for a random variable $Z$ and $I(U;X)<\infty$, then $I(U;X\mid Y,Z)\leq I(U;X\mid Y)$.
\end{lemma}
\begin{proof}
The density of $P_{UXY}$ w.r.t.\ $P_U\otimes P_{XY}$ is the product of the density of $P_{UY}$ w.r.t.\ $P_U\otimes P_Y$ and the density of $P_{UX\mid Y}$ w.r.t.\ $P_{U\mid Y}\otimes P_{X\mid Y}$.
Taking the expectation of the logarithm gives the chain rule, see also \cite[Chapter~3]{PolyanskiyWuBook}.
If $U-X-Y$ is a Markov chain, then $I(U;Y\mid X)=0$ by \cref{def:markov}, so the chain rule with $X$ and $Y$ exchanged gives $I(U;(X,Y))=I(U;X)$, and the claim follows.

For the last claim, $U-X-Y$ is also a Markov chain, since the conditional law of $(U,Y)$ given $X$ is the image of the product measure $P_{U\mid X}\otimes P_{(Y,Z)\mid X}$ under the map $(u,(y,z))\mapsto(u,y)$, which is again a product measure.
Since $Y$ is a function of $(Y,Z)$, the sequence $U-(Y,Z)-Y$ is a Markov chain, so data processing (\cref{lem:data-processing}) gives $I(U;Y)\leq I(U;(Y,Z))$.
Hence the claim for the two Markov chains $U-X-Y$ and $U-X-(Y,Z)$ gives
\[
 I(U;X\mid Y)
 =I(U;X)-I(U;Y)
 \geq I(U;X)-I(U;(Y,Z))
 =I(U;X\mid Y,Z).
\]
\end{proof}

The golden formula computes mutual information through any reference law for the output.
\begin{lemma}[Golden formula]\label{lem:golden-formula}
Let $X\sim\rho$, let $Y$ have conditional law $K_X=K(X,\cdot)$ given $X$, and let $\eta$ be a probability measure with $K_x\ll\eta$ for $\rho$-almost every $x$.
Then
\[
 I(X;Y)+\KL(\rho K\Vert\eta)=\E\KL(K_X\Vert\eta),
\]
where both sides may be infinite.
\end{lemma}
\begin{proof}
For $\rho$-almost every $x$, we have $K_x\ll\rho K\ll\eta$ and $\mathrm{d}K_x/\mathrm{d}\eta=(\mathrm{d}K_x/\mathrm{d}(\rho K))\cdot h$, where $h=\mathrm{d}(\rho K)/\mathrm{d}\eta$, so
\[
 \KL(K_x\Vert\eta)=\KL(K_x\Vert\rho K)+\int\log h\,\mathrm{d}K_x.
\]
The negative part of $\log h$ is integrable under $\rho K$, since $\int(\log h)_-\,\mathrm{d}(\rho K)=\int h(\log h)_-\,\mathrm{d}\eta\leq1/e$.
Averaging over $x\sim\rho$ and using $\int K_x\,\rho(\mathrm{d}x)=\rho K$ and $\E\KL(K_X\Vert\rho K)=I(X;Y)$ (\cref{fact:mi-average}) gives the claim.
\end{proof}

For a mixture, the $\Phi$-entropy splits into the average $\Phi$-entropy of the components and the $\Phi$-entropy of the mixing weights.
\begin{lemma}[Entropy decomposition]\label{lem:entropy-decomposition}
Let $J$ be an interval, let $\Phi:J\to\R$ be convex, and let $\pi=\int\pi_y\,\rho(\mathrm{d}y)$ be a mixture of probability measures $\pi_y$ over a probability measure $\rho$.
Let $f$ be a $J$-valued function such that $f$ and $\Phi(f)$ are $\pi$-integrable, and let $H(y)=\E_{\pi_y}f$.
Then
\begin{equation}\label{eq:entropy-decomposition}
 \Ent_\pi^\Phi(f)=\int\Ent_{\pi_y}^\Phi(f)\,\rho(\mathrm{d}y)+\Ent_\rho^\Phi(H).
\end{equation}
\end{lemma}
\begin{proof}
We first check that all the terms are finite.
Since $\int\E_{\pi_y}\bigl(|f|+|\Phi(f)|\bigr)\,\rho(\mathrm{d}y)=\E_\pi\bigl(|f|+|\Phi(f)|\bigr)<\infty$, for $\rho$-almost every $y$ the functions $f$ and $\Phi(f)$ are $\pi_y$-integrable, and $H(y)\in J$.
A convex function lies above an affine function $z\mapsto a+bz$, so Jensen's inequality gives $a+bH(y)\leq\Phi(H(y))\leq\E_{\pi_y}\Phi(f)$, and hence $\Phi(H)$ is $\rho$-integrable.

Next, we decompose.
Since $\E_\pi\Phi(f)=\int\E_{\pi_y}\Phi(f)\,\rho(\mathrm{d}y)$ and $\E_\pi f=\E_\rho H$,
\[
 \Ent_\pi^\Phi(f)
 =\int\bigl(\E_{\pi_y}\Phi(f)-\Phi(H(y))\bigr)\,\rho(\mathrm{d}y)+\E_\rho\Phi(H)-\Phi(\E_\rho H),
\]
which is \eqref{eq:entropy-decomposition}.
\end{proof}

The following lemma decomposes the mutual information into the relative entropy lost by the conditional laws of the input given the label, for every channel.
\begin{lemma}[Information as an average entropy loss]\label{lem:information-average}
Let $K$ be a channel with input law $\pi$, let $X\sim\pi$, and let $Y$ be the output of $K$ for the input $X$.
Let $U$ be a random variable such that $U-X-Y$ is a Markov chain and $I(U;X)<\infty$, and let $\zeta_u=\Law(X\mid U=u)$ be the classes of $U$.
Then $\KL(\zeta_U\Vert\pi)<\infty$ almost surely, and
\begin{equation}\label{eq:information-average}
 I(U;X\mid Y)=\E_U\bigl[\KL(\zeta_U\Vert\pi)-\KL(\zeta_UK\Vert\pi K)\bigr].
\end{equation}
\end{lemma}
\begin{proof}
Since $U-X-Y$ is a Markov chain, the conditional law of $Y$ given $U=u$ is $\zeta_uK$, and the law of $Y$ is $\pi K$.
Hence \cref{fact:mi-average} gives $I(U;X)=\E_U\KL(\zeta_U\Vert\pi)$ and $I(U;Y)=\E_U\KL(\zeta_UK\Vert\pi K)$.
The first identity and $I(U;X)<\infty$ give $\KL(\zeta_U\Vert\pi)<\infty$ almost surely, and data processing (\cref{lem:data-processing}) gives $I(U;Y)\leq I(U;X)<\infty$.
The chain rule (\cref{lem:chain-rule}) gives $I(U;X\mid Y)=I(U;X)-I(U;Y)$, and subtracting the two identities gives \eqref{eq:information-average}.
\end{proof}

The following lemma is the characterization of the contraction coefficient of relative entropy through mutual information \cite[Theorem~4]{PolyanskiyWu}, for priors with bounded density and binary labels.
We include its short proof, which follows theirs, for completeness.
\begin{lemma}[{\cite[Theorem~4]{PolyanskiyWu}}]\label{lem:label-contraction}
Let $K$ be a channel with input law $\pi$, let $X\sim\pi$, and let $Y$ be the output of $K$ for the input $X$.
For every $b>0$, the following are equivalent.
\begin{enumerate}[label=(\roman*)]
\item Every probability measure $\zeta\ll\pi$ with bounded density $\mathrm{d}\zeta/\mathrm{d}\pi$ satisfies
\[
 \KL(\zeta\Vert\pi)-\KL(\zeta K\Vert\pi K)\geq b\KL(\zeta\Vert\pi).
\]
\item Every random variable $U\in\{0,1\}$ such that $U-X-Y$ is a Markov chain satisfies
\[
 I(U;X\mid Y)\geq bI(U;X).
\]
\end{enumerate}
\end{lemma}
\begin{proof}
We first show that (i) implies (ii), by applying (i) to the conditional laws of $X$ given $U$.
Let $U$ be as in (ii), and let $\zeta_u=\Law(X\mid U=u)$ for $u\in\{0,1\}$ with $\P(U=u)>0$.
By Bayes' theorem, $\mathrm{d}\zeta_u/\mathrm{d}\pi(x)=\P(U=u\mid X=x)/\P(U=u)\leq1/\P(U=u)$, so $\zeta_u$ has bounded density.
Hence \cref{lem:information-average} gives the first equality in
\[
 I(U;X\mid Y)
 =\E_U\bigl[\KL(\zeta_U\Vert\pi)-\KL(\zeta_UK\Vert\pi K)\bigr]
 \geq b\E_U\KL(\zeta_U\Vert\pi)
 =bI(U;X),
\]
where the inequality is (i) with $\zeta=\zeta_U$, and the last equality follows from \cref{fact:mi-average}.

Next, we show that (ii) implies (i), by applying (ii) to a label whose class $\{U=1\}$ has law $\zeta$ and is rare.
Let $\zeta\ll\pi$ have a bounded density $w=\mathrm{d}\zeta/\mathrm{d}\pi$, let $0<\varepsilon<1/\sup w$, and let $U\in\{0,1\}$ be drawn independently of $Y$ given $X$, with $\P(U=1\mid X)=\varepsilon w(X)$, so that $U-X-Y$ is a Markov chain.
Then $\P(U=1)=\E_\pi[\varepsilon w]=\varepsilon$, and by Bayes' theorem,
\[
 \zeta_1(\mathrm{d}x)=\frac{\P(U=1\mid X=x)\,\pi(\mathrm{d}x)}{\P(U=1)}=w(x)\,\pi(\mathrm{d}x)=\zeta(\mathrm{d}x).
\]
Similarly, $\mathrm{d}\zeta_0/\mathrm{d}\pi=(1-\varepsilon w)/(1-\varepsilon)$, so \cref{lem:information-average,fact:mi-average} give
\begin{align*}
 I(U;X\mid Y)&=\varepsilon\bigl(\KL(\zeta\Vert\pi)-\KL(\zeta K\Vert\pi K)\bigr)+(1-\varepsilon)\bigl(\KL(\zeta_0\Vert\pi)-\KL(\zeta_0K\Vert\pi K)\bigr),\\
 I(U;X)&=\varepsilon\KL(\zeta\Vert\pi)+(1-\varepsilon)\KL(\zeta_0\Vert\pi).
\end{align*}
The class $\zeta_0$ is close to $\pi$, since $\KL(\zeta_0K\Vert\pi K)\geq0$ and $\KL\leq\chi^2$ (\cref{lem:kl-chi2}) give
\begin{align*}
 \KL(\zeta_0\Vert\pi)-\KL(\zeta_0K\Vert\pi K)
 &\leq\KL(\zeta_0\Vert\pi)\leq\chi^2(\zeta_0\Vert\pi)
 =\E_\pi\Bigl[\Bigl(\frac{\varepsilon(1-w)}{1-\varepsilon}\Bigr)^2\Bigr]\\
 &=\frac{\varepsilon^2}{(1-\varepsilon)^2}\chi^2(\zeta\Vert\pi)
 \leq\frac{\varepsilon^2\sup w}{(1-\varepsilon)^2}.
\end{align*}
Therefore, (ii) gives
\begin{align*}
 \varepsilon\bigl(\KL(\zeta\Vert\pi)-\KL(\zeta K\Vert\pi K)\bigr)+\frac{\varepsilon^2\sup w}{1-\varepsilon}
 &\geq I(U;X\mid Y)\\
 &\geq bI(U;X)
 \geq b\varepsilon\KL(\zeta\Vert\pi).
\end{align*}
Dividing by $\varepsilon$ and letting $\varepsilon\to0$ gives (i).
\end{proof}

\subsection{Functional inequalities}\label{app:functional-inequalities}
We use the following log-Sobolev inequalities for strongly log-concave and Gaussian measures.
\begin{fact}[Bakry--\'Emery {\cite{BakryEmery}, \cite[Corollary~2.1]{Chafai2004}}]\label{fact:bakry-emery}
Let $\mu(\mathrm{d}x)=e^{-W(x)}\mathrm{d}x$ with $\nabla^2W\succeq\rho I_n$ for some $\rho>0$.
Then $\CP(\mu)\leq1/\rho$ and $\LS(\mu)\leq1/\rho$, and more generally $\mu$ satisfies the $\Phi$-Sobolev inequality \eqref{eq:phi-functional} with constant $1/\rho$ for every $\Phi$ as in \cref{thm:general-equivalence}.
\end{fact}

\begin{fact}[Gaussian log-Sobolev inequality {\cite{Gross1975}}]\label{fact:gaussian-lsi}
The standard Gaussian law satisfies $\LS(\gamma_n)=1$.
Equivalently, applying \eqref{eq:lsi-convention} to $f=\sqrt g$ and using $\nabla\sqrt g=\frac12\sqrt g\,\nabla\log g$, every positive $g$ satisfies
\[
 \Ent_{\gamma_n}(g)\leq\frac12\E_{\gamma_n}\bigl[g\|\nabla\log g\|_2^2\bigr].
\]
\end{fact}

The Gaussian log-Sobolev inequality implies that the relative entropy w.r.t.\ $\gamma_n$ decays along the Ornstein--Uhlenbeck semigroup.
\begin{fact}[Entropy decay along the Ornstein--Uhlenbeck semigroup {\cite[Section~5.2]{BakryGentilLedoux}}]\label{fact:ou-decay}
For $q\in(0,1]$ and a probability measure $\nu$ on $\R^n$, let $\nu O_q$ be the law of $\sqrt q\,V+\sqrt{1-q}\,G$, where $V\sim\nu$ and $G\sim\gamma_n$ are independent.
Then $\gamma_nO_q=\gamma_n$ and $\KL(\nu O_q\Vert\gamma_n)\leq q\KL(\nu\Vert\gamma_n)$.
\end{fact}

More generally, a $\Phi$-Sobolev inequality for $\mu$ implies that $\Phi$-divergences w.r.t.\ $\mu$ contract along the Gaussian channel.
\begin{fact}[Contraction of $\Phi$-divergences {\cite[Lemma~15]{MitraWibisono}}]\label{fact:divergence-contraction}
Let $J$ be an open interval, and let $\Psi:J\to\R$ be smooth with $\Psi''>0$, $1/\Psi''$ concave, and $\Psi(1)=0$.
If $\mu$ satisfies the $\Psi$-Sobolev inequality \eqref{eq:phi-functional}, with $\Psi$ in place of $\Phi$, with constant $C$, then for every $\nu\ll\mu$ and every $t>0$,
\[
 D_\Psi(\nu K_t\Vert\mu K_t)\leq\frac{Ct}{1+Ct}D_\Psi(\nu\Vert\mu).
\]
\end{fact}

A log-Sobolev inequality also bounds $\E_\eta e^{\lambda f}$ for Lipschitz functions $f$, which gives Gaussian concentration.
\begin{lemma}[Herbst's argument]\label{lem:herbst}
If $\LS(\eta)\leq S$, every integrable $L$-Lipschitz $f$ satisfies, for every $\lambda\in\R$,
\[
 \log\E_\eta e^{\lambda(f-\E_\eta f)}
 \leq\frac{SL^2\lambda^2}{2}.
\]
\end{lemma}
\begin{proof}
By truncation, mollification, and Fatou's lemma (\cref{fact:fatou}), we may assume that $f$ is bounded and smooth with $\|\nabla f\|_2\leq L$, and then the LSI applies to $e^{\lambda f/2}$ by \cref{rem:test-functions}.
Set $Z(\lambda)=\E_\eta e^{\lambda f}$ and $J(\lambda)=\log Z(\lambda)$.
Boundedness of $f$ permits differentiation under the integral (\cref{fact:differentiation}), so
\[
 J'(\lambda)=\frac{\E_\eta[f e^{\lambda f}]}{Z(\lambda)},
 \qquad J(0)=0,\qquad J'(0)=\E_\eta f.
\]
The entropy and Dirichlet energy of the exponential test function are
\begin{align*}
 \Ent_\eta(e^{\lambda f})
 &=\lambda\E_\eta[f e^{\lambda f}]-Z(\lambda)\log Z(\lambda)
 =Z(\lambda)\bigl(\lambda J'(\lambda)-J(\lambda)\bigr),\\
 \E_\eta\bigl\|\nabla e^{\lambda f/2}\bigr\|_2^2
 &=\frac{\lambda^2}{4}\E_\eta\bigl[e^{\lambda f}\|\nabla f\|_2^2\bigr]
 \leq\frac{\lambda^2L^2}{4}Z(\lambda).
\end{align*}
Applying the LSI and dividing by $Z(\lambda)>0$ therefore gives
\[
 \lambda J'(\lambda)-J(\lambda)\leq\frac{SL^2\lambda^2}{2}.
\]
For $\lambda>0$, this implies
\[
 \left(\frac{J(\lambda)}{\lambda}\right)'
 =\frac{\lambda J'(\lambda)-J(\lambda)}{\lambda^2}
 \leq\frac{SL^2}{2}.
\]
Integrating from $\varepsilon$ to $\lambda$ gives
\[
 \frac{J(\lambda)}{\lambda}-\frac{J(\varepsilon)}{\varepsilon}
 \leq\frac{SL^2}{2}(\lambda-\varepsilon).
\]
Letting $\varepsilon\to0$ and using $J(\varepsilon)/\varepsilon\to J'(0)=\E_\eta f$ yields
\[
 J(\lambda)-\lambda\E_\eta f\leq\frac{SL^2\lambda^2}{2}.
\]
For $\lambda<0$, apply the same calculation to $-f$ with parameter $-\lambda>0$.
The case $\lambda=0$ is immediate.
\end{proof}

Log-Sobolev inequalities also imply transport inequalities for the Wasserstein distance.
\begin{definition}[Wasserstein distance]\label{def:wasserstein}
For probability measures $\theta$ and $\eta$ on $\R^n$, the squared 2-Wasserstein distance is
\[
 W_2(\theta,\eta)^2=\inf\E\|Z_1-Z_2\|_2^2,
\]
where the infimum is over all couplings of $Z_1\sim\theta$ and $Z_2\sim\eta$, that is, pairs of random vectors with these marginal laws.
\end{definition}

\begin{fact}[Triangle inequality for $W_2$ {\cite[Chapter~6]{Villani2009}}]\label{fact:w2-triangle}
For probability measures $\theta_1$, $\theta_2$, and $\theta_3$ on $\R^n$,
\[
 W_2(\theta_1,\theta_3)\leq W_2(\theta_1,\theta_2)+W_2(\theta_2,\theta_3),
\]
where both sides may be infinite.
\end{fact}

\begin{fact}[Otto--Villani {\cite{OttoVillani},\cite[Chapter 22]{Villani2009}}]\label{fact:otto-villani}
If $\eta$ is absolutely continuous, then every probability measure $\theta\ll\eta$ satisfies
\begin{equation}\label{eq:transport}
 W_2(\theta,\eta)^2\leq2\LS(\eta)\KL(\theta\Vert\eta).
\end{equation}
\end{fact}

Finally, log-concavity is preserved under taking marginals.
\begin{fact}[Pr\'ekopa--Leindler {\cite{Prekopa1973}}]\label{fact:prekopa}
If $F:\R^n\times\R^m\to[0,\infty)$ is log-concave, that is, $\log F$ is concave with $\log0=-\infty$, then $x\mapsto\int F(x,y)\,\mathrm{d}y$ is log-concave on $\R^n$.
In particular, if $X$ has a log-concave density on $\R^n$ and $v\in\R^n$ is nonzero, then $\langle v,X\rangle$ has a log-concave density on $\R$.
\end{fact}

\section{Omitted proofs}\label{app:omitted}
This appendix contains the proofs deferred from the main text.

\subsection{Proofs of \texorpdfstring{\cref{lem:dual-poincare,lem:poisson-dual-norm}}{Lemmas \getrefnumber{lem:dual-poincare} and \getrefnumber{lem:poisson-dual-norm}}}\label{app:dual-poincare}
\begin{proof}[Proof of \cref{lem:dual-poincare}]
We first show that $\CP(\mu)\leq C$ implies \eqref{eq:dual-poincare}.
Let $f\in L^2(\mu)$ with $\E_\mu f=0$, and let $h\in H^1(\mu)$ with $\E_\mu\|\nabla h\|_2^2\leq1$.
Since $\E_\mu f=0$, the Cauchy--Schwarz inequality gives
\[
 \E_\mu[fh]
 =\E_\mu[f(h-\E_\mu h)]
 \leq\sqrt{\E_\mu f^2\,\var_\mu(h)}
 \leq\sqrt{C\E_\mu f^2\,\E_\mu\|\nabla h\|_2^2}
 \leq\sqrt{C\E_\mu f^2},
\]
where the second inequality follows from the Poincar\'e inequality for $h$.
Taking the supremum over $h$ proves \eqref{eq:dual-poincare}.

Next, we show that \eqref{eq:dual-poincare} implies $\CP(\mu)\leq C$.
Let $h\in H^1(\mu)$ and set $f=h-\E_\mu h$.
Then $\E_\mu f=0$, $\E_\mu f^2=\var_\mu(h)$, and $\E_\mu[fh]=\var_\mu(h)$.
If $\E_\mu\|\nabla h\|_2^2=0$, then $\lambda h$ is a valid choice in \eqref{eq:dual-sobolev} for every $\lambda>0$, so $\|f\|_{H^{-1}(\mu)}\geq\lambda\var_\mu(h)$ for every $\lambda>0$, and \eqref{eq:dual-poincare} forces $\var_\mu(h)=0$.
Otherwise, applying \eqref{eq:dual-sobolev} to $h/\sqrt{\E_\mu\|\nabla h\|_2^2}$ and then \eqref{eq:dual-poincare} gives
\[
 \var_\mu(h)
 =\E_\mu[fh]
 \leq\|f\|_{H^{-1}(\mu)}\sqrt{\E_\mu\|\nabla h\|_2^2}
 \leq\sqrt{C\var_\mu(h)\,\E_\mu\|\nabla h\|_2^2},
\]
and dividing by $\sqrt{\var_\mu(h)}$ gives the Poincar\'e inequality for $h$ with constant $C$.

Finally, by homogeneity, the least $C\geq0$ for which \eqref{eq:dual-poincare} holds for every centered $f\in L^2(\mu)$ is the right-hand side of \eqref{eq:dual-poincare-sup}, and by the equivalence just proved this least $C$ is $\CP(\mu)$.
\end{proof}

\begin{proof}[Proof of \cref{lem:poisson-dual-norm}]
We first construct $u$.
Write $C=\CP(\mu)<\infty$.
The Poincar\'e inequality gives $\var_\mu(h)\leq C\E_\mu\|\nabla h\|_2^2$ for every $h\in H^1(\mu)$.
Hence $(u,h)\mapsto\E_\mu\langle\nabla u,\nabla h\rangle$ is an inner product on the centered functions in $H^1(\mu)$ whose norm is equivalent to the $H^1(\mu)$ norm, so this space is a Hilbert space.
For centered $h\in H^1(\mu)$, the Cauchy--Schwarz inequality and the Poincar\'e inequality give
\[
 |\E_\mu[fh]|
 \leq\sqrt{\E_\mu f^2\,\var_\mu(h)}
 \leq\sqrt{C\E_\mu f^2}\,\sqrt{\E_\mu\|\nabla h\|_2^2},
\]
so $h\mapsto\E_\mu[fh]$ is a bounded linear functional on this Hilbert space, and the Riesz representation theorem gives a unique centered $u\in H^1(\mu)$ satisfying \eqref{eq:poisson-weak} for centered $h$.
Since both sides of \eqref{eq:poisson-weak} vanish for constant $h$, it holds for all $h\in H^1(\mu)$.

Next, we show \eqref{eq:poisson-dual-norm}.
Taking $h=u$ in \eqref{eq:poisson-weak} gives $\E_\mu[fu]=\E_\mu\|\nabla u\|_2^2$.
For $h\in H^1(\mu)$ with $\E_\mu\|\nabla h\|_2^2\leq1$, \eqref{eq:poisson-weak} and the Cauchy--Schwarz inequality give
\[
 \E_\mu[fh]
 =\E_\mu\langle\nabla u,\nabla h\rangle
 \leq\sqrt{\E_\mu\|\nabla u\|_2^2},
\]
with equality for $h=u/\sqrt{\E_\mu\|\nabla u\|_2^2}$ when $\nabla u\neq0$.
Taking the supremum over $h$ gives \eqref{eq:poisson-dual-norm}, and when $\nabla u=0$ both sides of \eqref{eq:poisson-dual-norm} vanish.
\end{proof}

\subsection{Proof of \texorpdfstring{\cref{lem:uniform-loss}}{Lemma \getrefnumber{lem:uniform-loss}}}\label{app:channel}
In this subsection, $K$ is a channel with input law $\pi$ that satisfies \eqref{eq:small-set-contraction}, and we prove \cref{lem:uniform-loss}.
For a finite-entropy $\zeta\ll\pi$, write $f=\mathrm{d}\zeta/\mathrm{d}\pi$.
We split its relative entropy into the contributions from $\{f<e^{20}\}$ and $\{f\geq e^{20}\}$.
The first contribution is bounded by a constant times $\var_\pi(\sqrt f)$; \cref{lem:conditional-variance} will show that this variance is controlled by the entropy loss.
For the second contribution, we apply \eqref{eq:small-set-contraction} to $\zeta$ conditioned on level sets of $f$.
This use of level sets is analogous to the measure--capacity criteria for Poincar\'e and log-Sobolev inequalities \cite{BobkovGotze1999,BartheCattiauxRoberto2006}.

\begin{lemma}\label{lem:conditional-variance}
Let $K$ be a channel with input law $\pi$.
Suppose every finite-entropy $\zeta\ll\pi$ supported on a set of $\pi$-mass at most $1/2$ satisfies \eqref{eq:small-set-contraction}.
Let $Y$ be the output of $K$ for the input $X\sim\pi$.
Then every $g\in L^2(\pi)$ satisfies
\begin{equation}\label{eq:conditional-variance-bound}
 \E\var(g(X)\mid Y)
 \geq\frac1{20}\var_\pi(g).
\end{equation}
\end{lemma}
\begin{proof}
We first show that observing $Y$ leaves a fixed fraction of the variance of $\mathbf1_A(X)$ for every set $A$ of $\pi$-mass at most $1/2$ and the use a Cheeger inequality to generalize to all functions in $L^2(\pi)$.

Fix a set $A$ with $p=\pi(A)\in(0,1/2]$, and let $q(Y)=\P(X\in A\mid Y)$ be its posterior probability.
We show that
\begin{equation}\label{eq:set-localization}
 \E[q(Y)(1-q(Y))]\geq\frac45p(1-p).
\end{equation}

To prove \eqref{eq:set-localization}, we apply \eqref{eq:small-set-contraction} to the law $\pi_A$ of $X$ conditioned on $X\in A$, that is, $\mathrm{d}\pi_A/\mathrm{d}\pi=\mathbf1_A/p$.
It is supported on $A$, and $\KL(\pi_A\Vert\pi)=\E_{\pi_A}\log(1/p)=\log(1/p)$, so
\[
 \KL(\pi_AK\Vert\pi K)\leq\frac1{10}\log\frac1p.
\]
By Bayes' theorem,
\[
 \frac{\mathrm{d}(\pi_AK)}{\mathrm{d}(\pi K)}(Y)=\frac{\P(X\in A\mid Y)}{\P(X\in A)}=\frac{q(Y)}p,
\]
and hence
\[
 \chi^2(\pi_AK\Vert\pi K)=\E\Bigl(\frac{q(Y)}p-1\Bigr)^2=\frac{\var(q(Y))}{p^2}.
\]
Writing $\pi_{A^c}$ for the law of $X$ conditioned on $X\notin A$, we have $\pi=p\,\pi_A+(1-p)\,\pi_{A^c}$, so $\pi K=p\,\pi_AK+(1-p)\,\pi_{A^c}K\geq p\,\pi_AK$, and thus $\mathrm{d}(\pi_AK)/\mathrm{d}(\pi K)\leq1/p$.
Hence the bound of Sason and Verd\'u (\cref{fact:sason-verdu}) with $M=1/p$ gives
\[
 \chi^2(\pi_AK\Vert\pi K)\leq\frac{(1-p)^2}{p(\log(1/p)-1+p)}\KL(\pi_AK\Vert\pi K).
\]
Combining the above gives
\[
 \frac{\var(q(Y))}{p(1-p)}
 =\frac p{1-p}\chi^2(\pi_AK\Vert\pi K)
 \leq\frac{(1-p)\KL(\pi_AK\Vert\pi K)}{\log(1/p)-1+p}
 \leq\frac{(1-p)\log(1/p)}{10(\log(1/p)-1+p)}
 \leq\frac15.
\]
Here the denominator $\log(1/p)-1+p$ is positive, since $\log(1/p)>1-p$ for $p<1$, and the last step is equivalent to $h(p):=(1+p)\log(1/p)-2(1-p)\geq0$, which holds for $p\in(0,1]$ because $h(1)=0$ and $h'(p)=\log(1/p)+1-1/p\leq0$ by $\log w\leq w-1$ with $w=1/p$.
Finally, since $\E q(Y)=\P(X\in A)=p$ by the tower property,
\[
 \E[q(Y)(1-q(Y))]=\E q(Y)-\E q(Y)^2=p-p^2-\var(q(Y))\geq\frac45p(1-p),
\]
which proves \eqref{eq:set-localization}.

Next, we pass from sets to functions, by Cheeger's inequality for a Markov kernel built from the channel.
Let $X'$ be a fresh sample from the posterior, that is, $X'$ is conditionally independent of $X$ given $Y$, with the same conditional law as $X$.
Then the pair $(X,X')$ is exchangeable, so the kernel $x\mapsto\Law(X'\mid X=x)$ is reversible w.r.t.\ $\pi$.
Given $Y$, the variables $X$ and $X'$ are independent with the same law, so $\E[(g(X)-g(X'))^2\mid Y]=2\var(g(X)\mid Y)$, and the Dirichlet form of this kernel is
\[
 \frac12\E\bigl(g(X)-g(X')\bigr)^2=\E\var(g(X)\mid Y).
\]
The spectral gap of the kernel is the largest $\lambda\geq0$ such that the Dirichlet form is at least $\lambda\var_\pi(g)$ for every $g\in L^2(\pi)$.
For a set $A$, given $Y$ the events $\{X\in A\}$ and $\{X'\notin A\}$ are independent with probabilities $q(Y)$ and $1-q(Y)$, so the kernel moves from $A$ to $A^c$ with probability
\[
 \P(X\in A,X'\notin A)=\E[q(Y)(1-q(Y))].
\]
This probability is symmetric under exchanging $A$ and $A^c$, since $q(Y)(1-q(Y))$ is, so applying \eqref{eq:set-localization} to whichever of $A$ and $A^c$ has $\pi$-mass at most $1/2$ shows that every $A$ with $0<\pi(A)<1$ satisfies $\P(X\in A,X'\notin A)\geq\frac45\pi(A)\pi(A^c)$.
Hence the conductance of the kernel satisfies
\[
 \mathfrak c:=\inf_{A:\,0<\pi(A)<1}\frac{\P(X\in A,X'\notin A)}{\pi(A)\pi(A^c)}\geq\frac45.
\]
Cheeger's inequality for reversible Markov kernels \cite[Theorem~2.1]{LawlerSokal1988} bounds the spectral gap from below by $\mathfrak c^2/8$, so every $g\in L^2(\pi)$ satisfies
\[
 \E\var(g(X)\mid Y)
 =\frac12\E\bigl(g(X)-g(X')\bigr)^2
 \geq\frac{\mathfrak c^2}8\var_\pi(g)
 \geq\frac2{25}\var_\pi(g)
 \geq\frac1{20}\var_\pi(g).
\]
\end{proof}

\begin{proof}[Proof of \cref{lem:uniform-loss}]
As in \cref{lem:conditional-variance}, let $Y$ be the output of $K$ for the input $X\sim\pi$.
Write \[f=\mathrm{d}\zeta/\mathrm{d}\pi\]for the input likelihood ratio, and $g(Y)=\E[f(X)\mid Y]$.
Then $g=\mathrm{d}(\zeta K)/\mathrm{d}(\pi K)$ by the output likelihood ratio (\cref{lem:output-density}).

Since $\E_\pi f=\E_{\pi K}g=1$, we have $\KL(\zeta\Vert\pi)=\Ent_\pi(f)$ and $\KL(\zeta K\Vert\pi K)=\Ent_{\pi K}(g)$, so the left-hand side of \eqref{eq:uniform-loss} we are aiming to lower bound is the entropy loss
\[
 R:=\Ent_\pi(f)-\Ent_{\pi K}(g).
\]

Let
\[
 v=\var_\pi(\sqrt f)=1-(\E_\pi\sqrt f)^2.
\]We start with proving a useful inequality for what follows.
Since $\E_\pi\sqrt f\leq\sqrt{\E_\pi f}=1$ by Jensen's inequality, we have $v=(1-\E_\pi\sqrt f)(1+\E_\pi\sqrt f)\geq1-\E_\pi\sqrt f$, and therefore
\begin{equation}\label{eq:hellinger-v}
 \E_\pi(\sqrt f-1)^2=2(1-\E_\pi\sqrt f)\leq2v.
\end{equation}
Note that if $v=0$, then $f=1$ and the claim is trivial, so assume $v>0$.

\paragraph{Lower bounding the entropy loss} We first use \cref{lem:conditional-variance} to show that \[R\geq v/20.\]
Since $\pi$ is the mixture of the posteriors $\Law(X\mid Y=y)$ over $y\sim\pi K$, and $g(y)$ is the average of $f$ under the posterior, the entropy decomposition (\cref{lem:entropy-decomposition}) gives $\Ent_\pi(f)=\E\Ent(f(X)\mid Y)+\Ent_{\pi K}(g)$, where $\Ent(f(X)\mid Y)=\Ent_{\Law(X\mid Y)}(f)$ is the entropy of $f$ under the posterior, so
\[
 R=\E\Ent(f(X)\mid Y).
\]
To compare entropy with variance, let \[\varphi(z)=z\log z-z+1,\] which is nonnegative.
For $u\geq0$, we have $\varphi(u^2)-(u-1)^2=2u\varphi(u)\geq0$, so $\varphi(z)\geq(\sqrt z-1)^2$ for $z\geq0$.
Hence, for a nonnegative function $F$ and a probability law $\sigma$ with $m=\E_\sigma F>0$,
\[
 \Ent_\sigma(F)
 =m\E_\sigma\varphi(F/m)
 \geq m\E_\sigma\bigl(\sqrt{F/m}-1\bigr)^2
 =\E_\sigma\bigl(\sqrt F-\sqrt m\bigr)^2
 \geq\var_\sigma(\sqrt F),
\]
where the first equality is a direct computation, and the last step uses that $c=\E_\sigma\sqrt F$ minimizes $\E_\sigma(\sqrt F-c)^2$.
If $m=0$, then both $\Ent_\sigma(F)$ and $\var_\sigma(\sqrt F)$ vanish.
Applying this with $F=f$ and $\sigma=\Law(X\mid Y)$, and then \cref{lem:conditional-variance} to the function $\sqrt f\in L^2(\pi)$, gives
\begin{equation}\label{eq:loss-hellinger}
 R\geq\E\var(\sqrt{f(X)}\mid Y)\geq\frac1{20}\var_\pi(\sqrt f)=\frac v{20}.
\end{equation}

\paragraph{Upper bounding the relative entropy} Next, we focus on the right-hand side of \eqref{eq:uniform-loss} $\KL(\zeta\Vert\pi)$ which we aim to bound from above.

We do this by splitting $\KL(\zeta\Vert\pi)$ according to the size of $f$.
Since $\E_\pi f=1$, we have $\KL(\zeta\Vert\pi)=\E_\pi[f\log f]=\E_\pi\varphi(f)$.
For $z\geq0$,
\[
 \varphi(z)\leq\bigl(2+\max\{\log z,0\}\bigr)(\sqrt z-1)^2.
\]
Indeed, in terms of $u=\sqrt z$, the difference $D(u)=(2+2\max\{\log u,0\})(u-1)^2-\varphi(u^2)$ vanishes at $u=1$, while $D'(u)=4(u-1-u\log u)\leq0$ for $0<u<1$ and $D'(u)=4(u-1-\log u)+2(u-1)^2/u\geq0$ for $u>1$, where both signs follow from $\log w\leq w-1$.

In particular, it holds $\varphi(z)\leq22(\sqrt z-1)^2$ for $0\leq z\leq e^{20}$, so \eqref{eq:hellinger-v} gives
\[
 \E_\pi\bigl[\varphi(f)\mathbf1_{\{f<e^{20}\}}\bigr]
 \leq22\E_\pi(\sqrt f-1)^2
 \leq44v.
\]
For larger values of $f$, for \(j \ge 0\), let \[s_j=20\cdot2^j, \] and split $\{f\geq e^{20}\}$ into the bands
\[
 A_j=\{e^{s_j}\leq f<e^{2s_j}\},
\]
which are disjoint and cover $\{f\geq e^{20}\}$ since $s_{j+1}=2s_j$.
For any $j \geq 0,$ let \[r_j=\zeta(A_j)=\int_{A_j}f\,\mathrm{d}\pi \] be the mass that $\zeta$ puts on the band $A_j$.
On $A_j$, we have $f\geq1$ and $\log f<2s_j$, so $\varphi(f)\leq f\log f\leq2s_jf$, and therefore
\begin{equation}\label{eq:band-entropy}
 \KL(\zeta\Vert\pi)
 =\E_\pi\bigl[\varphi(f)\mathbf1_{\{f<e^{20}\}}\bigr]+\sum_{j\geq0}\E_\pi\bigl[\varphi(f)\mathbf1_{A_j}\bigr]
 \leq44v+2\sum_{j\geq0}s_jr_j.
\end{equation}

\paragraph{Bands of small $\zeta$-mass} We next show that the bands that carry little mass under $\zeta$ contribute little to \eqref{eq:band-entropy}.

To do this, we call a band indexed by $j$ \emph{``good''} if \[r_j\geq ve^{-s_j}\]and decompose based on this property the sum.

Since $e^{s/2}\geq1+s/2+s^2/8\geq s$ for $s\geq0$, we have $s_je^{-s_j}\leq e^{-s_j/2}$, and since $s_j\geq20(j+1)$, the remaining bands satisfy
\begin{equation}\label{eq:bad-bands}
 \sum_{j\text{ not good}}s_jr_j\leq v\sum_{j\geq0}s_je^{-s_j}\leq v\sum_{j\geq0}e^{-s_j/2}\leq v\sum_{j\geq0}e^{-10(j+1)}<v.
\end{equation}

\paragraph{Lower bounding the entropy loss on good bands} We now lower bound the entropy loss $R$ by contributions of the good bands.
For a good band $j$, let $\zeta_j$ be the law of $X\sim\zeta$ conditioned on $X\in A_j$, that is, $\mathrm{d}\zeta_j/\mathrm{d}\pi=f\mathbf1_{A_j}/r_j$, and let
\[
 B_j=\{y:g(y)\geq e^{s_j/2}\}.
\]
By the tower property, $\E[f(X)\log g(Y)]=\E[g(Y)\log g(Y)]$ and $\E f(X)=\E g(Y)$, so
\[
 R=\E[f(X)\log f(X)]-\E[g(Y)\log g(Y)]=\E\Bigl[f(X)\log\frac{f(X)}{g(Y)}-f(X)+g(Y)\Bigr].
\]
The integrand is nonnegative, because $a\log(a/b)-a+b=b\varphi(a/b)\geq0$ for $a\geq0$ and $b>0$.
Where $g(Y)=0$, we have $f(X)=0$ almost surely, since $\E[f(X)\mathbf1_{\{g(Y)=0\}}]=\E[g(Y)\mathbf1_{\{g(Y)=0\}}]=0$, so the integrand vanishes there with the convention $0\log(0/0)=0$.

On the event $\{X\in A_j,\,Y\notin B_j\}$, we have $f(X)\geq e^{s_j}$ and $g(Y)<e^{s_j/2}$, so $f(X)/g(Y)>e^{s_j/2}$.
Since $g(Y)\geq0$ and $s_j\geq20$, the integrand on this event satisfies
\[
 f(X)\log\frac{f(X)}{g(Y)}-f(X)+g(Y)
 \geq f(X)\Bigl(\log\frac{f(X)}{g(Y)}-1\Bigr)
 \geq f(X)\Bigl(\frac{s_j}2-1\Bigr)
 \geq\frac{s_j}4f(X).
\]
The events $\{X\in A_j,\,Y\notin B_j\}$ are disjoint, and since $\mathrm{d}\zeta_j/\mathrm{d}\pi=f\mathbf1_{A_j}/r_j$ and $Y$ has conditional law $K(X,\cdot)$ given $X$, we have $\E[f(X)\mathbf1_{A_j}(X)\mathbf1_{B_j^c}(Y)]=r_j(\zeta_jK)(B_j^c)$.
Hence
\begin{equation}\label{eq:good-bands-loss}
\begin{aligned}
 R
 &\geq\sum_{j\text{ good}}\E\Bigl[\Bigl(f(X)\log\frac{f(X)}{g(Y)}-f(X)+g(Y)\Bigr)\mathbf1_{A_j}(X)\mathbf1_{B_j^c}(Y)\Bigr]
 \geq\frac14\sum_{j\text{ good}}s_j\E\bigl[f(X)\mathbf1_{A_j}(X)\mathbf1_{B_j^c}(Y)\bigr]\\
 &=\frac14\sum_{j\text{ good}}s_jr_j(\zeta_jK)(B_j^c).
\end{aligned}
\end{equation}
It therefore remains to show that every good band $j$ satisfies
\begin{equation}\label{eq:band-missed}
 (\zeta_jK)(B_j^c)\geq\frac1{10}.
\end{equation}

\paragraph{The output likelihood ratio on good bands} Finally, we prove \eqref{eq:band-missed}, that is, for every good band $j$, if $X\sim\zeta_j$, then the output likelihood ratio $g(Y)$ is below $e^{s_j/2}$ with probability at least $1/10$.

Fix a good band $j$.
The law $\zeta_j$ has density at most $e^{2s_j}/r_j$, so it has finite entropy.
Moreover, since $f\geq e^{s_j}$ on $A_j$, we have $r_j\geq e^{s_j}\pi(A_j)$, so $\pi(A_j)\leq e^{-s_j}r_j\leq e^{-20}<1/2$ by $r_j\leq1$ and $s_j\geq20$, and \eqref{eq:small-set-contraction} gives
\begin{equation}\label{eq:band-contraction}
 \KL(\zeta_jK\Vert\pi K)\leq\frac1{10}\KL(\zeta_j\Vert\pi).
\end{equation}
Moreover, since $f<e^{2s_j}$ on $A_j$ and $r_j\geq ve^{-s_j}$,
\[
 \KL(\zeta_j\Vert\pi)
 =\E_{\zeta_j}\log\frac{f\mathbf1_{A_j}}{r_j}
 \leq\log\frac{e^{2s_j}}{r_j}
 =2s_j+\log\frac1{r_j}
 \leq3s_j+\log\frac1v.
\]
On the other hand, Jensen's inequality gives $\sqrt{g(Y)}=\sqrt{\E[f(X)\mid Y]}\geq\E[\sqrt{f(X)}\mid Y]$, so $\E_{\pi K}\sqrt g\geq\E_\pi\sqrt f$, and thus, as in \eqref{eq:hellinger-v},
\[
 \E_{\pi K}(\sqrt g-1)^2=2(1-\E_{\pi K}\sqrt g)\leq2(1-\E_\pi\sqrt f)\leq2v.
\]
Since $e^{s_j/4}\geq e^5\geq2$, we have $e^{s_j/4}-1\geq e^{s_j/4}/2$, and Markov's inequality gives
\[
 (\pi K)(B_j)
 =(\pi K)\bigl(\sqrt g-1\geq e^{s_j/4}-1\bigr)
 \leq(\pi K)\bigl(\sqrt g-1\geq e^{s_j/4}/2\bigr)
 \leq\frac{\E_{\pi K}(\sqrt g-1)^2}{(e^{s_j/4}/2)^2}
 \leq8ve^{-s_j/2}.
\]
Now suppose, for contradiction, that $q_j=(\zeta_jK)(B_j)>9/10$, and let $p_j=(\pi K)(B_j)\leq8ve^{-s_j/2}$.
The map $y\mapsto\mathbf1_{B_j}(y)$ sends $\zeta_jK$ to $\Bern(q_j)$ and $\pi K$ to $\Bern(p_j)$, so the data processing inequality (\cref{lem:data-processing}) gives
\begin{align*}
 \KL(\zeta_jK\Vert\pi K)
 &\geq\KL\bigl(\Bern(q_j)\Vert\Bern(p_j)\bigr)
 =q_j\log\frac1{p_j}+(1-q_j)\log\frac1{1-p_j}-\Hb(q_j)
 \geq\frac9{10}\log\frac1{p_j}-\log2\\
 &\geq\frac9{10}\Bigl(\log\frac1v+\frac{s_j}2-\log8\Bigr)-\log2
 >\frac1{10}\Bigl(3s_j+\log\frac1v\Bigr)
 \geq\frac1{10}\KL(\zeta_j\Vert\pi),
\end{align*}
where the second inequality uses $q_j>9/10$, $(1-q_j)\log(1/(1-p_j))\geq0$, and $\Hb\leq\log2$, and the fourth inequality holds because the difference of its two sides is
\[
 \frac45\log\frac1v+\frac3{20}s_j-\frac{37}{10}\log2
 \geq3-\frac{37}{10}\log2
 >0,
\]
by $v\leq1$, $s_j\geq20$, and $\log2<7/10$.
This contradicts \eqref{eq:band-contraction}, which proves \eqref{eq:band-missed}.

Together with \eqref{eq:good-bands-loss}, \eqref{eq:band-missed} gives
\[
 R\geq\frac1{40}\sum_{j\text{ good}}s_jr_j.
\]

\medskip
As a result, combining the bounds above gives
\[
 \KL(\zeta\Vert\pi)
 \leq44v+2\sum_{j\text{ not good}}s_jr_j+2\sum_{j\text{ good}}s_jr_j
 \leq46v+80R
 \leq46\cdot20R+80R
 =10^3R,
\]
which is \eqref{eq:uniform-loss}.
\end{proof}

\subsection{Proof of \texorpdfstring{\cref{lem:null-log-concavity}}{Lemma \getrefnumber{lem:null-log-concavity}}}\label{app:null-law}
\begin{proof}
By \cref{lem:F-bounds}, the log-likelihood ratio $F_t=\log(\mathrm{d}\mu/\mathrm{d}\nu_t)$ is finite, smooth, and convex.
Writing $\mu(\mathrm{d}x)=e^{-V(x)}\mathrm{d}x$ with $\nabla^2V\succeq0$, we have $\nu_t(\mathrm{d}x)=e^{-F_t(x)}\mu(\mathrm{d}x)=e^{-W(x)}\mathrm{d}x$ with $W=V+F_t$, and $\nabla^2W=\nabla^2V+\nabla^2F_t\succeq0$.
\end{proof}

\begingroup
\emergencystretch=1em
\printbibliography
\endgroup
\end{document}